%% file: main_selfren.tex
\documentclass[prd,aps,twocolumn,preprintnumbers, showpacs, nofootinbib,superscriptaddress,notitlepage]{revtex4-1}

\usepackage{graphicx} 
\usepackage{amsmath}
\usepackage{color,xcolor}
\usepackage{array}
\usepackage{amssymb, amsthm}
\usepackage{amsmath}    
\usepackage{mathrsfs}   
\usepackage{graphicx}   
\usepackage{colortbl}
\usepackage{footnote}   
\usepackage{slashed}    
\usepackage{subfiles}
\usepackage[capitalize]{cleveref}

\begin{document}
\title{Parton distribution function of a deuteronlike dibaryon system from lattice QCD}


\author{\includegraphics[scale=0.2]
{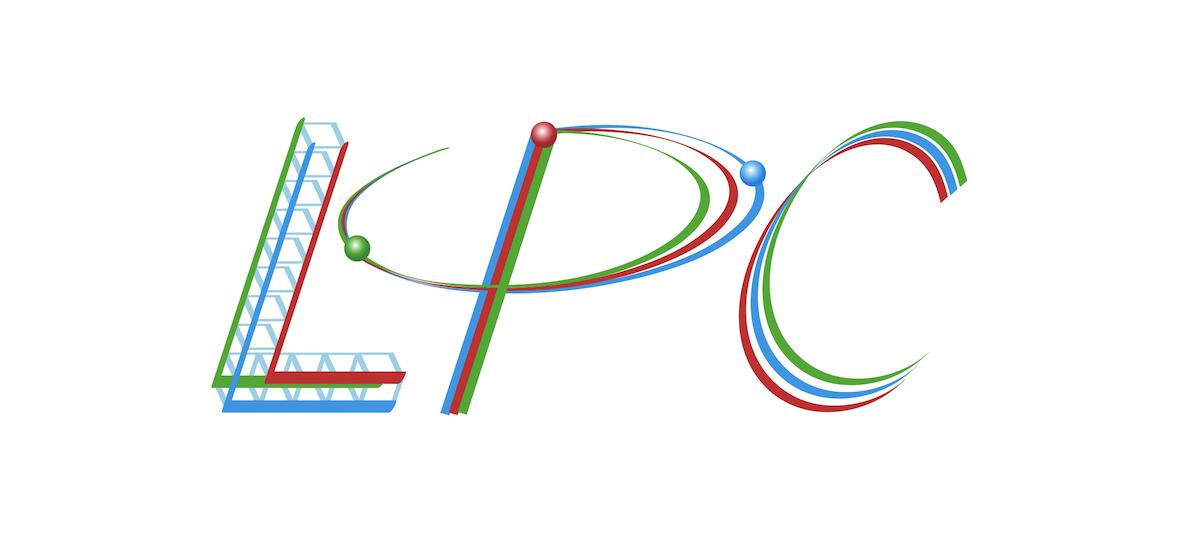}\\Chen Chen}
\affiliation{Institute of Modern Physics, Chinese Academy of Sciences, Lanzhou 730000, China}
\affiliation{University of Chinese Academy of Sciences, School of Nuclear Science and Technology, Beijing 100049, China}

\author{Yiqi Geng}
\affiliation{Institute of Modern Physics, Chinese Academy of Sciences, Lanzhou 730000, China}
\affiliation{Nanjing Normal University, Nanjing, Jiangsu 210023, China}

\author{Liuming Liu}
\email[Corresponding author: ]{liuming@impcas.ac.cn}
\affiliation{Institute of Modern Physics, Chinese Academy of Sciences, Lanzhou 730000, China}
\affiliation{University of Chinese Academy of Sciences, School of Nuclear Science and Technology, Beijing 100049, China}
\author{Peng Sun}
\email[Corresponding author: ]{pengsun@impcas.ac.cn}
\affiliation{Institute of Modern Physics, Chinese Academy of Sciences, Lanzhou 730000, China}
\affiliation{University of Chinese Academy of Sciences, School of Nuclear Science and Technology, Beijing 100049, China}

\author{Yi-Bo Yang}
\email[Corresponding author: ]{ybyang@itp.ac.cn}
\affiliation{University of Chinese Academy of Sciences, School of Physical Sciences, Beijing 100049, China}
\affiliation{CAS Key Laboratory of Theoretical Physics, Institute of Theoretical Physics, Chinese Academy of Sciences, Beijing 100190, China}
\affiliation{School of Fundamental Physics and Mathematical Sciences, Hangzhou Institute for Advanced Study, UCAS, Hangzhou 310024, China}
\affiliation{International Centre for Theoretical Physics Asia-Pacific, Beijing/Hangzhou, China}

\author{Fei Yao}
\affiliation{Center of Advanced Quantum Studies, Department of Physics, Beijing Normal University, Beijing 100875, China}
\affiliation{School of Science and Engineering, The Chinese University of Hong Kong, Shenzhen 518172, China}

\author{Jian-Hui Zhang}
\email[Corresponding author: ]
{zhangjianhui@cuhk.edu.cn}
\affiliation{School of Science and Engineering, The Chinese University of Hong Kong, Shenzhen 518172, China}


\author{Kuan Zhang}
\affiliation{University of Chinese Academy of Sciences, School of Physical Sciences, Beijing 100049, China}
\affiliation{CAS Key Laboratory of Theoretical Physics, Institute of Theoretical Physics, Chinese Academy of Sciences, Beijing 100190, China}

\begin{abstract}
We report a lattice QCD calculation of the parton distribution function (PDF) of a deuteronlike dibaryon system using large-momentum effective theory. The calculation is done on three Wilson-clover ensembles with a fixed lattice spacing $a=0.105$~fm and two pion masses. The lattice matrix elements are computed at proton momenta up to 2.46~GeV, with the signal of high-momentum modes being improved by applying the momentum smearing technique. The state-of-the-art renormalization, matching, and extrapolation are then applied to obtain the final result of the light-cone PDF. A comparison between the result of the dibaryon system and the sum of the proton and neutron PDFs is also given. 
\end{abstract}

\maketitle

\section{Introduction}
Parton distribution functions (PDFs) are widely applied in high-energy physics. On the one hand, they are necessary inputs for computing cross sections in electron-hadron or hadron-hadron collisions. On the other hand, they contain important information about the intrinsic properties of nucleons and nuclei. To date, the PDFs of single-baryon systems like the proton or neutron have been extensively studied both experimentally and theoretically.
In the meantime, the PDFs of multibaryon systems like the deuteron and $^3$He have also attracted a lot of attention. Comparing the structure functions between single-baryon and multibaryon systems reveals important insights into the interaction between baryons within a nucleus. For example, the European Muon Collaboration (EMC) first observed in 1983 that the structure functions of the nucleus differ from the sum of the structure functions of individual nucleons inside the nucleus~\cite{EuropeanMuon:1983wih}, which is now known as the ``nuclear EMC effect.'' It can be characterized by the ratio 
\begin{align}
 R(D)=\frac{F^D_2}{F^p_2+F^n_2} ,
\end{align}
with $F_2(x)=x\sum\limits_{i} f_i(x)Q_i^2 $ at the leading order,  where $f_i(x)$ is the PDF of the $i$-th quark flavor. The observation of the EMC effect has opened up new avenues of research in nuclear physics, leading to extensive further study and investigation through many experiments~\cite{ASHMAN1988603,PhysRevLett.50.1431,Accardi:2016qay,Cocuzza:2021rfn,Segarra:2020gtj}. 
On the theoretical side,  substantial efforts have also been made to constrain  nucleon-nucleon interactions by calculating basic properties of multinucleon systems using the lattice QCD (LQCD) approach~\cite{Amarasinghe:2021lqa,Detmold:2020snb,Illa:2021clu,Beane:2006mx,Beane:2006gf,NPLQCD:2010ocs,Beane:2011zpa,NPLQCD:2011naw,NPLQCD:2012mex,Orginos:2015aya,Wagman:2017tmp,Fukugita:1994ve,Yamazaki:2011nd,Yamazaki:2012hi,Yamazaki:2015asa,Nemura:2008sp,Inoue:2011pg,Berkowitz:2015eaa,Francis:2018qch,Junnarkar:2019equ,Horz:2020zvv,NPLQCD:2020lxg,Green:2021qol,HALQCD:2018qyu,Murakami:2022cez}. Among them, the NPLQCD group 
has pioneered the calculation of moments of the proton, diproton, and  $^3$He  systems~\cite{Detmold:2020snb, Illa:2021clu}.

In addition, recent theoretical developments~\cite{Braun:2007wv,Ji:2013dva,Ji:2014gla,Ma:2017pxb,Lin:2017snn,Radyushkin:2017cyf} have also enabled the direct calculation of the Bjorken $x$-dependence of PDFs from LQCD. In the past few years, various LQCD calculations for nucleon PDFs have been done~\cite{LatticeParton:2018gjr,Chen:2018xof,Lin:2018pvv,Lin:2017ani,Liu:2018hxv,LatticeParton:2022xsd, Orginos:2017kos,Alexandrou:2015rja,Alexandrou:2020zbe,Gao:2022uhg,Karpie:2021pap,HadStruc:2022nay}, based on either the large-momentum effective theory(LaMET)~\cite{Ji:2013dva,Ji:2014gla,Ji:2020ect} or the short-distance expansion (pseudo-PDF) approach~\cite{Radyushkin:2017cyf}. 
Applying such calculations to multibaryon systems will help to further elucidate the origin of nuclear effects. 

The deuteron is the simplest multibaryon bound system. It is composed of a proton and a neutron and has a small binding energy  
of $2.224575(9)$ MeV~\cite{VanDerLeun:1982bhg}. 
Such a low binding energy poses a significant challenge for identifying a bound-state deuteron through LQCD calculations of two-point correlation functions.  
Utilizing three-point correlation functions, such as those employed in the calculation of PDFs faces the same problem. However, it could still serve as a potentially useful and complementary tool to study nuclear effects,  as revealed by the EMC effect.  

In this work, we calculate the PDF of a deuteronlike dibaryon system using the LaMET approach. The ultraviolet (UV) divergences are removed by a nonperturbative renormalization in the hybrid scheme~\cite{Ji:2020brr}, which has been widely used in previous studies~\cite{LatticeParton:2022xsd,LatticeParton:2022zqc,Hua:2020gnw}.
While only one single lattice spacing is considered in this work, we perform our calculations on ensembles with two different pion masses at $941$ MeV and $293$ MeV. 
 To assess the impact of finite-volume effects, we conduct simulations at the lower pion mass of $293$ MeV on two sets of ensembles with lattice sizes of  $32^3\times64$ and $24^3\times72$. In order to study the nuclear effect, we make a comparison between the PDF of the dibaryon system and the sum of the proton and neutron PDFs. 

The rest of the paper is organized as follows. In Sec.~\ref{section B}, we describe the calculation of bare matrix elements in lattice simulations. In Sec.~\ref{section 
C}, we outline the procedure of renormalization and matching. In Sec.~\ref{section D}, we present the numerical results of the $x$-dependent PDF and the ratio of the dibaryon system PDF and the sum of the proton and neutron PDFs. Finally, a summary is given in Sec.~\ref{section E}.

\section{Lattice Simulation}
\label{section B}

\subsection{Lattice setup}
Throughout the paper, we use the $2+1$ flavor QCD ensembles with stout-smeared clover fermion action and Symanzik gauge actions generated by  CLQCD collaboration~\cite{Hu:2023jet,Du:2024wtr}. Three ensembles, named  C32P29, C24P29 and  C24P90 are used to calculate the bare matrix elements of the nucleon and dibaryon systems. The detailed information is given  in Table~\ref{Tab:setup}.

\renewcommand{\arraystretch}{1.5}
\begin{table}
\footnotesize
\centering
\begin{tabular}{cclcccccc}

\hline
\hline
Ensemble ~&$a$(fm) ~& \ \!$L^3\times T$  ~& $m_\pi$(MeV) ~& $m_\pi L$  ~& $m_{\eta_s}$(MeV)  \\

\arrayrulecolor{black}

\hline

C32P29  ~& 0.105~& $32^3\times 64$  ~& 292.9(1.2)  ~&4.9   &659.1(1.3)            \\

\hline
C24P29  ~& 0.105~& $24^3\times 72$  ~& 293.1(1.3)  ~&3.7 &659.7(1.3)                  \\

\hline
C24P90  ~& 0.105~& $24^3\times 72$  ~& 940.7(1.2)  ~&12.0 ~& 940.7(1.2)                  \\

\arrayrulecolor{gray!50}

\arrayrulecolor{black}
\hline

\end{tabular}
 \caption{The simulation setup, including lattice spacing $a$, lattice size $L^3\times T$ , $\pi$ mass $m_{\pi}$ and $\eta_s$ mass $m_{\eta_s}$.}
 \label{Tab:setup}
\end{table}

\subsection{Light-cone PDF vs quasi-PDF }

The unpolarized PDF is defined as
\begin{align}
 q(x,\mu)=\int \frac{d\xi^-}{4\pi} e^{-ix P^+ \xi^-} \langle P|\bar{\psi}(\xi^-)\gamma^{+} 
U(\xi^-,0) \psi(0)|P \rangle,
\end{align}
where $|P\rangle$ denotes the ground state of a nucleon or a multinucleon system with momentum $P$, $x$ is the momentum fraction carried by the quark, $\mu$ is the renormalization scale and $\xi^\pm=(\xi^t\pm \xi^z)/\sqrt{2}$ are light-cone coordinates. The Wilson line along the light-cone direction 
\begin{align}
U(\xi^-,0)= \mathcal{P}\,{\rm exp}\big[ig\int_{0}^{\xi^-} du ~ {n}\cdot A(un)\big]
\end{align}
is introduced to ensure gauge invariance of the nonlocal quark bilinear correlator.


The PDF involves real-time dependence and cannot be directly accessed on the lattice. However, according to LaMET, one can start from the calculation of the quasi-PDF, which takes the following form:
\begin{align}
 \tilde{q}\left(x, P_z\right)=\int \frac{dz}{4\pi} e^{-ixz P_z } \langle P|\bar \psi(z) \Gamma U(z,0) \psi(0)| P \rangle,
\end{align}
where the matrix element on the rhs is often called the quasi-light-front (quasi-LF) correlation. $P_z$ represents the momentum of a nucleon or a multinucleon system along the $z$ direction. $\Gamma$ can be chosen as $\gamma_t$ or $\gamma_z$ for unpolarized quark quasi-PDFs. In this work, we choose $\Gamma=\gamma_t$, which is free from operator mixing with a scalar quark quasi-PDF operator~\cite{Constantinou:2017sej, Green:2017xeu,Chen:2017mie}. 
The relation between the quasi-PDF and the light-cone PDF is given by a factorization formula

\begin{align}\label{mtcheq}
 \tilde{q}\left(x, P_z\right)= & \int_{-1}^1 \frac{d y}{|y|} C\left(\frac{x}{y}, \frac{\mu}{y P_z}\right)  q(y, \mu)  \nonumber \\
& +\mathcal{O}\left(\frac{\Lambda_{\mathrm{QCD}}^2}{\left(x P_z\right)^2}, \frac{\Lambda_{\mathrm{QCD}}^2}{\left((1-x) P_z\right)^2}\right) ,
\end{align}
where $q(y, \mu)$ is the light-cone PDF depending on the renormalization scale, $C\left(\frac{x}{y}, \frac{\mu}{y P_z}\right)$ is the matching kernel relevant to the longitudinal partonic momentum 
 ~\cite{Izubuchi:2018srq},  and $\mathcal{O}\left(\frac{\Lambda_{\mathrm{QCD}}^2}{\left(x P_z\right)^2}, \frac{\Lambda_{\mathrm{QCD}}^2}{\left((1-x) P_z\right)^2}\right)$ denote higher-twist contributions suppressed by $P_z$.

\subsection{Two-point and three-point correlators }
The momentum-projected two-point correlator $C^{\text{2pt}}(\vec{P};t_{sep})$ and three-point correlator $C_{\Gamma}^{\text{3pt}}(\vec{P};t,t_{sep})$ can be defined as 
\begin{align}\label{eq:correlator}
C^{\text{2pt}}(\vec{P};t)&=\langle0\mid N_{\sigma }(\vec{P};t)  N'^{\dagger}_{{\sigma}}(\vec{P};0) \mid0\rangle,  \\     \nonumber 
C_{\Gamma}^{\text{3pt}}(\vec{P};z;t,t_{sep})&=\langle0\mid  N_{\sigma } (\vec{P};t_{sep}) O_{\Gamma}(z;t)  N'^{\dagger}_{{\sigma}}(\vec{P};0)\mid0  \rangle.
\end{align}
Here, $N_{\sigma }(\vec{P},t) $ denotes the nucleon or multinucleon interpolating operators projected to certain momentum $\vec{P}$,  $t_{sep}$ denotes the time interval between the source and the sink, with the source being placed at $t=0$, and $\sigma$ labels the rows of a certain irreducible representation, under which the operator transforms.  $O_{\Gamma}(z;t)=\bar{\psi}(z;t)\Gamma U(z,0;t)\psi(0;t)$, where $\Gamma=\gamma_t$ is the current operator located at time $t$ with quark  and antiquark separated along the $z$ direction. 

We construct the interpolating operators of  the nucleon and deuteron following the method proposed in Ref.~\cite{Amarasinghe:2021lqa}. Since a nucleon has quantum numbers $J^P = \frac{1}{2}^+$, the operator should transform in the $G_1^+$ irrep  of the cubic group. The proton interpolating operator can be written as
\begin{align}
\begin{split}
  p_{\sigma}(\vec{x};t) = & \varepsilon^{a b c}   u_{\zeta}^{a}(\vec{x};t) \left( C \gamma_{5} P_{+}  \right)_{\zeta \xi} d_{\xi}^{b}(\vec{x};t) \\ & \times	\left( P_{+} \right)_{\sigma \rho} u_{\rho}^{c}(\vec{x};t), 
\end{split}
 \label{eq:Ninterp}
\end{align}
where $C$ is the charge conjugation matrix and $P_+$ is the positive-parity projection operator $P_{+}=\frac{1+\gamma_4}{2}$. Note that the gamma matrices here are in Dirac-Pauli representation and $\sigma \in \{0,1\}$. Neutron operator $n_{\sigma}(\vec{x};t)$ can be obtained by swapping  $u$ and $d$ in Eq. (\ref{eq:Ninterp}). 

In order to efficiently compute the Wick contractions of the correlation functions, one can rewrite the nucleon interpolating operators above as the following~\cite{Amarasinghe:2021lqa}: 
\begin{align}
p_{\sigma}(\vec{x};t) &= \sum_{\alpha} w_\alpha^{[N]\sigma} u^{i(\alpha)}(\vec{x};t) d^{j(\alpha)}(\vec{x};t) u^{k(\alpha)}(\vec{x};t)  ,\nonumber \\
 n_{\sigma}(\vec{x};t) &= \sum_{\alpha} w_\alpha^{[N]\sigma} d^{i(\alpha)}(\vec{x};t) u^{j(\alpha)}(\vec{x};t) d^{k(\alpha)}(\vec{x};t), 
  \label{eq:Ninterp_qk}
\end{align}
where quark fields are labeled with indices $i,j,k,...,$ which are a combination of spinor and color indices. For example, $i=(\zeta, a)$ with spinor index $\zeta\in\{0,1,2,3\}$ and color index $a \in\{0,1,2\}$, and $w_\alpha^{[N]\sigma}$ denotes the weights of different spin-color combinations, with $\alpha \in \{ 
 1,..., {\cal{N}}_\omega^{ \left[ N\right]} \}$ running over the ${\cal{N}}_\omega^{ \left[ N\right]}$ combinations of $u^{i\left(\alpha\right)}d^{j\left(\alpha\right)}u^{k\left(\alpha\right)}$. For each
$\sigma$, only $12$ terms appear due to the sparseness of gamma matrices and antisymmetric tensor $\epsilon^{abc}$, which significantly reduces the computation time for contractions.

Nucleon operators making up momentum-projected correlators in Eq. (\ref{eq:correlator}) can be obtained by the Fourier transform 
\begin{align}
p_{\sigma}(\vec{P};t)=\sum_{\vec{x} \in{\Lambda}} e^{i\vec{P} \cdot \vec{x}} p_{\sigma}(\vec{x};t), \nonumber \\
n_{\sigma}(\vec{P};t)=\sum_{\vec{x} \in{\Lambda}} e^{i\vec{P} \cdot \vec{x}} n_{\sigma}(\vec{x};t).
\end{align}
In our calculation, $\vec{P}=(0,0,(2 \pi n)/{L})$ is directed along the $z$-axis. In principle, the operators should be summed over all spatial lattice sites for both the sink and the source. However, this approach would necessitate the calculation of all-to-all propagators, which is infeasible given our limited computational resources.  As a practical compromise, we sum over all spatial lattice sites for the sink and a few points evenly distributed in the $z=0$ plane for the source.  Specifically, we contract point-to-all propagators with sources fixed at one point, enhance the measurements by shifting the source position, and repeat the calculation. This method is also applied to the calculation of deuteronlike dibaryon correlators.

Deuteron has quantum numbers $I(J^P) = 0(1^+)$. Therefore, we build the following two-nucleon interpolating operators:
\begin{align}
\begin{split}
D_1(\vec{P};t)&=\sum_{\vec{x}_{1},\vec{x}_{2} \in \Lambda} e^{i\frac{\vec{P}}{2} \cdot \left(\vec{x}_{1}+\vec{x}_{2}\right)}\\& \frac{1}{\sqrt{2}}\left[p_{0 }(\vec{x}_{1};t) n_{0 }(\vec{x}_{2};t)-n_{0 }(\vec{x}_{1};t) p_{0 }(\vec{x}_{2};t)\right]  ,  \\ 
 D_2(\vec{P};t)&=\sum_{\vec{x}_{1},\vec{x}_{2} \in \Lambda} e^{i\frac{\vec{P}}{2} \cdot \left(\vec{x}_{1}+\vec{x}_{2}\right)} \\& \frac{1}{2}\left[p_{0 }(\vec{x}_{1};t) n_{1 }(\vec{x}_{2};t)+p_{1 }(\vec{x}_{1};t) n_{0 }(\vec{x}_{2};t)\right.  \\
	&\left.-n_{0 }(\vec{x}_{1};t) p_{1 }(\vec{x}_{2};t)-n_{1 }(\vec{x}_{1};t) p_{0 }(\vec{x}_{2};t)\right]  ,\\ 
 D_3(\vec{P};t)&=\sum_{\vec{x}_{1},\vec{x}_{2} \in \Lambda} e^{i\frac{\vec{P}}{2} \cdot \left(\vec{x}_{1}+\vec{x}_{2}\right)} \\& \frac{1}{\sqrt{2}}\left[p_{1 }(\vec{x}_{1};t) n_{1 }(\vec{x}_{2};t)-n_{1 }(\vec{x}_{1};t) p_{1 }(\vec{x}_{2};t)\right]  ,
\end{split} 
\label{eq:D123}
\end{align}
where the subscripts of  $D_{\{1,2,3\}}$ represent the three components of $J=1$, and $\vec{P}$ is the center-of-mass momentum of the system.  
For simplicity, we only consider the case in which both the proton and the neutron carry half of the total momentum. $\vec{x}_{1}$ and $\vec{x}_{2}$ are space coordinates of the two nucleons. These operators are used at the sink, while at the source, the two nucleons are put at the same space position since we compute point-to-all quark propagators, as explained above. 

The quark-level representation of dibaryon operators can be constructed by inserting Eq.~(\ref{eq:Ninterp_qk})
 into Eq.~(\ref{eq:D123}),

\begin{align}
D_{\rho}(\vec{P};t)&=\sum_{\vec{x}_{1},\vec{x}_{2} \in \Lambda} e^{i\frac{\vec{P}}{2} \cdot \left(\vec{x}_{1}+\vec{x}_{2}\right)} \nonumber\\
&\sum_{\alpha}  w_\alpha^{[D]\rho}  u^{i(\alpha)}_{g}(\vec{x}_1,t) d^{j(\alpha)}_{g}(\vec{x}_1,t) u^{k(\alpha)}_{g}(\vec{x}_1,t) \nonumber\\
&    \times  d^{l(\alpha)}_{g}(\vec{x}_2,t) u^{m(\alpha)}_{g}(\vec{x}_2,t) d^{n(\alpha)}_{g}(\vec{x}_2,t) .
\end{align}
 where the weights $w_\alpha^{[D]\rho}$, with $\alpha \in \{1,\ldots,\mathcal{N}_w^{[D]\rho}\}$, with $\mathcal{N}_w^{[D]1} = \mathcal{N}_w^{[D]3} = 144$ and $\mathcal{N}_w^{[D]2} = 288$, are obtained from products of weights of nucleon $w_\alpha^{[N]\sigma}$, as shown in ~\cite{Amarasinghe:2021lqa}.
 
\begin{figure}[tbp]
\includegraphics[width=.32\textwidth]{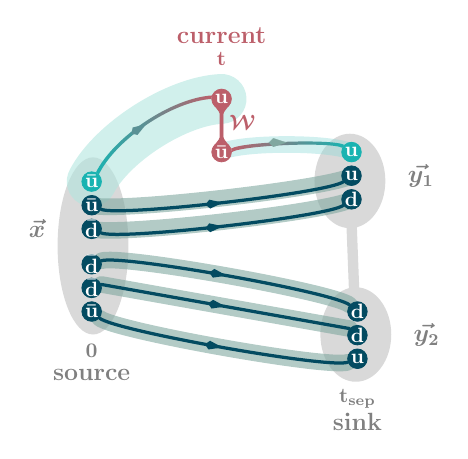}
\caption{ Illustration of the three-point correlator of a deuteronlike system before momentum projection (one possible way of contraction). }
\label{fig:eq-time corr}
\end{figure}

The three-point correlation function can be obtained by summing over all possible contractions of connected diagrams. Our calculation does not account for the disconnected diagram contributions, which are not expected to have a significant impact on our qualitative findings~\cite{Alexandrou:2021oih}.
 In Fig.~\ref{fig:eq-time corr} we give an illustration of one possible contraction of the three-point correlator of the deuteronlike dibaryon system.    

Three ensembles, namely, C32P29, C24P29,  and  C24P90, are used to calculate two-point and three-point correlators of the deuteronlike dibaryon system and the nucleon. The calculations are performed with the nucleon carrying different momenta $P_z^{n}=(2\pi n)/L$ 
 along the $z$ direction, and the total momentum of the dibaryon system is $2P_z^{n}$ .  For each momentum, several sets of source-sink separations $t_{sep}$ are calculated.    For larger momenta, correlators are averaged over several sources evenly distributed on the $z=0$ plane to control the statistical uncertainties.  The detailed information is shown in Table~\ref{Tab:qgME}.

\renewcommand{\arraystretch}{1.5}
\begin{table}
\footnotesize
\centering
\begin{tabular}{ccccc}

\hline
Ensemble ~&$P_z^{n}$(GeV) ~& $t_{sep}/a$  ~& $N_{conf}$ ~& $nsrc(time \times space) $ \\
\arrayrulecolor{black}

\hline

C32P29  ~&1.48 ~& {6,7,8}  ~ &$870$ ~& $4 \times 4$                 \\
  ~&1.85 ~& {6,7,8}  ~ & ~& $4 \times 4$     \\
    ~&2.21 ~& {6,7,8}  ~ & ~& $4 \times 4$     \\

\arrayrulecolor{gray!50}

\hline
\arrayrulecolor{gray!50}
C24P29  ~&1.48 ~& {6,7,8} ~&$759$   ~& $3 \times 2$                 \\
 ~&1.97 ~& {6,7,8} ~&    ~& $3 \times 8$    \\
  ~&2.46 ~& {6,7,8} ~&    ~& $6 \times 8$    \\

\hline

C24P90  ~&1.48 ~& {6,7,8,9,10}  ~ &$750$ ~& $3 \times 1$                 \\
  ~&1.97 ~& {6,7,8,9}  ~ & ~& $3 \times 1$                 \\
  ~&2.46 ~& {6,7,8}  ~ & ~& $3 \times 4$         \\

\arrayrulecolor{black}
\hline

\end{tabular}
 \caption{Information on the  nucleon's momentum $P_z^{n}$ (the center-of-mass momentum of  the dibaryon system is $2P_z^{n}$), time separation between source and sink($t_{sep}$) and number of sources($nsrc$) on the $z=0$ plane used to calculate two-point and three-point correlators. }
 \label{Tab:qgME}
\end{table}

To improve the signal of high-momentum states, we applied momentum-Gaussian smearing on sources and sinks~\cite{Bali:2016lva}, which takes the  form

\begin{align}
    \psi(x) &\to S_{mom}\psi(x) \nonumber \\
    &=(1-\alpha) \psi(x)+\frac{\alpha}{6}\sum_{j}U_{j}(x)e^{ik\hat{e}_{j}}\psi(x+\hat{e}_{j}) \nonumber \\
    &=(1+\frac{\alpha}{6}a^2D_jD_j)\psi(x)+\mathcal{O}(a^3),
\end{align} 
where $k$ represents the momentum smearing parameter and
$U_{j}$ denotes a stout-smeared gauge link along a specified direction. The above operation should be iterated by $n_{iter}=\mathcal{O}(2 \sigma ^2)$ times to achieve the desired smearing size $\sigma a$, which corresponds to $n_{iter}=50$ and $\sigma a=0.5$~fm in our calculation. 
 This process shifts the smearing center to momentum  $\mathcal{O}(k)$, thereby significantly improving the signal of high-momentum states. In our calculation we set $k$ equal to half of the nucleon momentum,  which is the optimized choice to improve the signal according to previous experience~\cite{Bali:2016lva}. In addition, we applied one-step hypercubic(HYP) smearing on the Wilson link of the current operator.

\subsection{Joint fitting}

The two-point correlator (2pt) $C^{2\text{pt}}(P_z,t)$ and three-point correlator (3pt) $C_\Gamma^{3\text{pt}} (P_z, t, t_\text{sep})$   can be decomposed as
\begin{align}
\label{eq:twostate}
C^\text{2pt}(P_z;t) &=|{\cal A}_0|^2 e^{-E_0t}+|{\cal A}_1|^2 e^{-E_1t}+\cdots,  \nonumber \\ 
C^\text{3pt}_{\Gamma}(P_z;z;t,t_\text{sep}) &=
   |{\cal A}_0|^2 \langle 0 | {O}_\Gamma(z;t) | 0 \rangle  e^{-E_0t_\text{sep}} \nonumber \\
   &+|{\cal A}_1|^2 \langle 1 | {O}_\Gamma(z;t)  | 1 \rangle  e^{-E_1t_\text{sep}} \nonumber \\
   &+{\cal A}_1{\cal A}_0^* \langle1 | {O}_\Gamma(z;t)  | 0 \rangle  e^{-E_1 (t_\text{sep}-t)} e^{-E_0 t} \nonumber\\
   &+{\cal A}_0{\cal A}_1^* \langle 0 | {O}_\Gamma (z;t) | 1 \rangle  e^{-E_0 (t_\text{sep}-t)} e^{-E_1 t}+\cdots ,
\end{align}
where  $\left\langle 0 \left| \mathcal{O}_\Gamma \right|  0\right\rangle$ are the desired ground-state quasimatrix elements, $| n \rangle$ with $n>0$ represents the excited states, ${\cal A}_{0,1}$ are amplitudes depending on smearing parameters, $E_{0,1}$ are the ground-state and excited-state energy, respectively, and the ellipse denotes the contribution from $n>1$ excited states which decays faster than the ground and first excited states.

The above expressions can be further simplified into the fitting form

\begin{align}
   &C^\text{2pt}(P_z,t)\approx c_4e^{-E_0 t}(1+c_5 e^{-\Delta E t})\,,\nonumber \\
  & R^{\text{3pt}}_{\Gamma}(P_z,t,t_{\mathrm{sep}})  \equiv  \frac{C^{\text{3pt}}_{\Gamma}(t,t_{\mathrm{sep}})}{C^{\text{2pt}} (t_{\mathrm{sep}})}\,,\nonumber  \\  
  & \approx \frac{c_0 
+c_1 e^{-\Delta E (t_{\mathrm{sep}}-t)}+c_2 e^{-\Delta E t}\nonumber+c_3 e^{-\Delta E t_{\mathrm{sep}}}}{1+c_5 e^{-\Delta E t_{\mathrm{sep}}}} \,, \\
\label{eq:joint_fit}
\end{align}
where $c_0$ corresponds to the bare quasimatrix elements and  $\Delta E=E_1-E_0$ denotes the energy gap between the ground state and the first excited state.

\begin{figure*}[htbp]
\includegraphics[width=.31\textwidth]{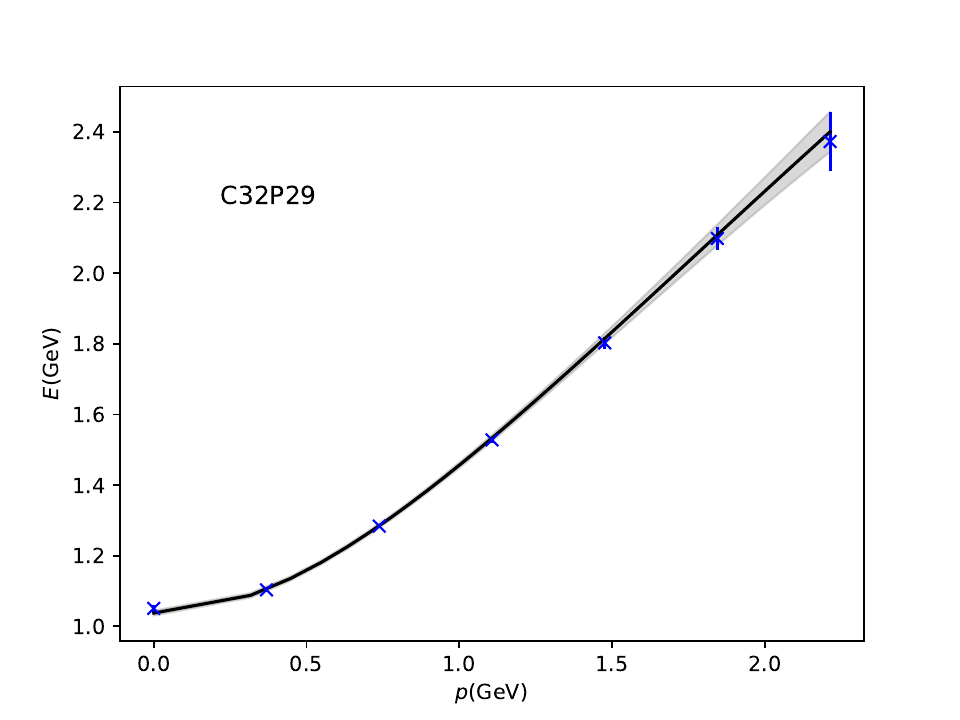}
\includegraphics[width=.31\textwidth]{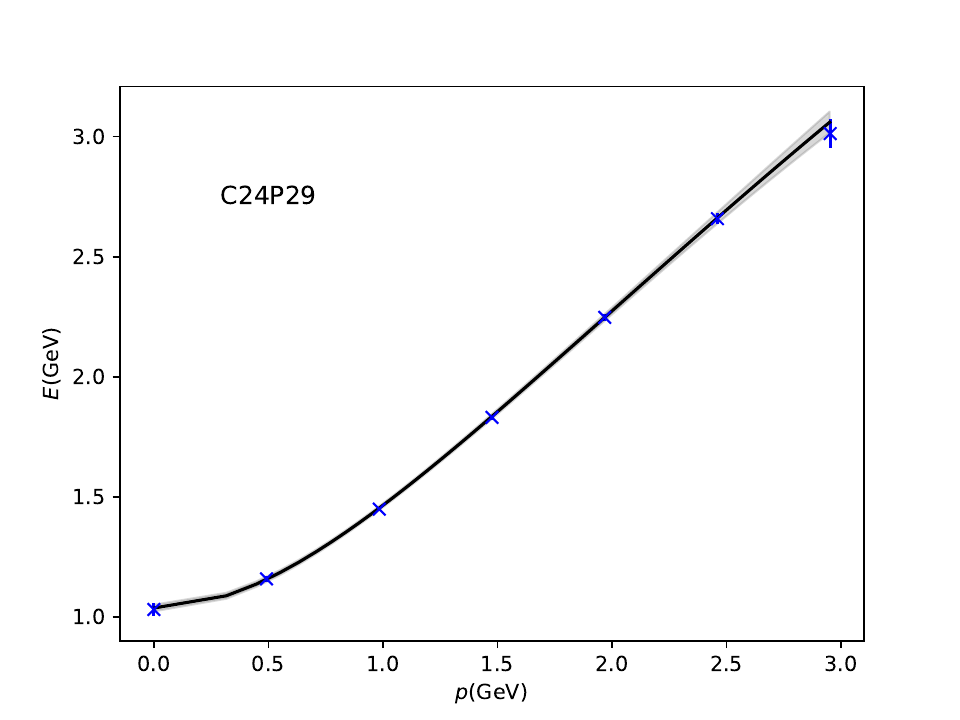}
\includegraphics[width=.31\textwidth]{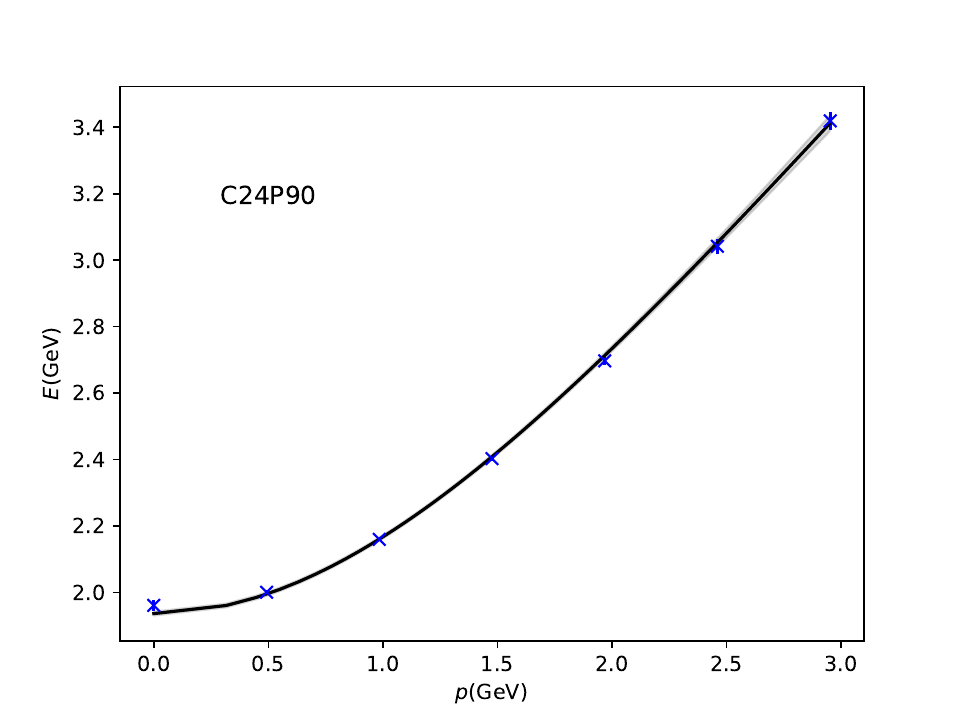}

\caption{ The dispersion relation $E_0(P_z)^2=m^2+k_2P_z^2+k_3P_z^4a^2$ of each ensemble used in this work.  }
\label{fig:disp_rel}
\end{figure*}

The ground-state effective energy $E_0$ is obtained by fitting the 2pt function in Eq.~(\ref{eq:joint_fit}). After obtaining the ground-state effective energy $E_0 (P_z)$  under different momenta, we fit for the dispersion relation 
\begin{align}
    E_0(P_z)^2=m^2+k_2P_z^2+k_3P_z^4a^2.
\end{align}
The quadratic term with lattice spacing $a$ is included to parametrize the discretization effects. The fitting results of each ensemble are shown in
Fig.~\ref{fig:disp_rel}.
The result of $k_2$ and $k_3$ is shown in Table ~\ref{Tab:disp_rel}.

\renewcommand{\arraystretch}{1.5}
\begin{table}
\footnotesize
\centering
\begin{tabular}{cccc}

\hline
\hline
Ensemble ~&$k_2$ & $k_3$  ~& $\chi^2$/d.o.f.  \\

\arrayrulecolor{black}

\hline
C32P29  ~& $1.067(29)$ ~& $-0.022(50)$ ~& 1.43               \\
\arrayrulecolor{gray!50}
\hline

C24P29  ~& $1.085(28)$ ~& $-0.015(19)$ ~& 0.33   \\
\arrayrulecolor{gray!50}
\hline
C24P90  ~& $0.953(14)$ ~& $-0.0054(99)$ ~& 0.81    \\


\arrayrulecolor{black}
\hline

\end{tabular}
 \caption{Fitting result of dispersion relation $E(p)^2=E_0^2+k_2p^2+k_3p^4a^2$, including $k_2$, $k_3$ and $\chi^2/d.o.f.$.
 }
 \label{Tab:disp_rel}
\end{table}

To obtain bare matrix elements $c_0$, we apply the joint fit of 2pt and the ratio of 3pt to 2pt at several $t_{\mathrm{sep}}$s. For the ensembles C24P29 and C32P29, the fit range of 2pt starts from $t=2a$, while for the ensemble C24P90, the excited states decay faster, and the fit range starts from $t=1a$. The correlators are jackknife resampled before fitting. For each $t_{sep}$ of the ratio, we choose the fit range as  $t\in [ 1,t_{sep}-1]$ to eliminate the contribution of higher excited states. 

\begin{figure*}[htbp]
\includegraphics[width=.3\textwidth]{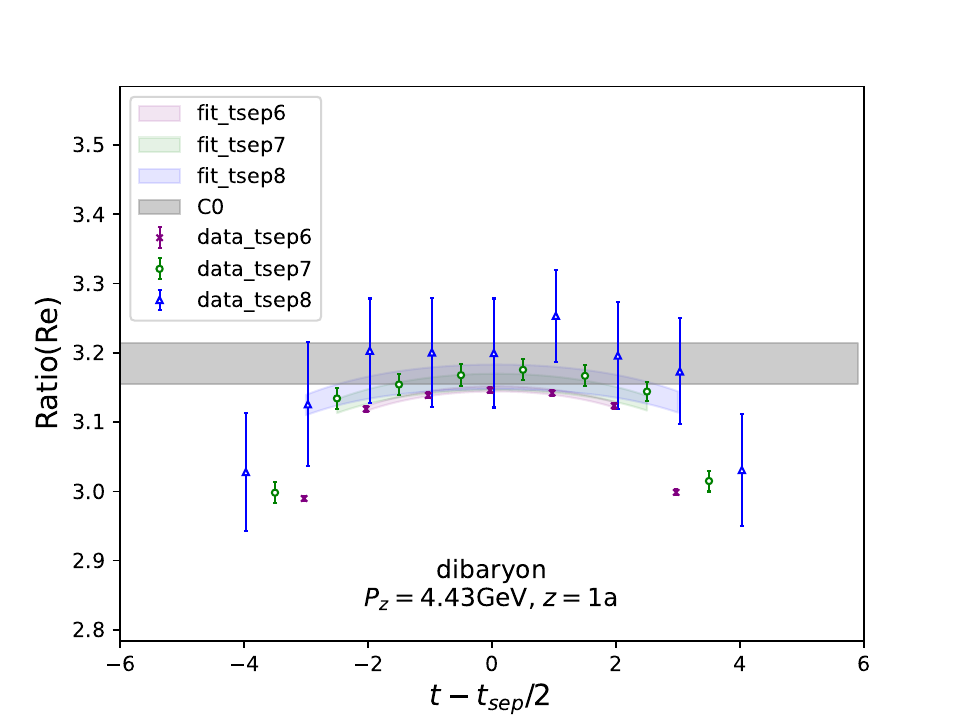}
\includegraphics[width=.3\textwidth]{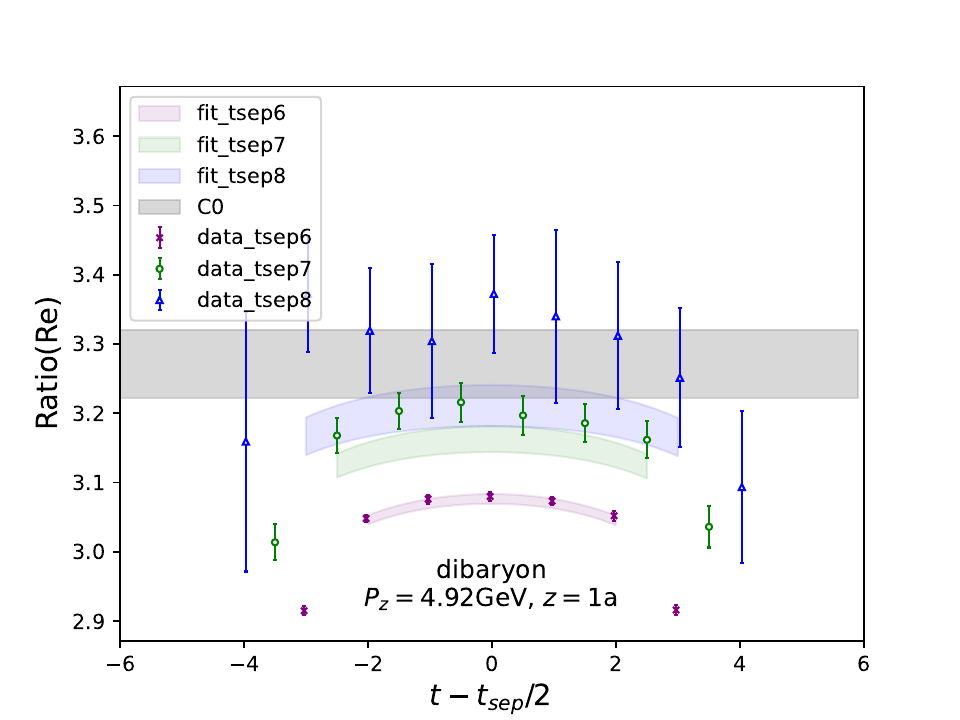}
\includegraphics[width=.3\textwidth]{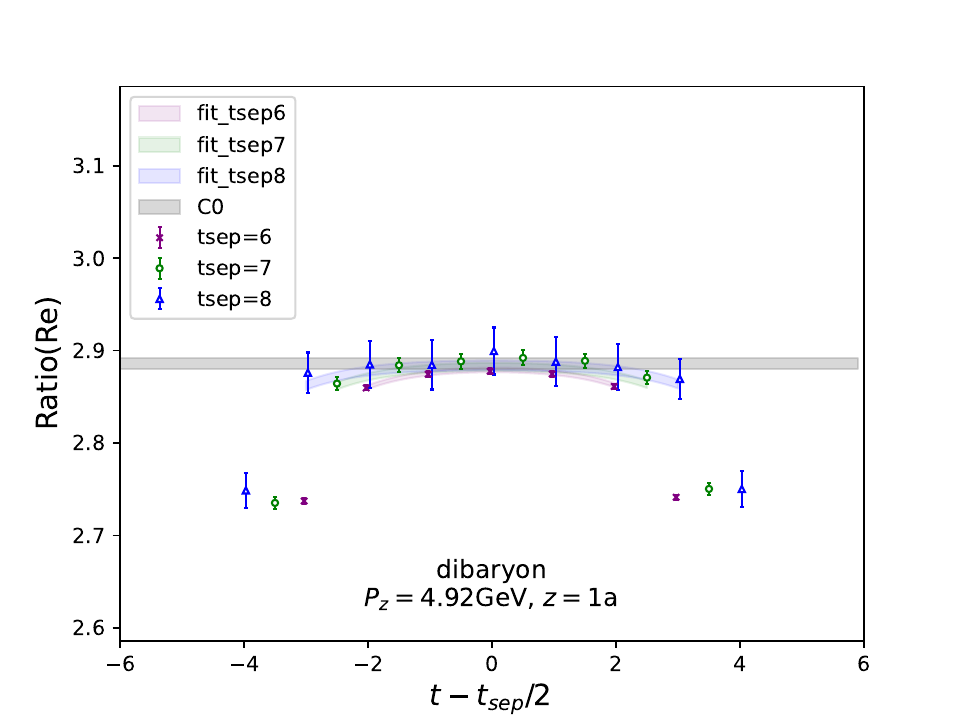}\\
  
\includegraphics[width=.3\textwidth]{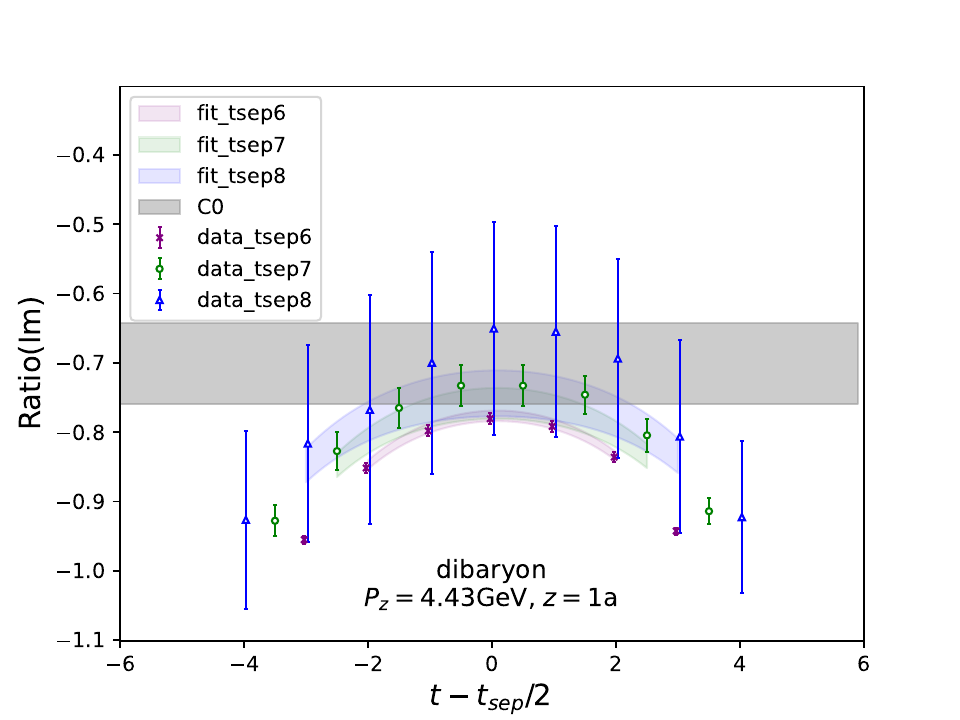}
\includegraphics[width=.3\textwidth]{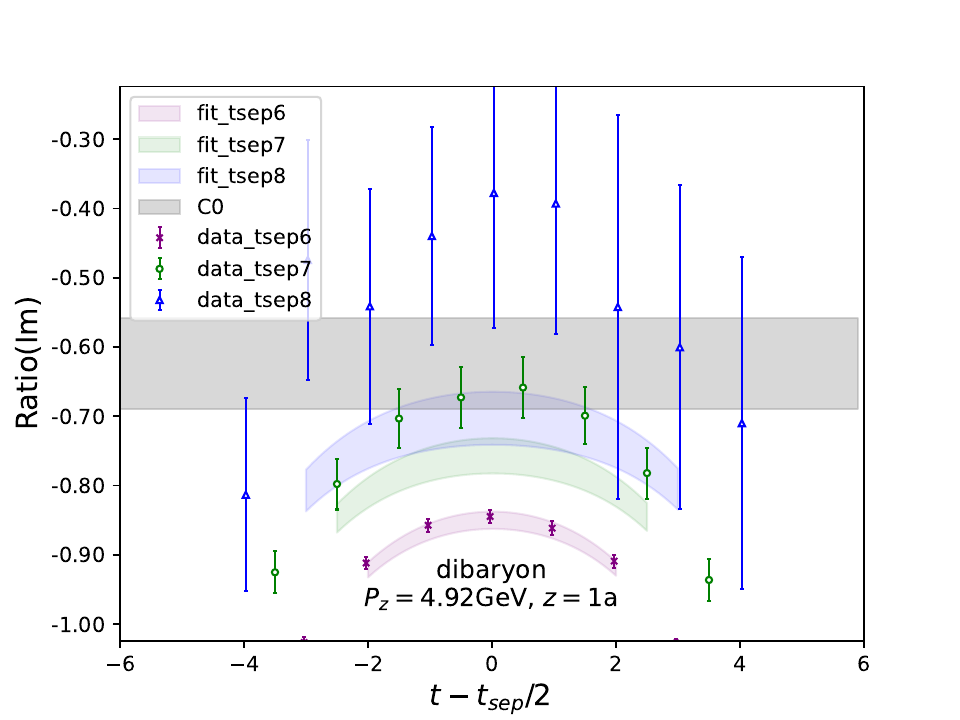}
\includegraphics[width=.3\textwidth]{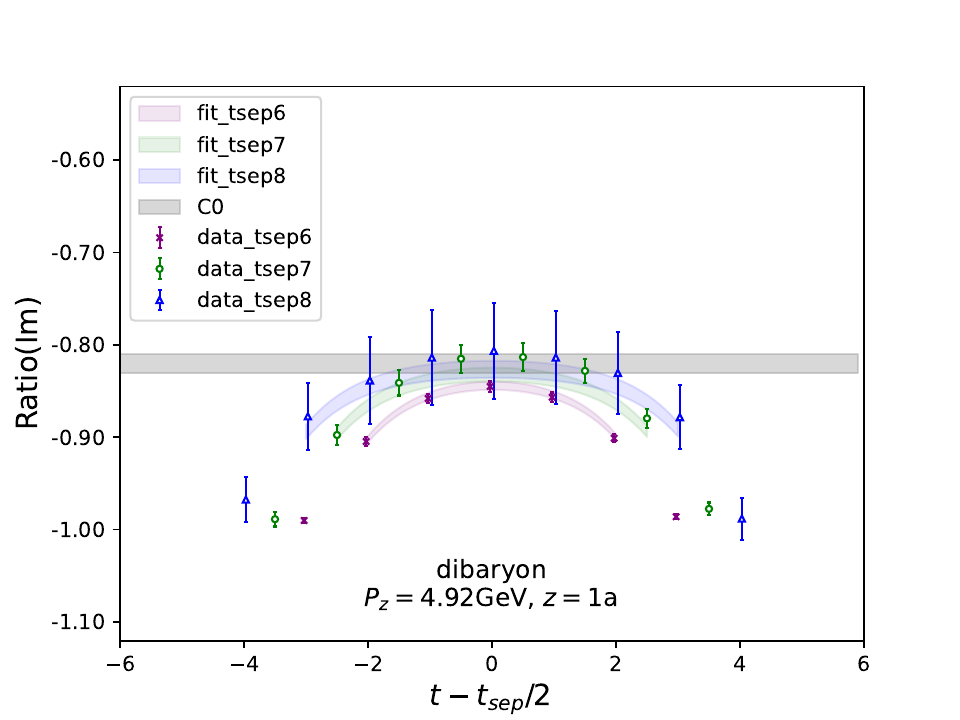}

\caption{ Demonstration of fitting the correlation function on each ensemble: C32P29(left), C24P29(middle) and C24P90(right). Here, we present the $z=1a$ dibaryon result of the largest-momentum case of each ensemble as an example. We compare the lattice data of the ratio (error bars)  and  results predicted by the fitting (colored bands). The ground-state contribution $c_0$ obtained by the fitting is shown as the gray band.}
\label{fig:fit_examp}
\end{figure*}

For the nucleon case, the source and sink are symmetric, allowing us to set $c_1=c_2$. However, for the dibaryon system,  the correlator consists of a hexaquark source and a dibaryon sink, so we treat $c_1$ and $c_2$ as two independent variables.  Given the significant correlation between the 2pt and the ratio,  we combine the data of the 2pt and the ratio into one correlation matrix in the fitting. In Fig.~\ref{fig:fit_examp}, we plot the ratio of 3pt to 2pt, taking the largest-momentum case of the dibaryon system as an example. Fitting results of other momenta and for the nucleon are shown in Appendix~\ref{subsec：data_ana}. The ground-state contribution $c_0$ obtained by the fit is shown as the gray band. Our fitting results are compared with the original data, demonstrating that the data for each $t_{sep}$ can be well described by the fit.

\section{Renormalization and Matching}

\label{section C}

\subsection{Hybrid renormalization  }
The bare quasi-LF correlation can be multiplicatively renormalized as
\begin{align}
\tilde{h}(z,P_z,1/a)=Z_L(a) e^{-\delta m(a)z} \,\tilde{h}_R(z,P_z,1/a),
\end{align}
where $\tilde{h}_R(z, P_z,1/a)$ is the renormalized quasi-LF correlation, $Z_L(a)$ contains all $z$-independent logarithmic UV divergences, and $\delta m(a)$ includes the $z$-dependent linear UV divergences from Wilson-line self-energy.
To remove these divergences, some nonperturbative renormalization approaches have been proposed and implemented in the literature~\cite{Chen:2016fxx,Izubuchi:2018srq,Alexandrou:2017huk,Radyushkin:2018cvn,Braun:2018brg,Li:2018tpe}. However, they suffer from the fact that the renormalization factor contains undesired nonperturbative contributions, which distort the IR behavior of the original quasi-LF correlation. 
%
In our work, we adopt the so-called hybrid renormalization to avoid this problem~\cite{Ji:2020brr}. More precisely, we renormalize the quasi-LF correlations at short and long distances differently. At short distances, the renormalization is carried out by dividing by the same hadron matrix element in the rest frame, as is done in the ratio scheme~\cite{Radyushkin:2018cvn}.  At long distances, the renormalized factor for quasi-LF correlations is determined by the self-renormalization~\cite{LatticePartonCollaborationLPC:2021xdx} by fitting the bare matrix elements at multiple lattice spacings to a parametrization formula containing discretization effects and linear divergences. The renormalized quasi-LF correlation then takes the following form:
\begin{align}\label{eq:hybridscheme}
\tilde{h}_R(z,P_z)
=&\frac{\tilde{h}(z,P_z,1/a)}{\tilde{h}(z,P_z=0,1/a)}\theta(z_s-|z|)\notag\\
&+	\eta_s\frac{\tilde{h}(z,P_z,1/a)}{Z_R(z,1/a)} \theta(|z|-z_s), 
\end{align}
where $z_s$ is introduced to divide the short and long distances, and $Z_R(z,1/a)$ indicates the renormalization factor extracted from the self-renormalization procedure, which includes the linear divergence part $\delta m(a)$ and some discretization effects. The scheme conversion factor $\eta_s$ is to ensure the continuity of the renormalized quasi-LF correlation at $z=z_s$. 

\renewcommand{\arraystretch}{1.5}
\begin{table}
\footnotesize
\centering
\begin{tabular}{cclcccccc}

\hline
\hline
Ensemble ~&$a$(fm) ~& \ \!$L^3\times T$  ~& $m_\pi$(MeV) ~& $m_\pi L$  ~& $m_{\eta_s}$(MeV)  \\

\arrayrulecolor{black}
\hline
C32P29  ~& 0.105~& $32^3\times 64$  ~& 292.9(1.2)  ~&4.9   &659.1(1.3)            
\\

\hline
H48P32  ~& 0.0519 ~& $48^3\times 144$  ~& 319.0(1.7)  ~& 4.0 &695.6(3.5)                
\\

\hline
F32P30  ~& 0.0775~& $32^3\times 96$  ~& 303.6(1.3)  ~&3.8 &677.67(90)               
\\
\arrayrulecolor{black}
\hline

\end{tabular}
 \caption{The lattice setup for simulating zero-momentum pion hadron matrix elements.}
 \label{Tab:setup_ren}
\end{table}

Since the renormalization follows from the UV property of the quark bilinear operator, we can choose the pion hadron matrix element in the rest frame to extract the renormalization factor. 
The corresponding lattice setup is shown in Table.~\ref{Tab:setup_ren}. We start from the normalized pion matrix elements in the rest frame
\begin{align}
\label{}
\tilde{h}^{\pi}_B(z,P_z=0,1/a)=Z_R (z,1/a)\,\tilde{h}^{\pi}_R(z,P_z=0)
\end{align}
with 
\begin{align}
\label{rnmlfctr}
Z_R &(z,1/a)=\exp\bigg\{\frac{kz}{a\ln[a\Lambda_{\text{QCD}}]}+m_0 z+f(z)a^2\notag\\
&+\frac{3C_F}{b_0}\ln\left[\frac{\ln[1/(a\Lambda_{\text{QCD}})]}{\ln[\mu/\Lambda_{\text{QCD}}]}\right]+\frac{1}{2}\text{ln}\bigg[1+\frac{d}{\text{ln}[a\Lambda_{\text{QCD}}]}\bigg]^2 \bigg\},
\end{align}
where the first term on the rhs is the linear divergence, and $m_0z$ denotes a finite-mass contribution from
the renormalization ambiguity. $f(z)a^2$ represents the
discretization effects, and the last two terms come from the
resummation of leading and subleading logarithmic divergences. According to Ref.~\cite{LatticePartonCollaborationLPC:2021xdx}, the fitting parameters $k$ and $f(z)$ are related to the specific
lattice action, while $d$ and $m_0$ are determined by matching to the continuum scheme at short distances. 
In the $\overline{\text{MS}}$ scheme, the short-distance result takes the following form at the next-to-leading order (NLO)\cite{Izubuchi:2018srq}:
\begin{align}
\label{Zms}
C_{0,\text{NLO}}\left( z,\mu\right)&=1+\frac{\alpha_s C_F}{4\pi}\left(3\ln\frac{z^2\mu^2}{4e^{-2\gamma_E}}+5\right).
\end{align}

In our calculation, we choose to determine the parameters in $Z_R$ through a global fit, as follows:
\begin{align}
\text{ln}& \tilde{h}^{\pi}_B(z,P_z=0,1/a)  =\frac{kz}{a\text{ln}[a\Lambda_{\text{QCD}}]}+f(z)a^2 \nonumber \\ &+\frac{3C_F}{b_0}\text{ln}\bigg[\frac{\text{ln}(1/a\Lambda_{\text{QCD}})}{\text{ln}(\mu/\Lambda_{\text{QCD}})}\bigg]+\frac{1}{2}\text{ln}\bigg[1+\frac{d}{\text{ln}[a\Lambda_{\text{QCD}}]}\bigg]^2\nonumber \\
\quad &+\begin{cases}
\text{ln}\big[C_{0,\text{NLO}}\left( z,\mu\right) \big] +m_0z & \text{if } z_0 \le z \le z_1\\
g(z) & \text{if } z_1<z
\end{cases},\label{eq:self1_quark}
\end{align}
where $k$, $\Lambda_{\text{QCD}}$, $f(z)$, $m_0$, $d$ and $g(z)$ are treated as free-fitting parameters. We choose $z_0=0.05$ fm , $z_1=0.2$ fm and the maximum value of $z$ is $1.55$ fm. 

According to Ref.~\cite{LatticePartonCollaborationLPC:2021xdx}, to account for  additional systematic uncertainties related to $a\mu$,  the input error of the uncorrelated fit can be modified as
\begin{equation*}
(\sigma_{\text{ln}\tilde{h}})_{new}=\sqrt{(\sigma_{\text{ln}\tilde{h}})^2_{\text{old}}+(\delta_{\text{sys}}a \mu)^2} ,    
\end{equation*}
where $\delta_{\text{sys}}$ is set to be $0.0005$. By doing this more weights can be given to the small lattice spacing data during the fitting process.  
The  fitting results of parameters and the $\chi^2$/d.o.f.  are shown in Table~\ref{Tab:renor_para}.

\renewcommand{\arraystretch}{1.5}
\begin{table}
\footnotesize
\centering
\begin{tabular}{cc}

\hline
\hline
parameter ~& fit result  \\

\arrayrulecolor{black}

\hline

$k$  ~& $0.674(0.008)$        \\

\hline
$\Lambda_\text{QCD}$(GeV)  ~& $0.16(0.03)$                \\

\hline
$m_0$(GeV)  ~& $0.14(0.06)$               \\

\hline
$d$  ~& $-0.04(0.02)$               \\
\hline
$\chi^2$/d.o.f.  ~& $1.03$              \\

\arrayrulecolor{gray!50}

\arrayrulecolor{black}
\hline

\end{tabular}
 \caption{The parameters and $\chi^2$/d.o.f. are determined by doing a global fit for Eq.~(\ref{eq:self1_quark}). }
 \label{Tab:renor_para}
\end{table}

\begin{figure}[tbp]
\includegraphics[width=.45\textwidth]{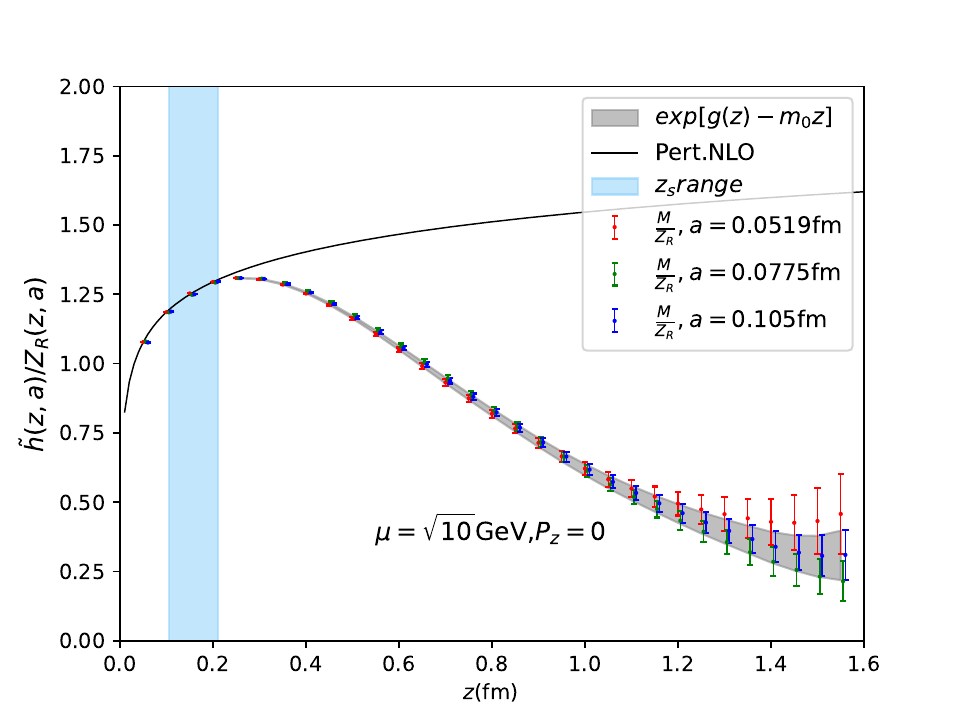}
\caption{ Comparison of renormalized matrix elements   with the perturbative one-loop $\overline{\text{MS}}$ results (black curve). Error bars represent the renormalized  matrix elements of each lattice spacing. $g(z)$ is the
residual from the fitting process~\cite{LatticePartonCollaborationLPC:2021xdx} and $\rm{Exp}[g(z)-m_0 z]$ (gray band) indicates the $a$-independent renormalized matrix element. $z_s$ (introduced to divide short and long distances in the hybrid scheme) is taken as $0.21$ fm and varied down to $0.105$ fm to estimate systematic uncertainties, as shown by the blue band.}
\label{fig:fit_MSbar}
\end{figure}

The comparison of the
renormalized matrix element with the perturbative one-loop $\overline{\text{MS}}$ result is given in Fig.~\ref{fig:fit_MSbar}.
This figure shows a good agreement between the renormalized matrix element and the continuum one-loop result at short distances, except at very small $z$, where higher-order corrections become important. 
In this way, we take $z_s= 0.21$ ~fm in the hybrid renormalization process and vary it down to $0.105$~fm to estimate the systematic uncertainty related to the choice of $z_s$. 

\begin{figure*}[htbp]

 \centering
\includegraphics[width=.31\textwidth]{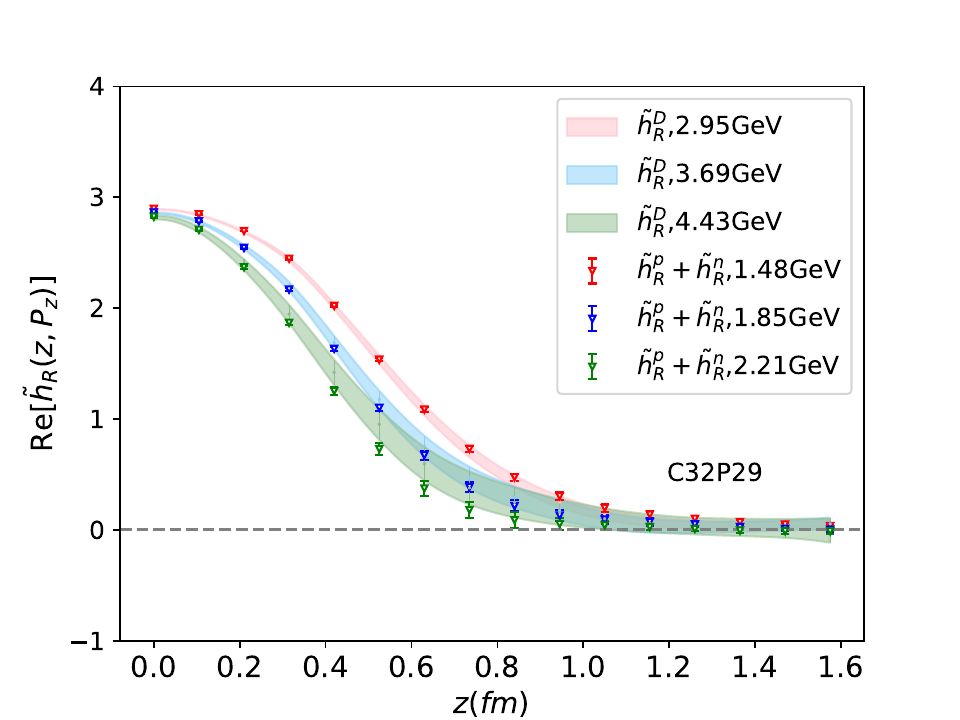}
\includegraphics[width=.31\textwidth]{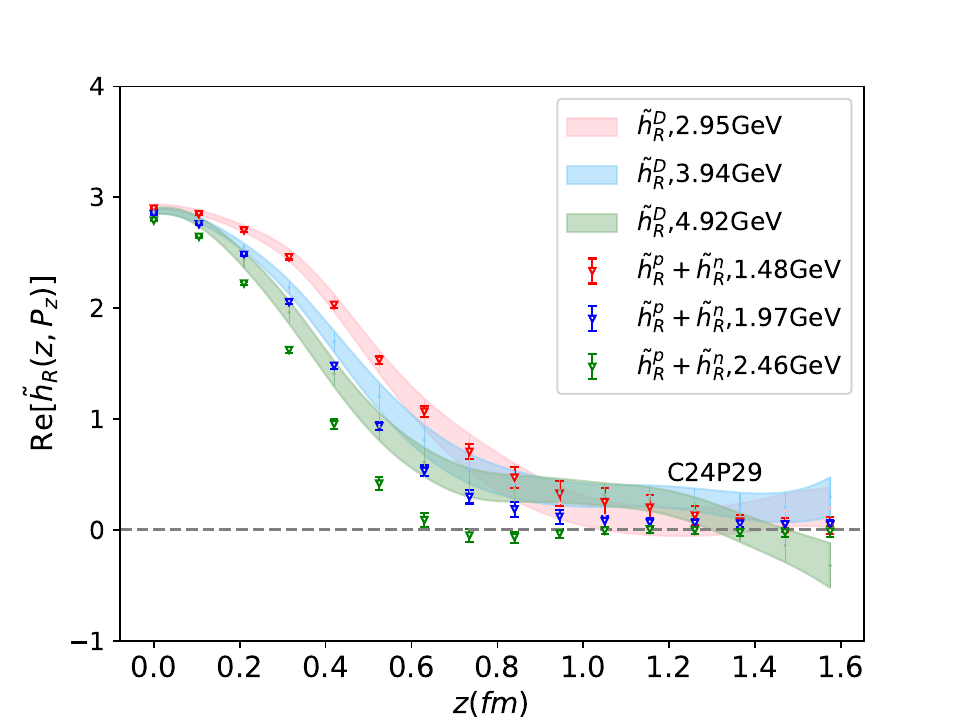}
\includegraphics[width=.31\textwidth]{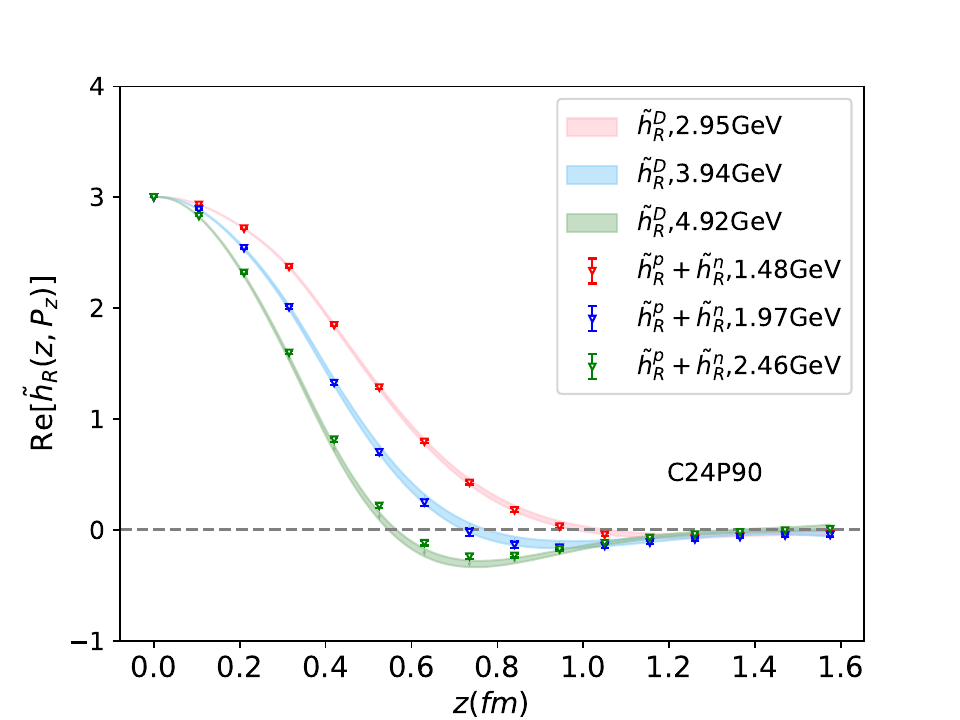}
\\
\includegraphics[width=.31\textwidth]{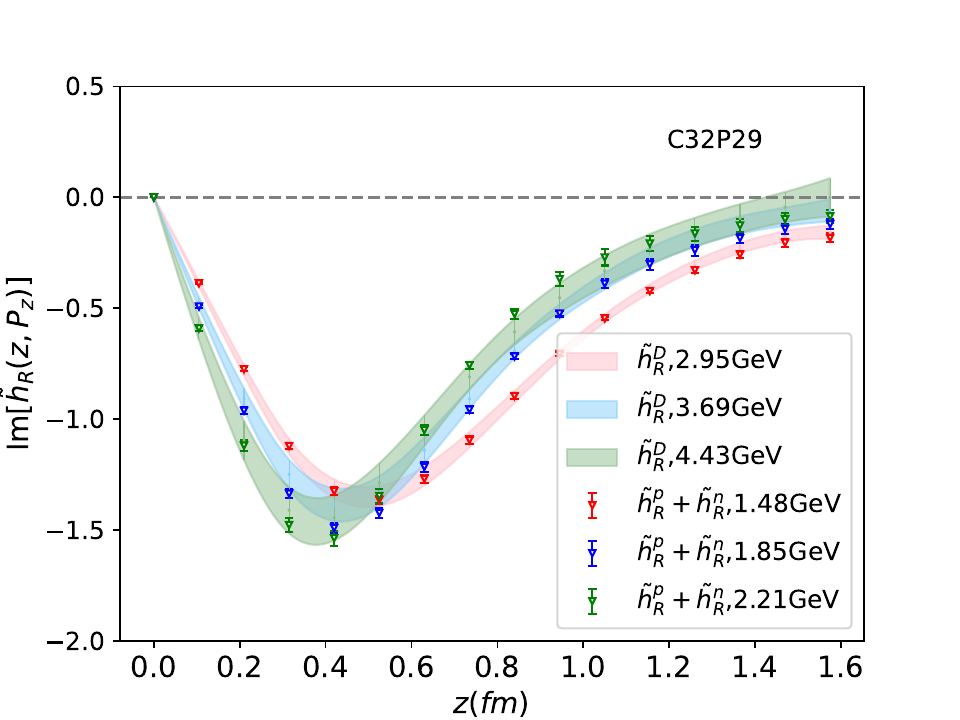}
\includegraphics[width=.31\textwidth]{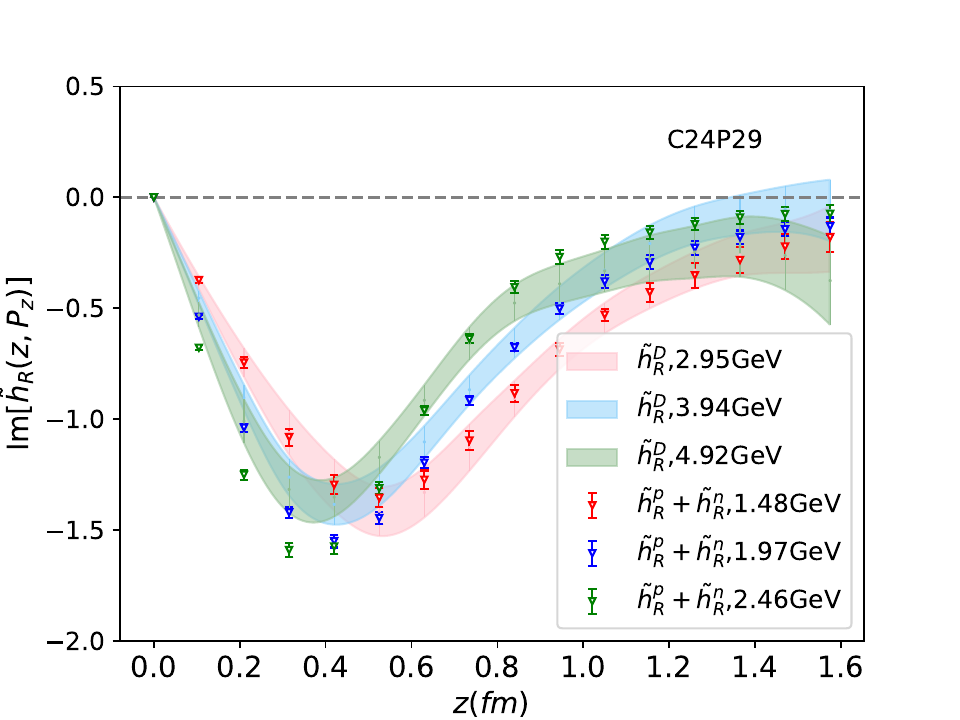}
\includegraphics[width=.31\textwidth]{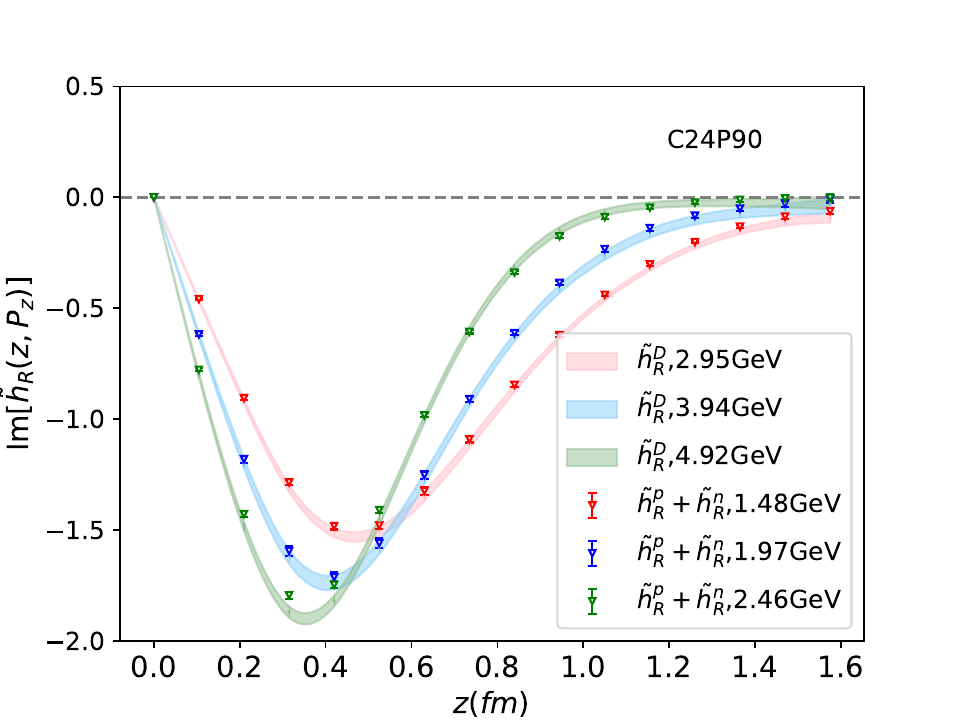}\\

\caption{ The real  and imaginary parts of  renormalized matrix elements of C32P29 (left)  , C24P29 (middle), and C24P90(right) as a function of $z$ at scale $\mu=\sqrt{10}$~GeV. Different colors represent different momentums of the external state. The colored bands and error bars represent the renormalized matrix elements for a dibaryon and for the sum of a proton and a neutron, respectively. }
\label{fig:ren_ma}
\end{figure*}

As a function of $z$, the renormalized matrix elements are plotted in Fig.~\ref{fig:ren_ma}. We present the results of the dibaryon system and of free nucleons (proton and neutron) simultaneously. As shown in the figure, the results exhibit a convergence trend with increasing $P_z$ for different ensembles. For larger momentum, statistical uncertainties are controlled by increasing the number of sources in space.

\vspace*{1.5em}

\subsection{Large-$\lambda$ extrapolation and Fourier transform}

\begin{figure*}[htbp]
\includegraphics[width=.45\textwidth]{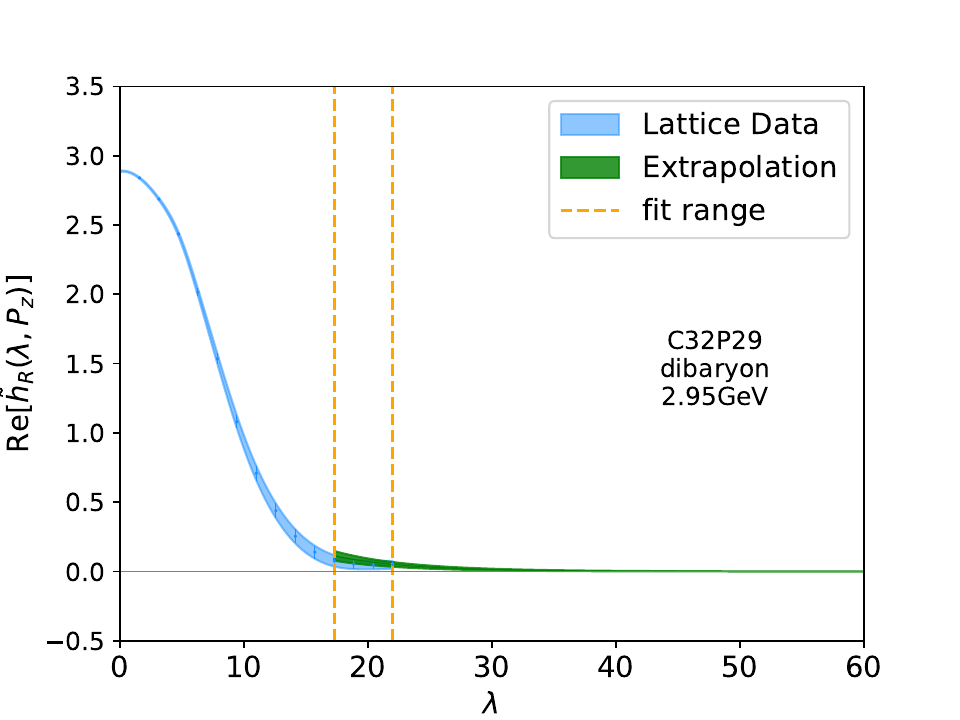}
\includegraphics[width=.45\textwidth]{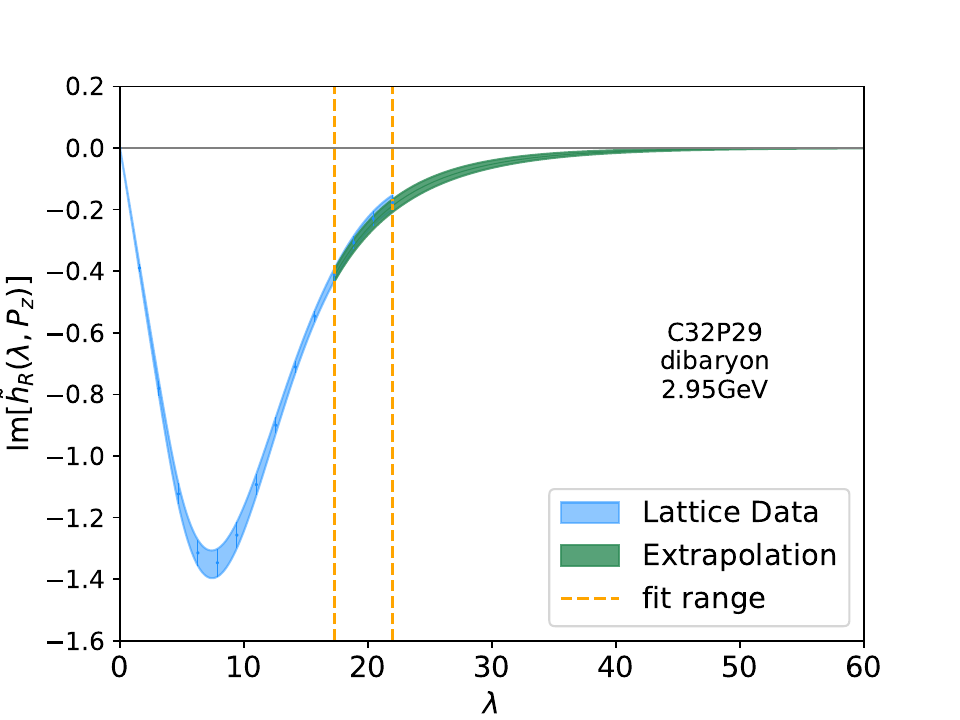}

\caption{The large-$\lambda$ extrapolation of renormalized matrix elements for the case of a dibaryon system with $P_z=2.95$~GeV for ensemble C32P29. The blue and green bands represent lattice data and extrapolated results respectively. The orange dashed lines mark off the fit range for Eq. \ref{eq:extrap}. As can be seen from the plot, our fitting results  agree well with lattice data within the fit range.}
\label{fig:lambda_extrap}
\end{figure*}

As can be seen from the figures of renormalized matrix elements, the uncertainties become worse with increasing $\lambda= z P_z$. However, a Fourier transform to momentum space requires the quasicorrelator at all distances. To resolve this issue, 
we follow~\cite{Ji:2020brr} and adopt an
extrapolation with the following form in the large $\lambda$ region:

\begin{align}
	  \tilde{h}_R(\lambda) &= \Big[\frac{l_1}{(i\lambda)^{a_1}}  +  e^{-i\lambda}\frac{l_2}{(-i \lambda)^{b_1}} \Big]e^{-\lambda/\lambda_0},
	  \label{eq:extrap}
\end{align} 
where the algebraic terms in the square bracket are associated with a power law behavior of the unpolarized PDFs in the end point region, and the exponential term comes from the expectation that at finite momentum the correlation function has a finite correlation length (denoted as $\lambda_0$)~\cite{Ji:2020brr}, which becomes infinite with the momentum approaching infinity. In Fig.~\ref{fig:lambda_extrap}, we show the extrapolation result of the dibaryon system with $P_z=2.95$~GeV for the ensemble C32P29. After the $\lambda$ extrapolation, we can perform a Fourier transform and obtain the $x$-dependent quasi-PDF in momentum space.



\subsection{Matching and infinite-momentum extrapolation}

As mentioned before, we neglect the contribution of disconnected diagrams and mixing with gluons. Therefore, the dibaryon and nucleon PDFs can be extracted through a perturbative matching shown in Eq.~(\ref{mtcheq}), where the corresponding perturbative kernel up to the NLO in the hybrid scheme reads~\cite{Chou:2022drv,Yao:2022vtp}
\begin{widetext}
\begin{align}\label{eq:mommatchingkernelratio}
 C_{\text{ratio}}&\left(\xi,\frac{\mu}{yP_z}\right)=\delta\left(1-\xi\right)
 +\frac{\alpha_s C_F}{2\pi}
 \begin{cases}
\left[\frac{1+\xi^2}{1-\xi}\ln\frac{\xi}{\xi-1}+1-\frac{3}{2(1-\xi)}\right]_+ & \xi > 1 \\
\left[\frac{1+\xi^2}{1-\xi}\left(-\ln\frac{\mu^2}{y^2P_z^2}+\ln 4\xi (1-\xi)-1\right)+1+\frac{3}{2(1-\xi)}\right]_+ & 0<\xi<1 \\
\left[-\frac{1+\xi^2}{1-\xi}\ln\frac{\xi}{\xi-1}-1+\frac{3}{2(1-\xi)}\right]_+ & \xi<0 ,
\end{cases}
\end{align}

\begin{align}
C_{\text{hybrid}}(\xi,\lambda_s,\frac{\mu}{yP_z})=C_{\text{ratio}}&\left(\xi,\frac{\mu}{yP_z}\right)+\frac{3\alpha_s C_F}{4\pi}\left
( -\frac{1}{|1-\xi|}+\frac{2\text{Si}((1-\xi)|y|\lambda_s)}{\pi(1-\xi)}\right)_+,
\end{align}
\end{widetext}
with $\xi=\frac{x}{y}$.

We also need to extrapolate to infinite momentum to extract the physical unpolarized PDF. Here we use the following ansatz which takes into account the leading power contribution
\begin{align}
q_0(x,\mu,P_z)=q(x,\mu)+\frac{d_0(x)}{(P_z)^2},
\end{align}
where $q_0(x,\mu,P_z)$ on the lhs denotes the PDF results obtained under different momenta, and $q(x,\mu)$ is the physical PDF in the infinite-momentum limit.

\section{Numerical results}
\label{section D}





\begin{figure}[tbp]
\includegraphics[width=.45\textwidth]{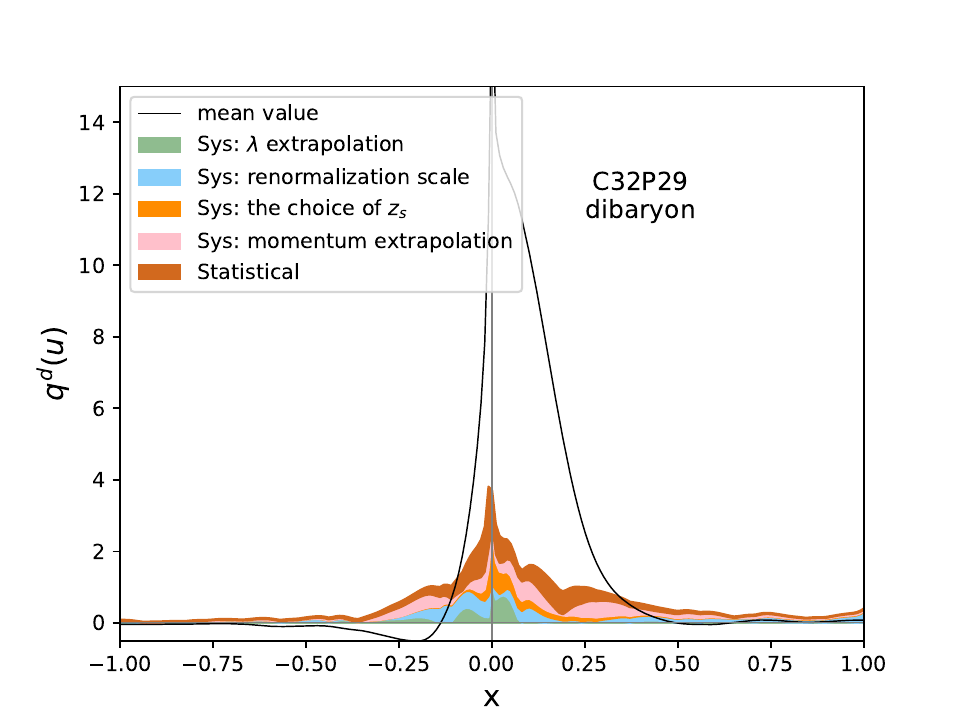}
\caption{ Estimation of statistical and systematic uncertainties, taking the dibaryon result of the ensemble C32P29 as an example. The width of the (nonoverlapping) coloured
bands denotes the size of each uncertainty (these uncertainties
are added in quadrature and  the square root is taken to obtain the full  uncertainty), and the black curve is the mean value obtained at $z_s = 0.21$~fm, $\mu=\sqrt{10}$~GeV and $\lambda$ extrapolation range  $\{P_z=2.95~\mathrm{GeV}:\lambda \geq 11aP_z, P_z=3.69~\mathrm{GeV}:\lambda \geq 11aP_z,P_z=4.43~\mathrm{GeV}:\lambda \geq 11aP_z,\}$. }
\label{fig:err_com_deu}
\end{figure}

In Fig.~\ref{fig:err_com_deu}, we show an estimate of systematic and statistical uncertainties of the light-cone PDF, taking the dibaryon result on the ensemble C32P29 as an example. The systematic uncertainties mainly have four sources. The first is the $\lambda$ extrapolation. We take the difference between the results with two fit ranges, $\{P_z=2.95~\mathrm{GeV}:\lambda \geq 11aP_z, P_z=3.69~\mathrm{GeV}:\lambda \geq 11aP_z,P_z=4.43~\mathrm{GeV}:\lambda \geq 11aP_z\}$
and $\{P_z=2.95~\mathrm{GeV}:\lambda \geq 9aP_z, P_z=3.69~\mathrm{GeV}:\lambda \geq 10aP_z,P_z=4.43~\mathrm{GeV}:\lambda \geq 10aP_z\}$, as an estimate of the systematic uncertainty due to the $\lambda$ extrapolation. The second is the renormalization scale dependence, estimated by varying the scale from $\sqrt{10}$~GeV to $2$~GeV and taking their difference. 
The third is the uncertainty from momentum extrapolation and we take the difference between results extrapolated to infinite momentum and that given by largest momentum as an estimation. The last contribution is from the choice of $z_s$ in the hybrid scheme. The error  is estimated by calculating the difference between
the results at $z_s=0.21$~fm and $z_s=0.105$~fm. 

\begin{figure}[tbp]
\includegraphics[width=.45\textwidth]{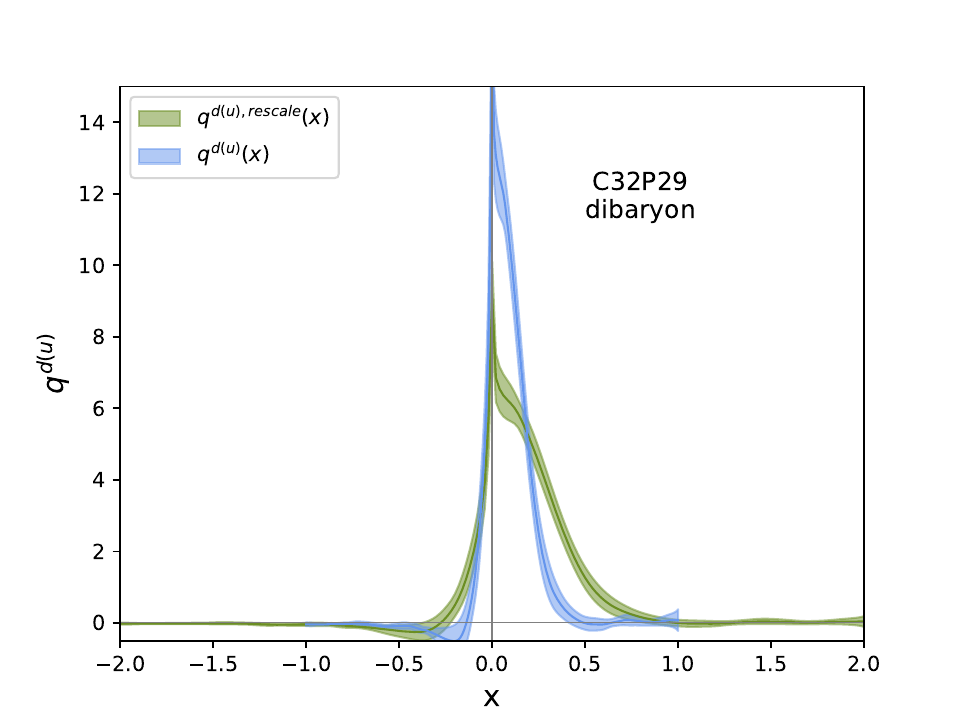}
\caption{ An example of the deuteronlike dibaryon PDF before (blue band) and after rescaling  (green band), taking the result of  ensemble C32P29 as an example. }
\label{fig:rescale_comp}
\end{figure}

\begin{figure*}[htbp]
\includegraphics[width=.31\textwidth]{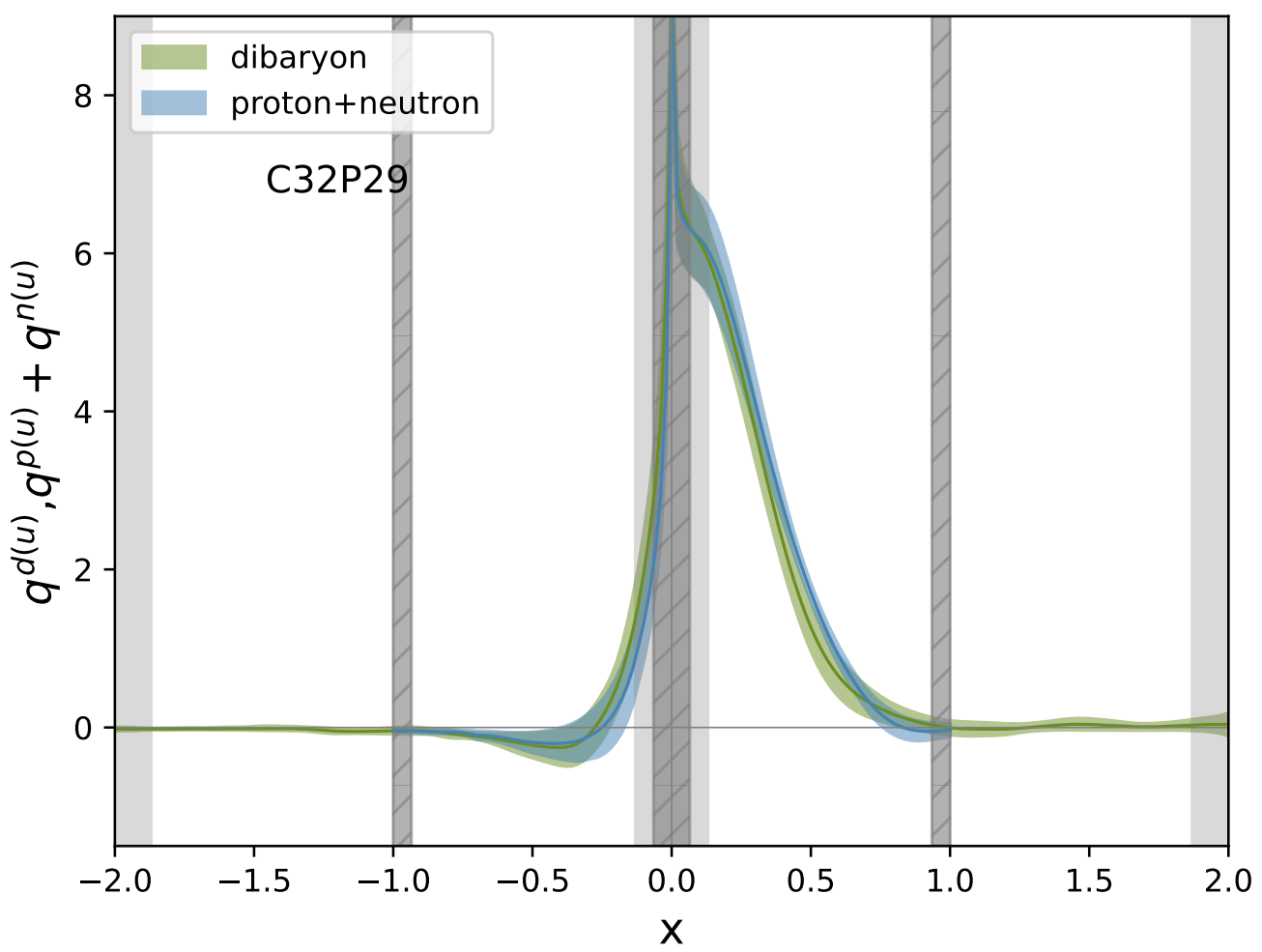}
\includegraphics[width=.31\textwidth]{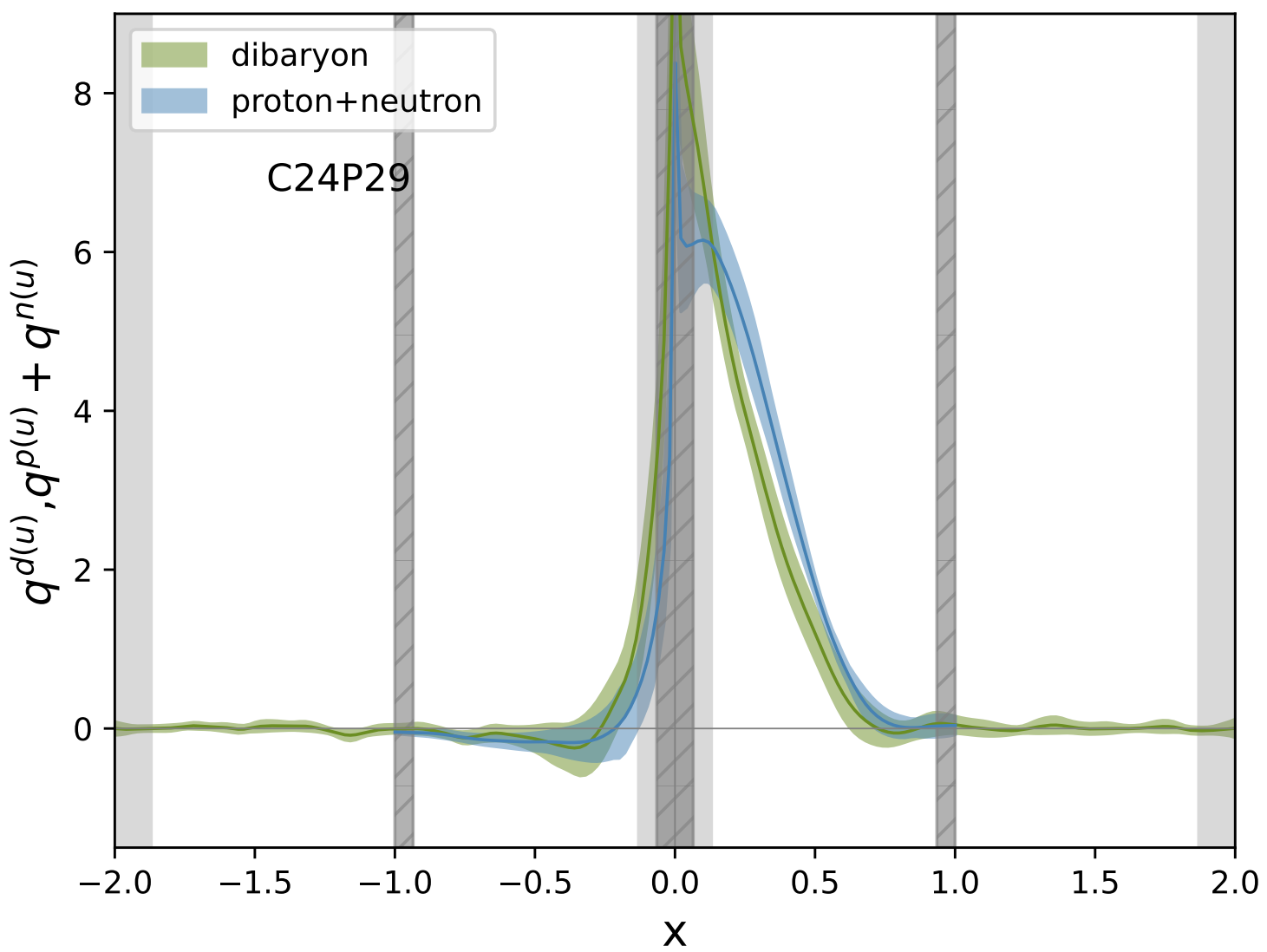}
\includegraphics[width=.31\textwidth]{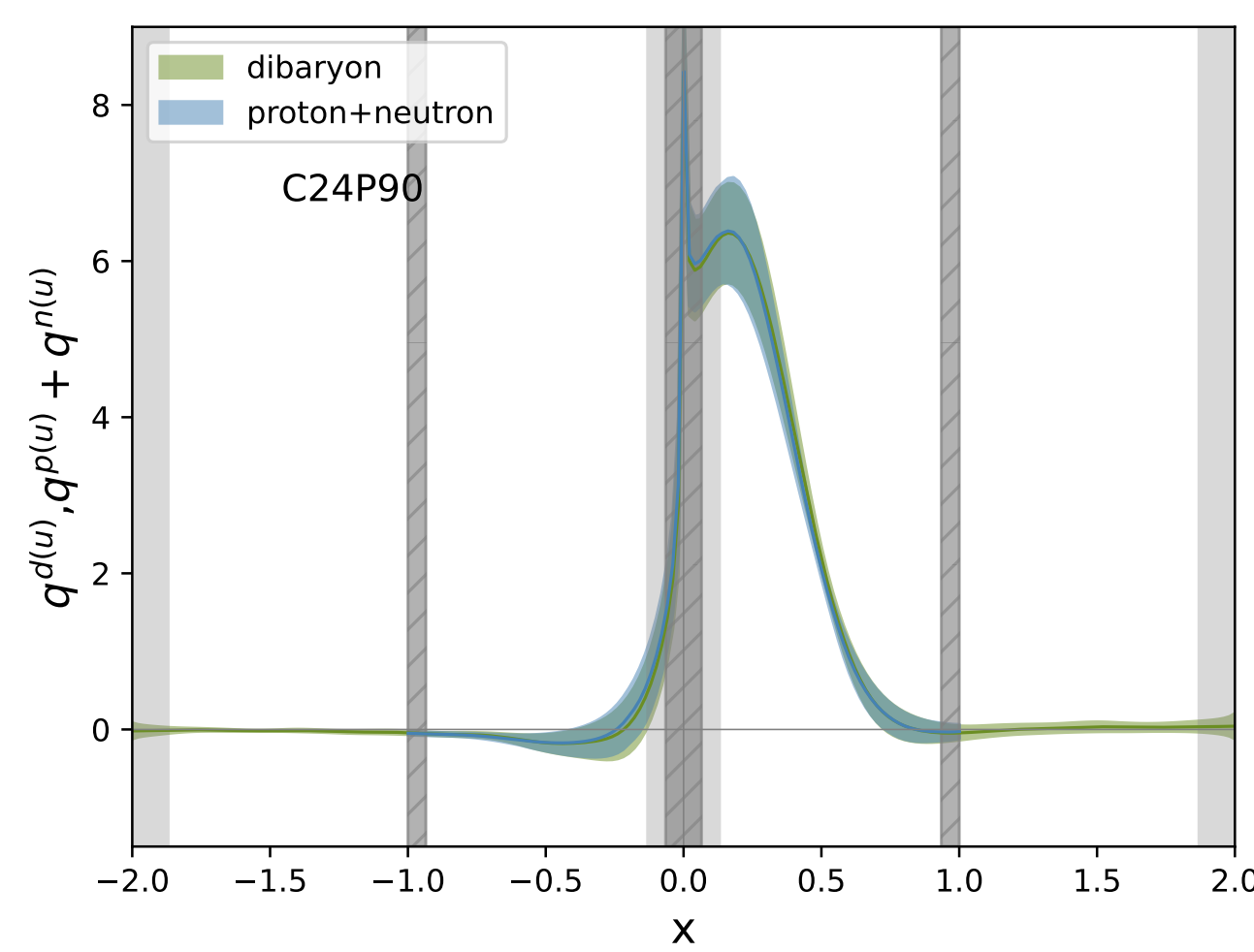}

\caption{ Our final results on the light-cone PDF at renormalization scale $\mu=\sqrt{10}$~GeV, extrapolated to the infinite-momentum limit. The result of the dibaryon(green bands) and the sum of the proton and neutron (blue bands) are both plotted for comparison purposes. The light gray bands ( $-2<x<-1.87$,  $-0.13<x<0.13$, and $1.87<x<2$ ) and shaded   deep gray bands ($-1<x<-0.935$, $-0.065<x<0.065$, and  $0.935<x<1$) denote the regions where LaMET results are considered unreliable for the dibaryon and nucleon, respectively. When $-0.13<x<0.13$,  the LRR+RGR-improved  results of the dibaryon blow up, indicating that higher-twist contributions cannot be neglected. Similarly, at  large 
$x$ regions, the same instability occurs.}
\label{fig:lc_EMC}
\end{figure*}

By comparing the differences between the PDFs of the multibaryon system and the sum of a proton and a neutron,  we can obtain information about the interaction between the baryons inside a nucleus. For a free nucleon the distribution $q(x)$ is a function of $x=\frac{p}{P_z}=\frac{p}{P_{z}^{n}}$ , where $p$ and $P_{z}^{n}$ represent the momentum of the parton and the nucleon, respectively. For a multinucleon system with each of the $N$ nucleons carrying the same momentum, the distribution is a function of $x=\frac{p}{P_z}=\frac{p}{NP_{z}^{n}}$. The different definitions of $x$ in a single nucleon and in a multinucleon system prevent us from comparing their distributions directly.   To facilitate the comparison, we rescale the distribution of multinucleon system into $q^r(x')$, where $x'=Nx=\frac{p}{P_z^n}$. The distribution after rescaling $q^{r}(x')$ can be related by the previous one $q(x)$ by 

\begin{align}
q^{r}(x')=Cq(x)=Cq(\frac{x'}{N})
\end{align}
where $C$ is a normalization factor.  From that 
    \begin{align}
        \int x'q^{r}(x')dx' = N\int xq(x)dx,
    \end{align}
which means that the average momentum fraction after rescaling is $N$ times that before rescaling, we conclude $C=\frac{1}{N}$.
For the PDF of a deuteronlike dibaryon system,  we have 
\begin{align}
q^{r}(x')=\frac{1}{2}q(\frac{x'}{2}) .
\end{align}
In the following paragraphs we will replace $x'$ with $x$ for convenience. It is worth noting that in the previous section, $x$ is equal to $\frac{p}{NP_z^n}$ for a multibaryon system. Now, the definition has changed to $x=\frac{p}{P^n_z}$  after we do the rescaling. In Fig.~\ref{fig:rescale_comp},  we give an example of the dibaryon PDF before and after rescaling. For a deuteronlike dibaryon system  with nucleon-nucleon interaction,
 the upper limit of $x$ can be $2$ after rescaling.  The contribution in the region $1<x<2$ is from nuclear effects. In Fig.~\ref{fig:lc_EMC},  we obtain a clear signal for the dibaryon PDF in the region of $0<x<1$.  
However, in the region of $x>1$ the dibaryon PDF is zero within uncertainty.  
 
\begin{figure*}[htbp]
\includegraphics[width=0.31\textwidth]{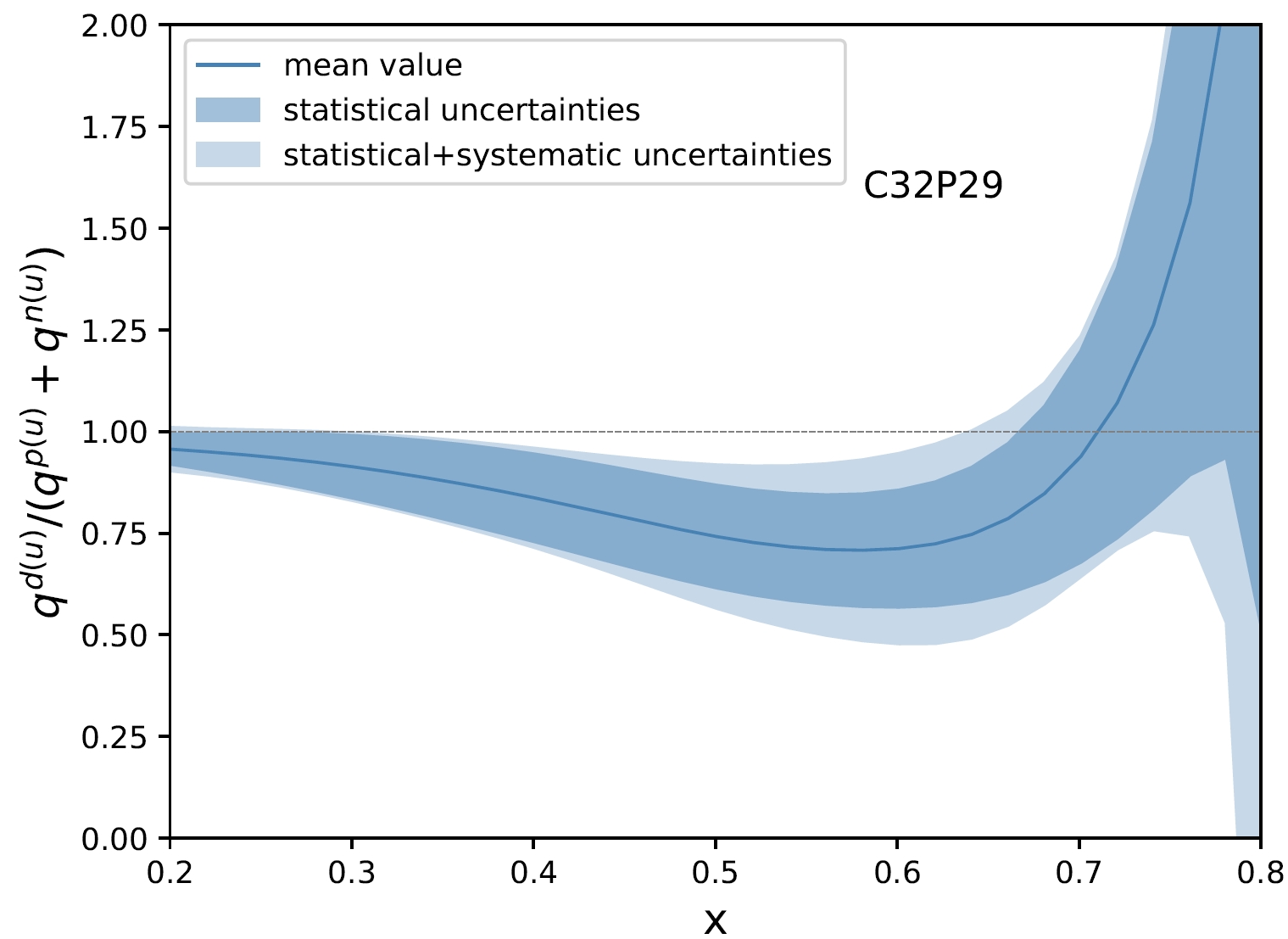}
\includegraphics[width=0.31\textwidth]{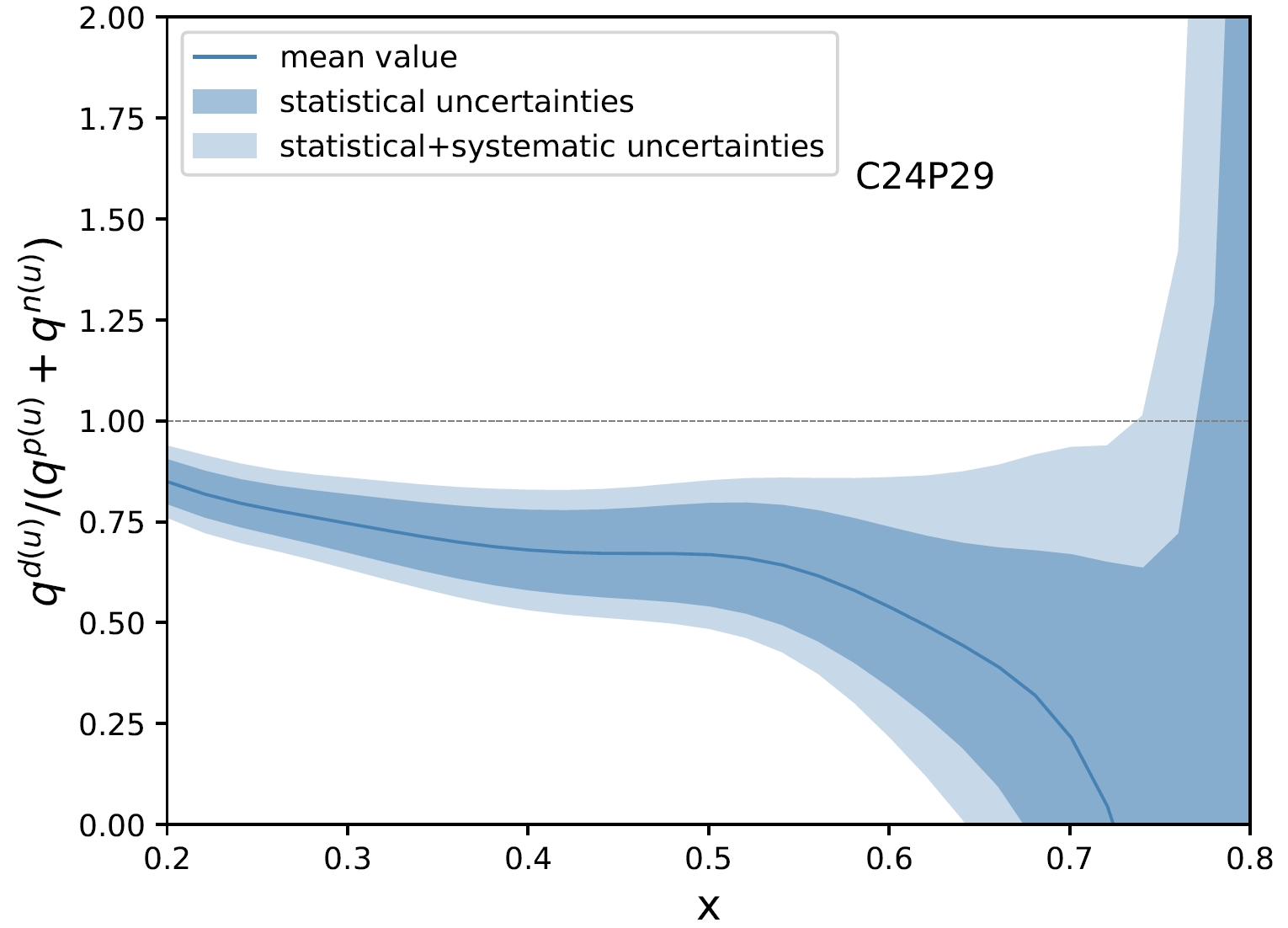}
\includegraphics[width=0.31\textwidth]{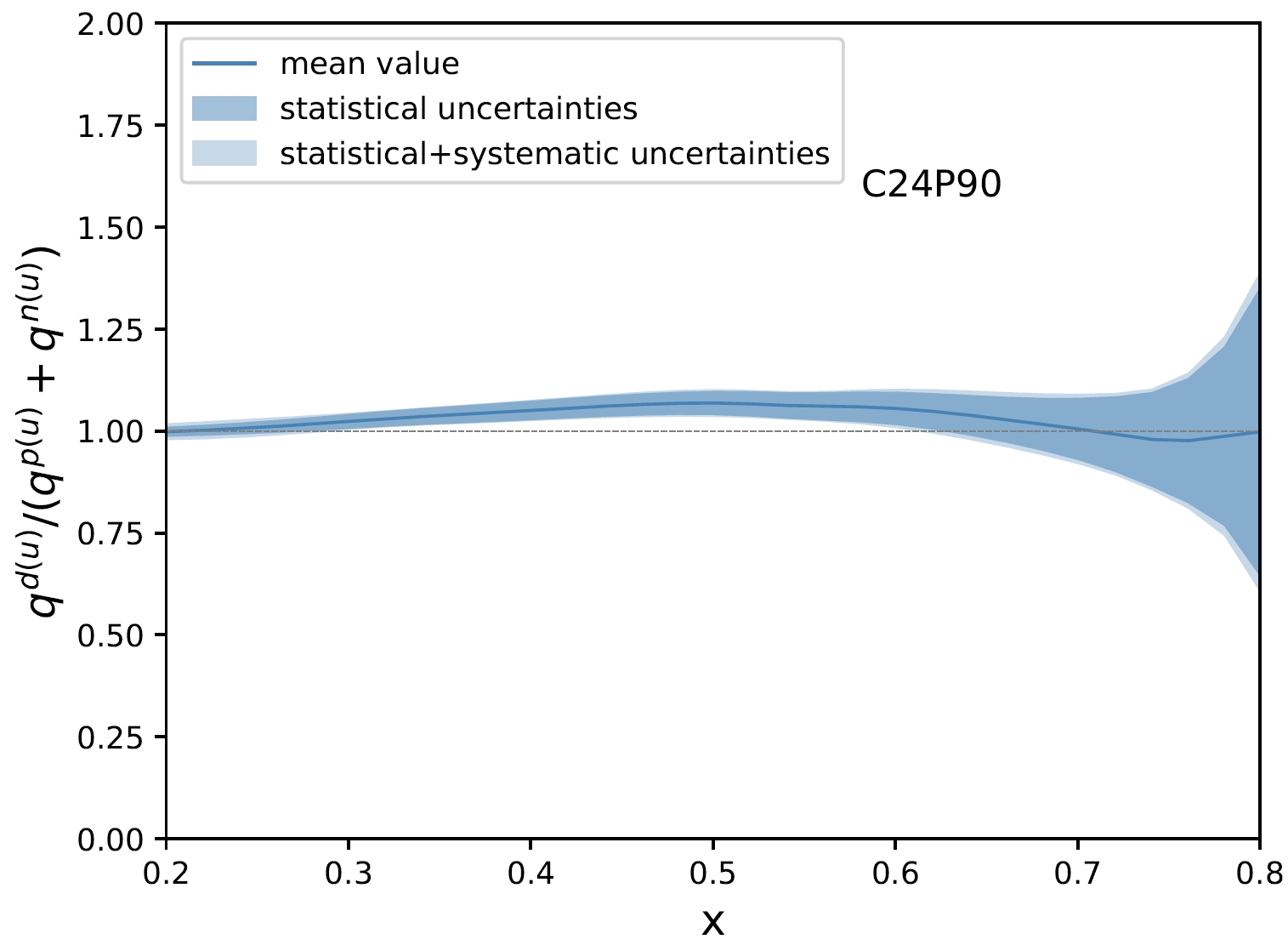}

\caption{The ratio of the deuteronlike dibaryon PDF to the sum of proton and neutron PDFs, with $x$ ranging from $0.2$ to $0.8$,  where the  LaMET theory is reliable. The deep blue bands indicate statistical uncertainties, while the light blue bands represent the combined effects of both statistical and systematic uncertainties. When $x$ is larger than $0.7$ divergences appear since nucleon results as denominators are very close to zero.    }
\label{D_P_ratio}
\end{figure*}

In Fig.~\ref{fig:lc_EMC}, we also show the comparison of the rescaled dibaryon PDF and the sum of proton and neutron PDFs, where the uncertainties include both statistical and systematic uncertainties. The light gray bands and shaded deep gray bands denote the regions where LaMET results are not reliable for  the dibaryon and nucleon, respectively, as revealed by the leading infrared renormalon(LRR)~\cite{Zhang:2023bxs} and renormalization group resummation(RGR)~\cite{Su:2022fiu} analysis in ~Appendix B.  At small $x$ regions, the LRR+RGR-improved results blow up, indicating that the higher twist contribution cannot be neglected. Similarly, at large 
$x$ regions, the same instability occurs. In reliable regions, the NLO results are consistent with those incorporating the LRR and RGR effects. For illustration purposes, in Fig.~\ref{D_P_ratio} we plot the ratio of the rescaled dibaryon PDF and the sum of proton and neutron PDFs, with $x$ ranging from $0.2$ to $0.8$, where the LaMET theory is reliable. When $x$ is larger than 0.7 divergence appears since nucleon results as denominators are very close to zero. One can see that at $m_\pi \sim 293$ MeV, the dibaryon PDF is generally smaller than the sum of proton and neutron PDFs, which is consistent with experiments. 
At $m_\pi \sim 941$ MeV, the dibaryon PDF is larger than the sum of proton and neutron PDFs. This may indicate a different binding nature of the two-nucleon system at different pion masses. Nevertheless, 
more precise results are required in future studies to provide reliable information on the nuclear effects in the deuteron.



\begin{figure}[htbp]
\includegraphics[width=.438\textwidth]{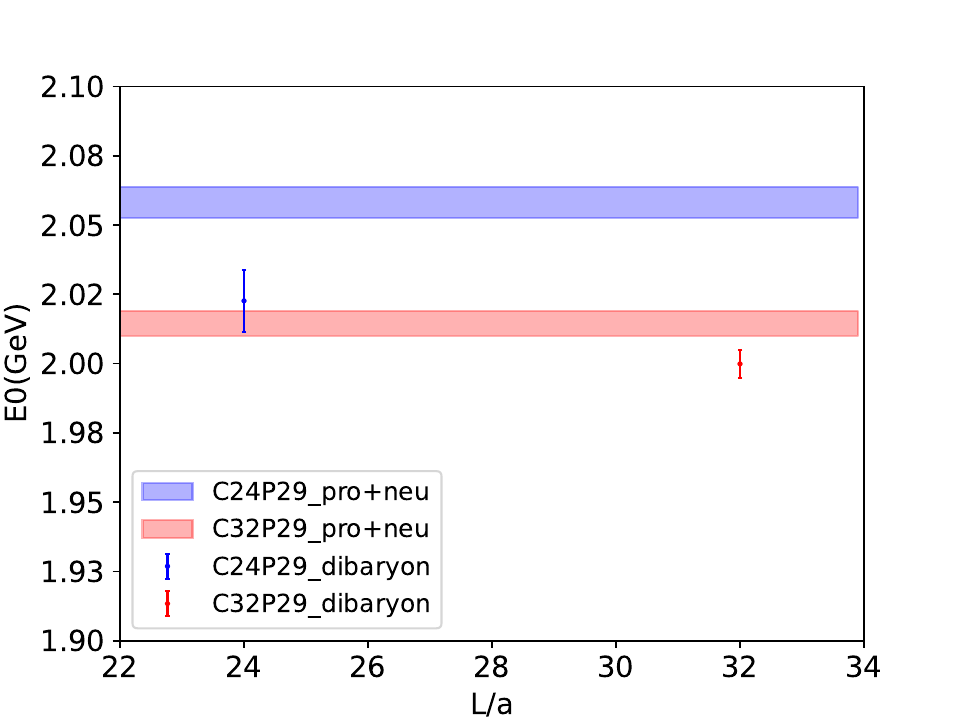}
  \includegraphics[width=.438\textwidth]{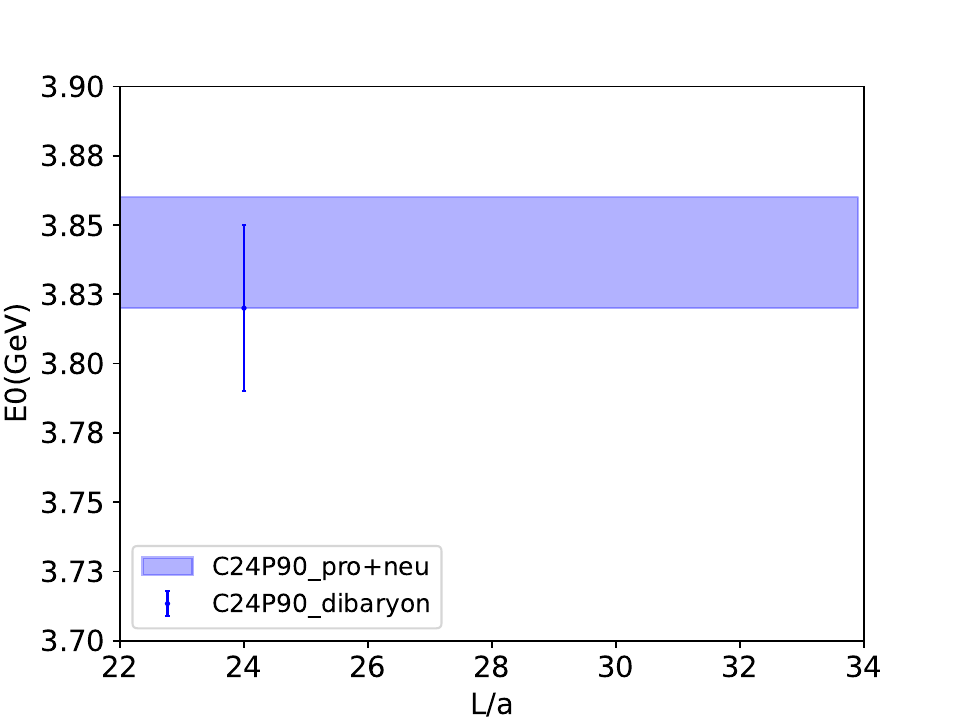}
\caption{ Comparison of the energy of the two-nucleon system (data points) and the free energy of the two nucleons (bands). Results of ensembles C24P29 and C32P29 ($m_\pi = 293$ MeV) are plotted in the upper panel and  results of ensemble C24P90 
 ($m_\pi = 941$ MeV) are plotted in the lower panel. }
\label{fig:eff_mass}
\end{figure}

Before ending this section, we should emphasize that determining whether a bound state exists for a multinucleon system is a critical issue.  Various studies have utilized either the L\"uscher's finite-volume formalism~\cite{Beane:2009py,NPLQCD:2011naw,NPLQCD:2012mex,Orginos:2015aya,Berkowitz:2015eaa,Wagman:2017tmp,Horz:2020zvv,Amarasinghe:2021lqa,Green:2021sxb} or the HALQCD method~\cite{Ishii:2006ec,Aoki:2009ji,Ishii:2012ssm,Inoue:2011ai}  with pion masses ranging from $300$ to $1200$~MeV,  to explore the existence of a bound state in multinucleon systems. However, despite extensive discussions and research, a consensus on the existence of bound states remains elusive. A rigorous determination of bound states on our ensembles is beyond the scope of this study. We compute the spectrum of the two-nucleon system and compare it with the energy of two free nucleons, as shown in Fig.~\ref{fig:eff_mass}. For the ensembles C24P29 and C32P29, the distillation quark smearing method~\cite{HadronSpectrum:2009krc} is utilized to improve the precision. It can be seen that the energy of the two-nucleon system is slightly lower than the free energy for the ensembles C24P29 and C32P29, indicating attractive interactions. However, we cannot conclude whether there is a bound state. To answer this question, we will calculate the scattering amplitude of two nucleons using L\"uscher's finite-volume method in a follow-up study.  

\begin{figure}[htbp]
\includegraphics[width=.438\textwidth]{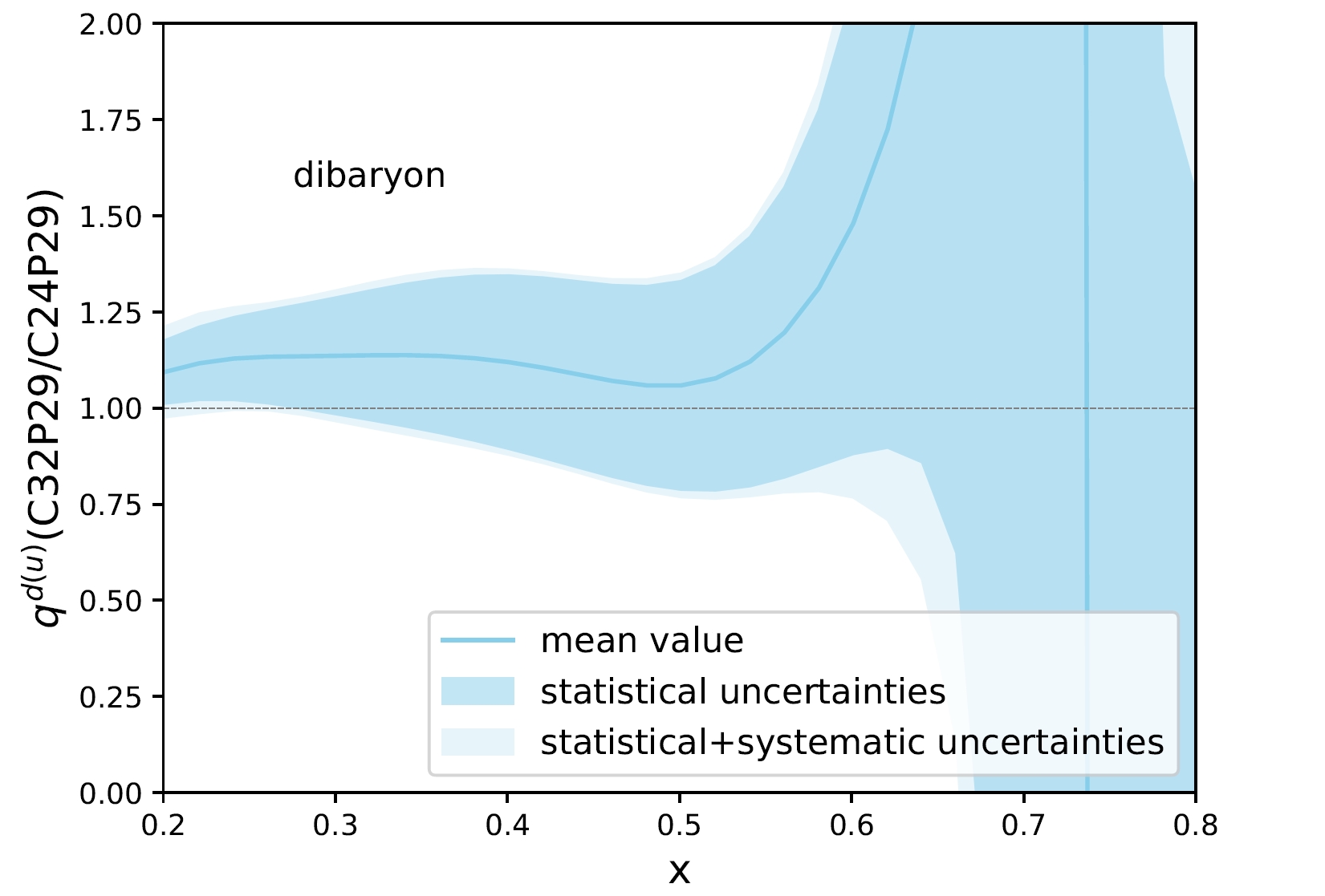}
\includegraphics[width=.438\textwidth]{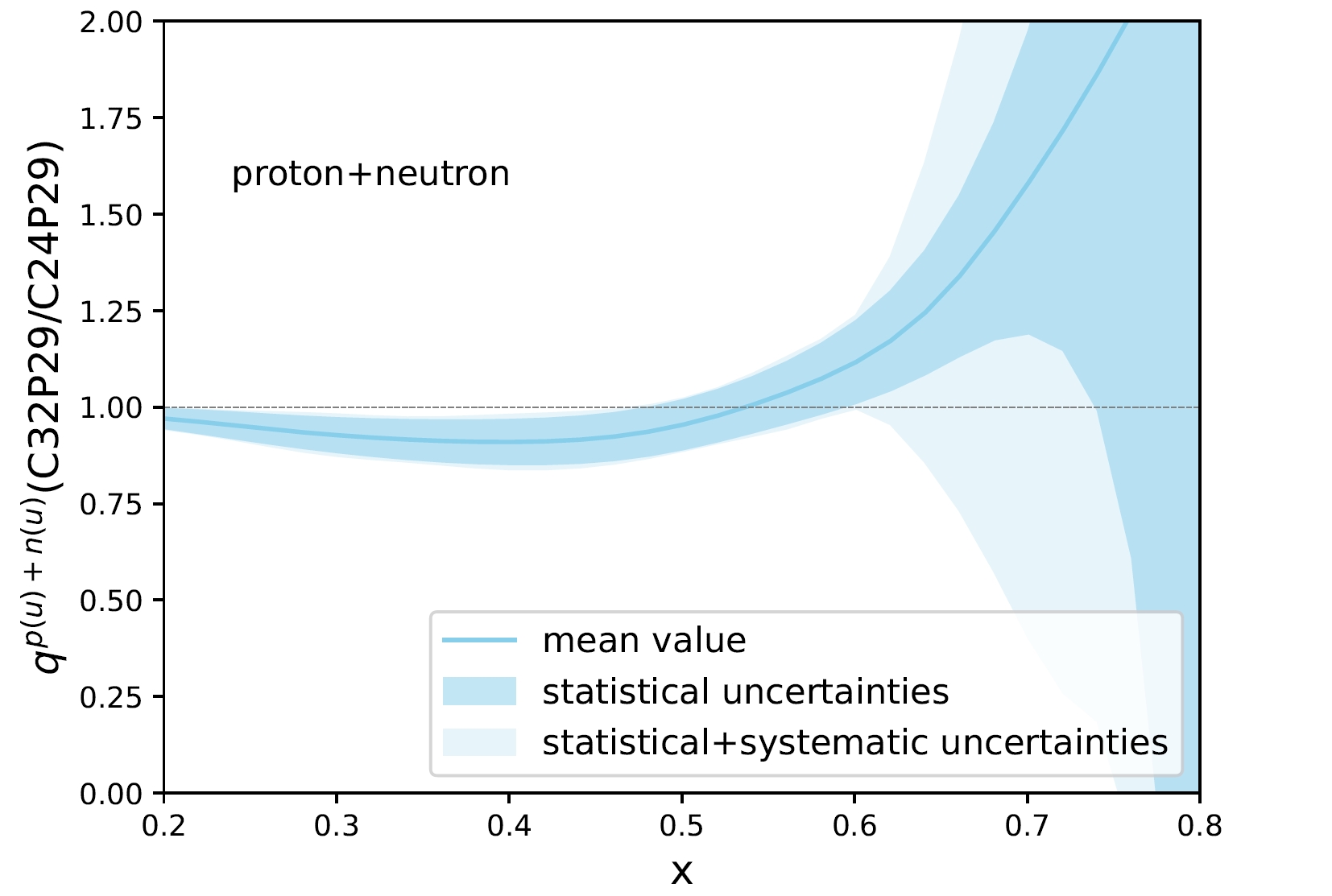}

\caption{The the ratio of  light-cone PDF of C32P29 to that of  C24P29. The results of the dibaryon system are plotted in the upper panel, while the results of the sum of the proton and neutron are plotted in the lower panel.  The deep bands indicate statistical uncertainties, while the light bands represent the combined effects of both statistical and systematic uncertainties. }
\label{fig:IVE}
\end{figure}

On the other hand, it is also important to estimate the impact of finite-volume effects in lattice QCD calculations. There has been research on the effects of finite-volume on the hadron spectrum~\cite{Luscher:1986pf,Kim:2005gf,Christ:2005gi} and hadron structure ~\cite{Beane:2004rf,Green:2012ej,Bhattacharya:2015wna,Bhattacharya:2016zcn,Gupta:2018qil,Lin:2019ocg}.  
Though no obvious finite-volume effect has been discovered for nucleon PDF, as seen in~\cite{Lin:2019ocg}, the situation could be ambiguous for multinucleon systems. In Fig.\ref{fig:IVE} we plot the ratio of the PDF of C32P29 to that of C24P29 to estimate the finite-volume effects. As illustrated in the plot, the results are roughly equal to one within uncertainties. 
In order to determine the finite-volume effects, we need many more statistics and larger-volume ensembles. We will address such issues in our future work.

\section{Summary}
\label{section E}
In this paper, we presented the first lattice QCD calculation of the quark PDF of the deuteronlike dibaryon system via the LaMET method.
Our calculation was done on ensembles with a single lattice spacing. With high statistics, 
we have used multiple source-sink separations to control the excited-state contamination and applied the state-of-the-art renormalization, matching, and extrapolation procedure. We have also studied the momentum dependence, and extrapolated the result to the infinite-momentum limit.  
Finally, we compared the PDF of a deuteronlike dibaryon system with that of free nucleons. 
The current precision of our calculation was not sufficient enough to draw definitive conclusions on the nuclear effects of dibaryon systems.   



In the upcoming EIC and EicC experiments, 
detecting the nuclear effects related to nuclear PDFs will be one of the most important physical goals. An $ab~initio$ calculation of deuteron PDFs will help us understand the related phenomenology.  As the first-ever calculation of deuteronlike dibaryon PDFs on lattices, our study requires improvements in various aspects, which will be addressed in future work.

Our analysis was based on a single lattice spacing of  $0.105$ fm, which means that discretization errors have not been accounted for.  Additionally, the pion masses of the ensembles used were $293$ MeV and $941$ MeV, creating challenges in performing a reliable chiral extrapolation due to the significant gap between these values.  Furthermore, at the nonphysical pion
mass, it is still an open question whether a bound-state
deuteron exists or not on the lattice, complicating the determination of finite-volume effects.   
Nevertheless, in this work, we studied, for the first time, the PDFs of a multinucleon system. Our final results demonstrated the potential to detect nuclear effects by investigating PDFs through the LQCD approach, with the intention of addressing these challenges in future studies.

\section*{ACKNOWLEDGEMENTS}
We thank the CLQCD collaborations for providing us with their gauge configurations with dynamical fermions~\cite{Hu:2023jet}, which are generated on HPC cluster of ITP-CAS, the Southern Nuclear Science Computing Center(SNSC), the Siyuan-1 cluster supported by the Center for High-Performance Computing at Shanghai Jiao Tong University, and the Dongjiang Yuan Intelligent Computing Center. We are grateful to Ying Chen for the valuable discussion and comments. We thank Wei Wang for his constructive feedback on our work. We would like to thank Xiaonu Xiong for providing us a template for plotting the three-point correlator.  We would also like to thank Yushan Su
for providing the $Mathematica$ notebook to confirm the correct implementation of RGR and LRR. The calculations were performed using the Chroma software suite~\cite{Edwards:2004sx} with QUDA~\cite{Clark:2009wm,Babich:2011np,Clark:2016rdz} through the HIP programming model~\cite{Bi:2020wpt} and QCU software. 
The numerical calculations in this paper were carried out at the Dongjiang Yuan Intelligent Computing Center,  the HPC cluster of ITP-CAS and  the Southern Nuclear Science Computing Center(SNSC).  This work is supported in part by National Natural Science Foundation of China (NSFC) under Grant No. 12293060, No. 12293061, No. 12293062, No. 12175279, No. 12293065, No. 12047503, No. 12435002, No. 12375080 and No. 11975051, the Strategic Priority Research Program of the Chinese Academy of Sciences with Grant No. XDB34030301, No. XDB34030303 and No. YSBR-101, the Guangdong Major Project of Basic and Applied Basic Research No. 2020B0301030008,  the Education Integration Young Faculty Project of University of Chinese Academy of Sciences, CUHK-Shenzhen under grant No. UDF01002851, and a NSFC-DFG joint grant under Grant No. 12061131006 and SCHA 458/22 and the GHfund A No.\ 202107011598. 

\clearpage

\subfile{APPENDIX.tex}

\clearpage

\bibliographystyle{apsrev}
\bibliography{ref}

\end{document}

%% file: APPENDIX.tex
\section{APPENDIX }\label{sec:appdx}
\subsection{Details on data analysis}
\label{subsec：data_ana}

\begin{figure*}[htbp]
\includegraphics[width=.3\textwidth]{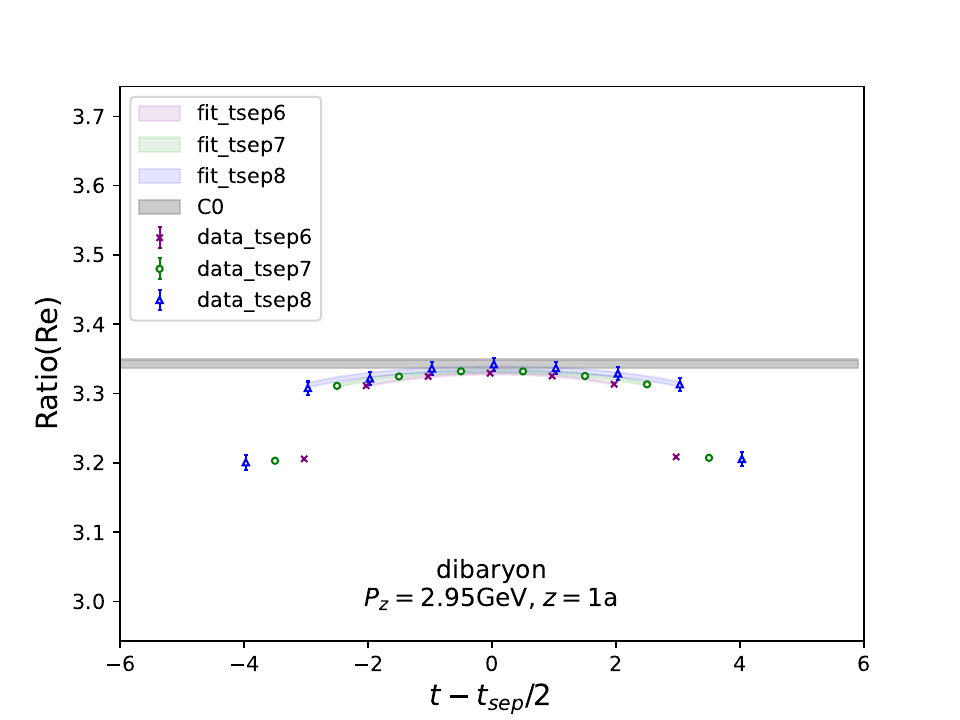}
\includegraphics[width=.3\textwidth]{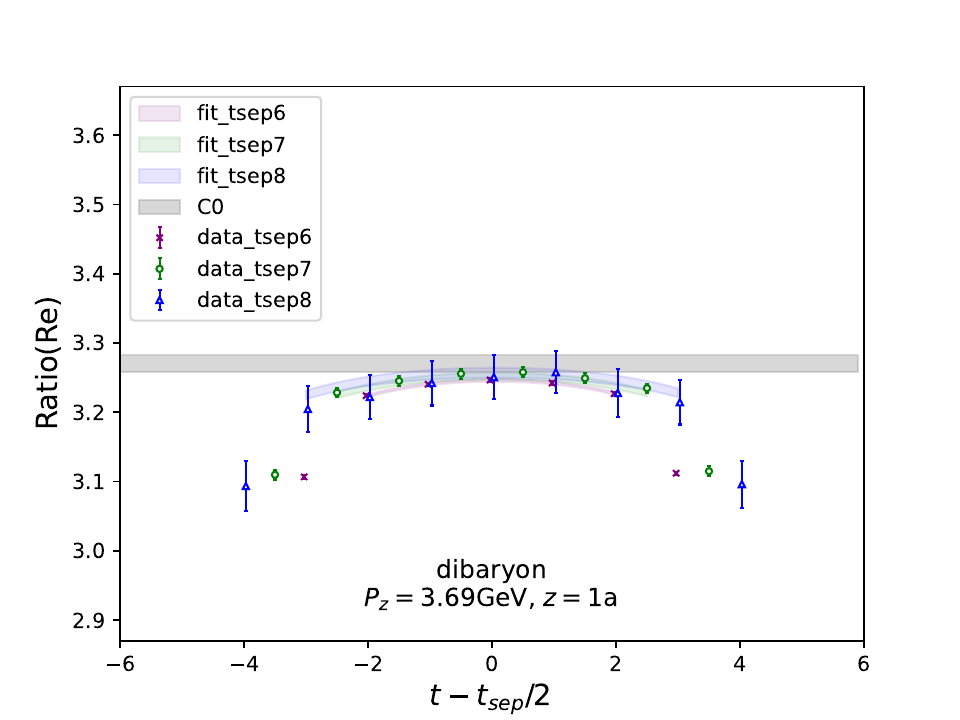}
\includegraphics[width=.3\textwidth]{figures/hybrid/P29M/fit_deu_mom6_z1_re.pdf}\\
  
\includegraphics[width=.3\textwidth]{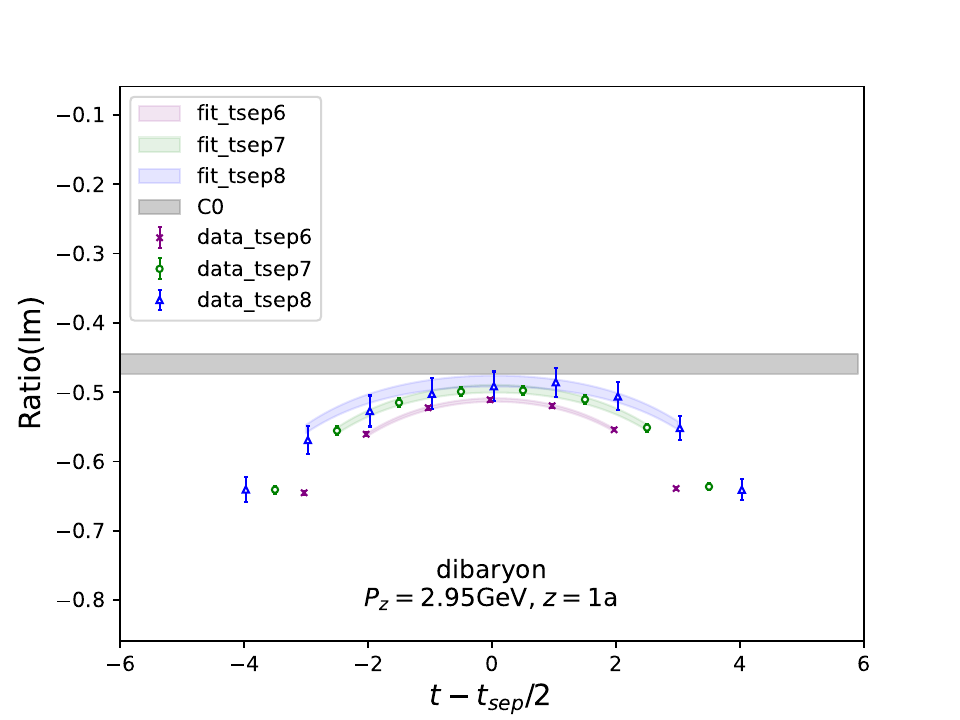}
\includegraphics[width=.3\textwidth]{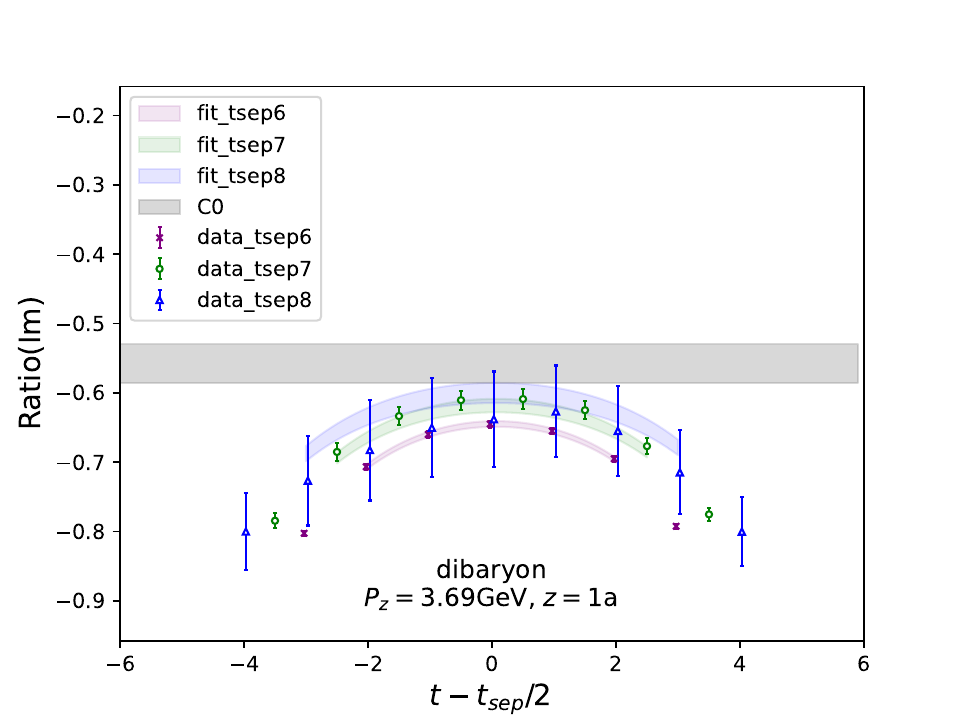}
\includegraphics[width=.3\textwidth]{figures/hybrid/P29M/fit_deu_mom6_z1_im.pdf}
\\
\includegraphics[width=.3\textwidth]{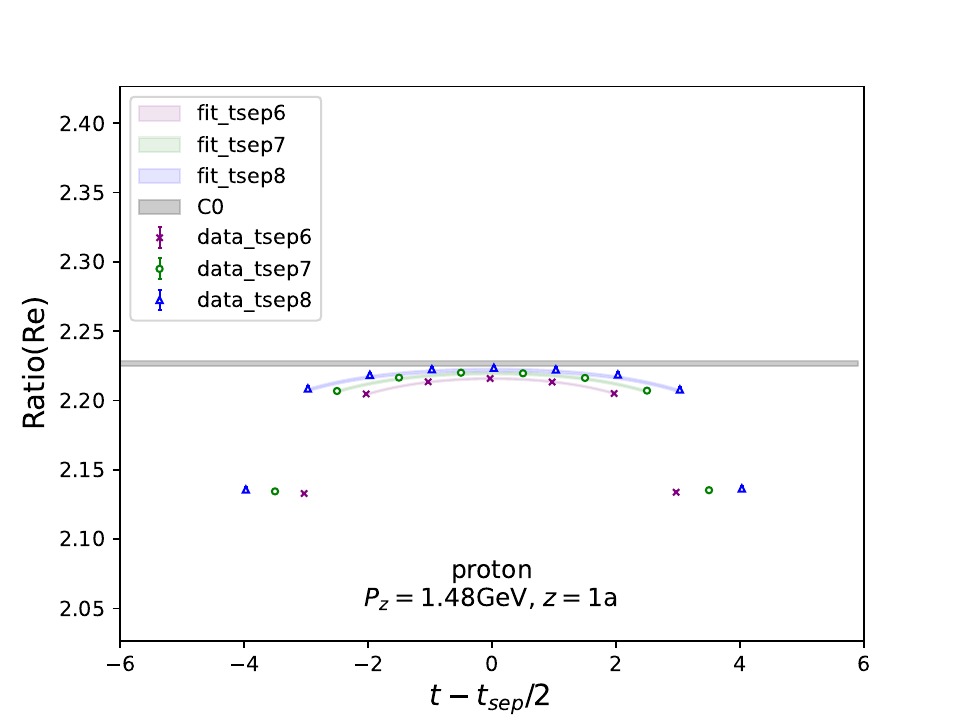}
\includegraphics[width=.3\textwidth]{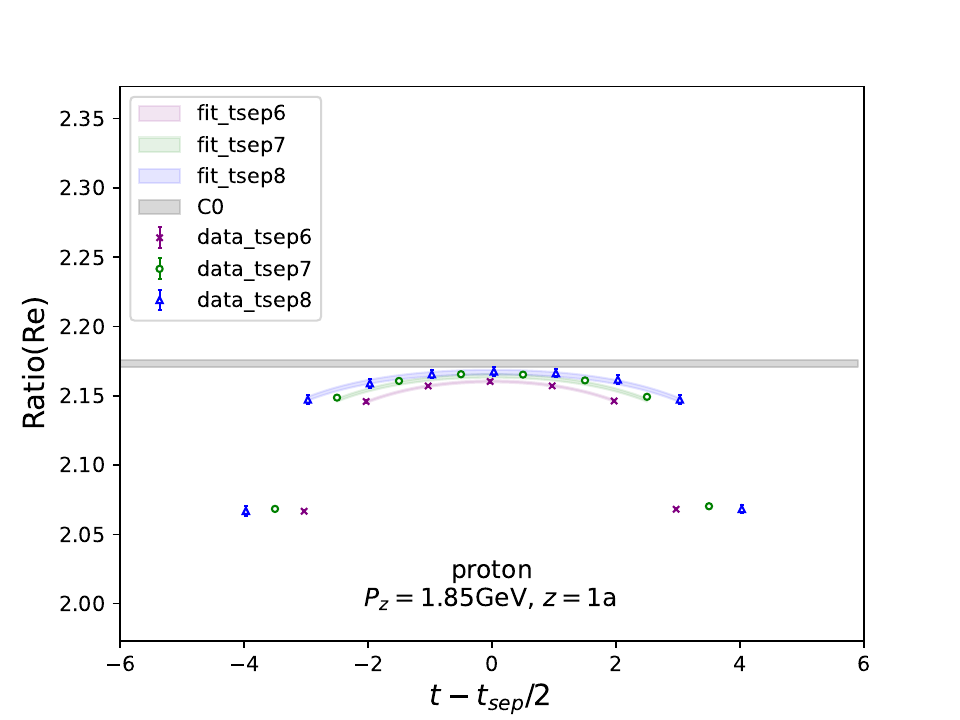}
\includegraphics[width=.3\textwidth]{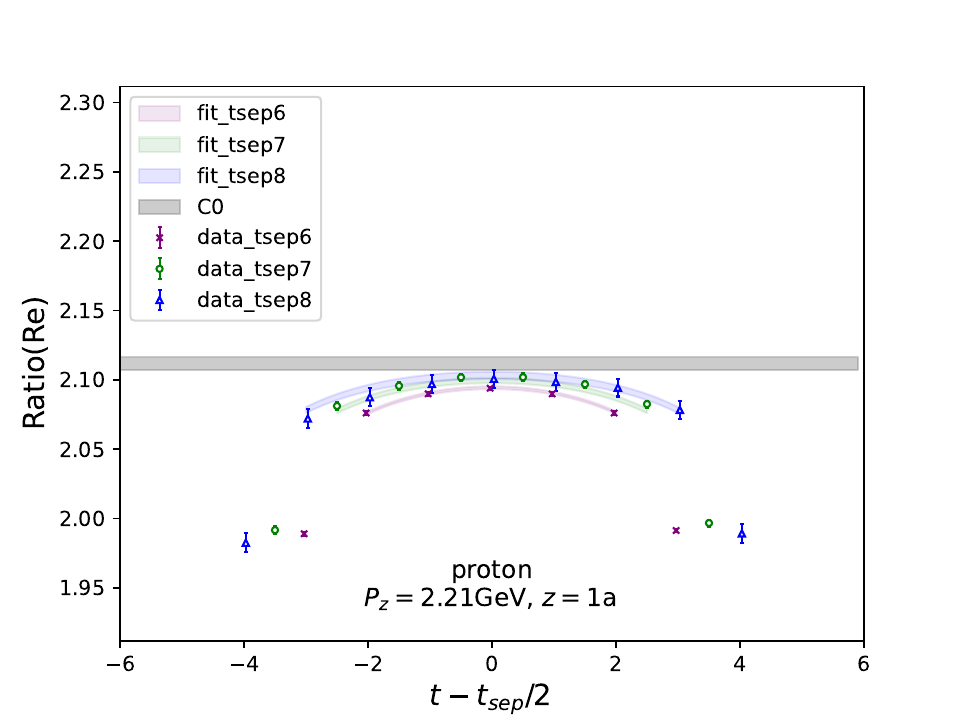}\\
\includegraphics[width=.3\textwidth]{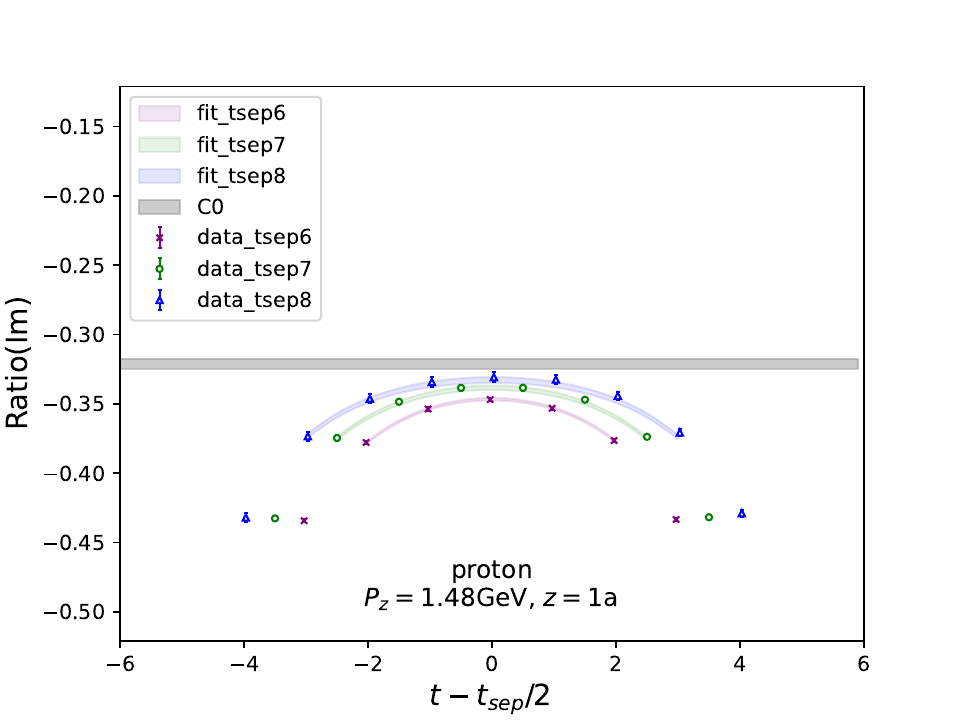}
\includegraphics[width=.3\textwidth]{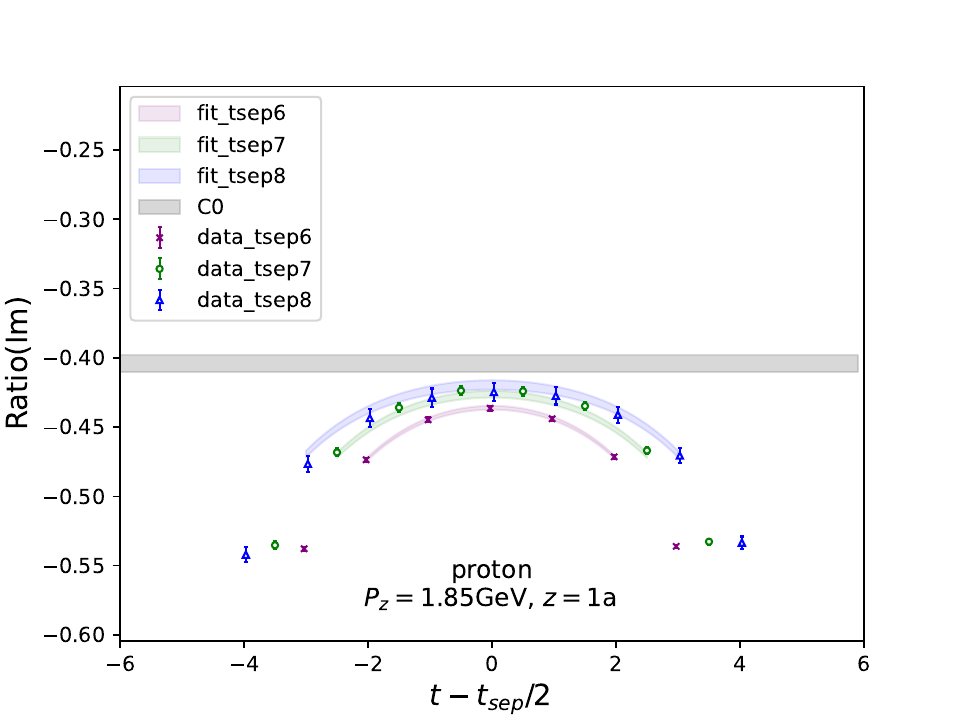}
\includegraphics[width=.3\textwidth]{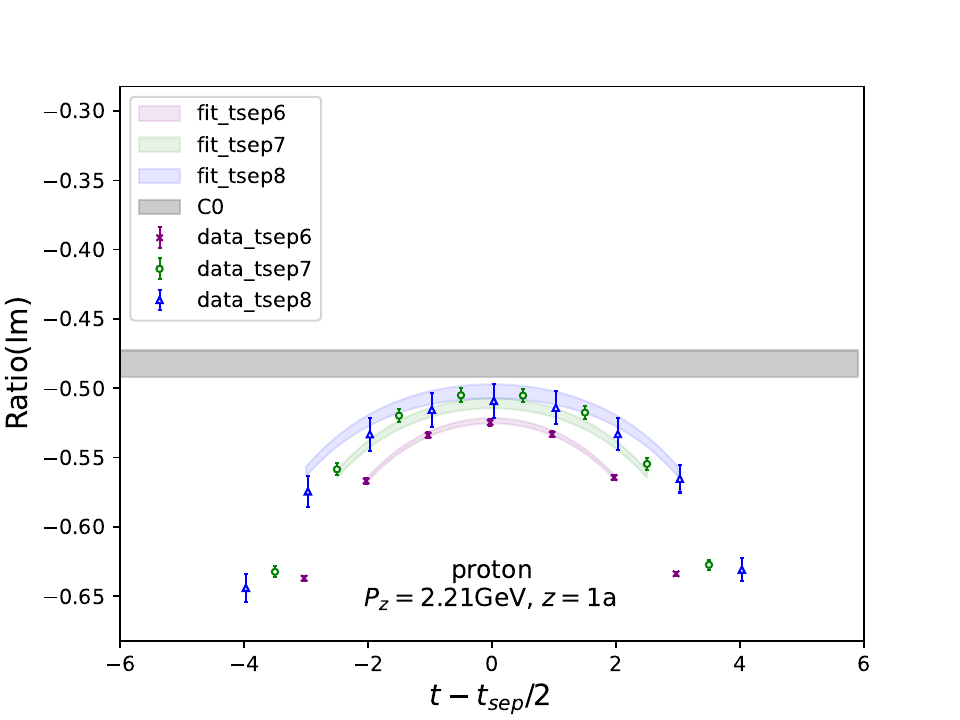}
\\
\includegraphics[width=.3\textwidth]{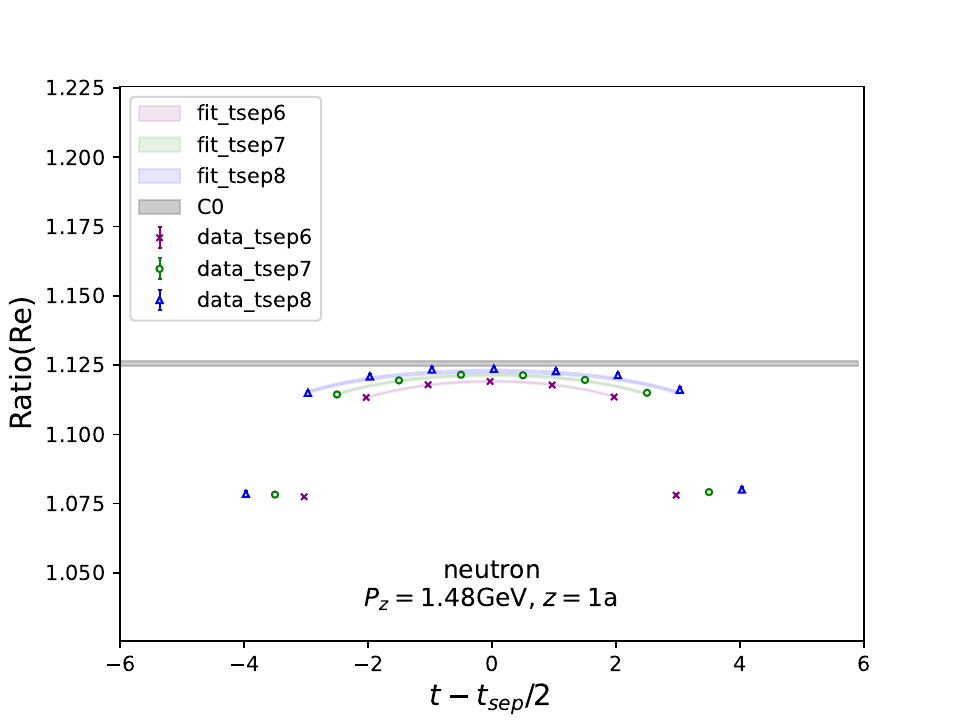}
\includegraphics[width=.3\textwidth]{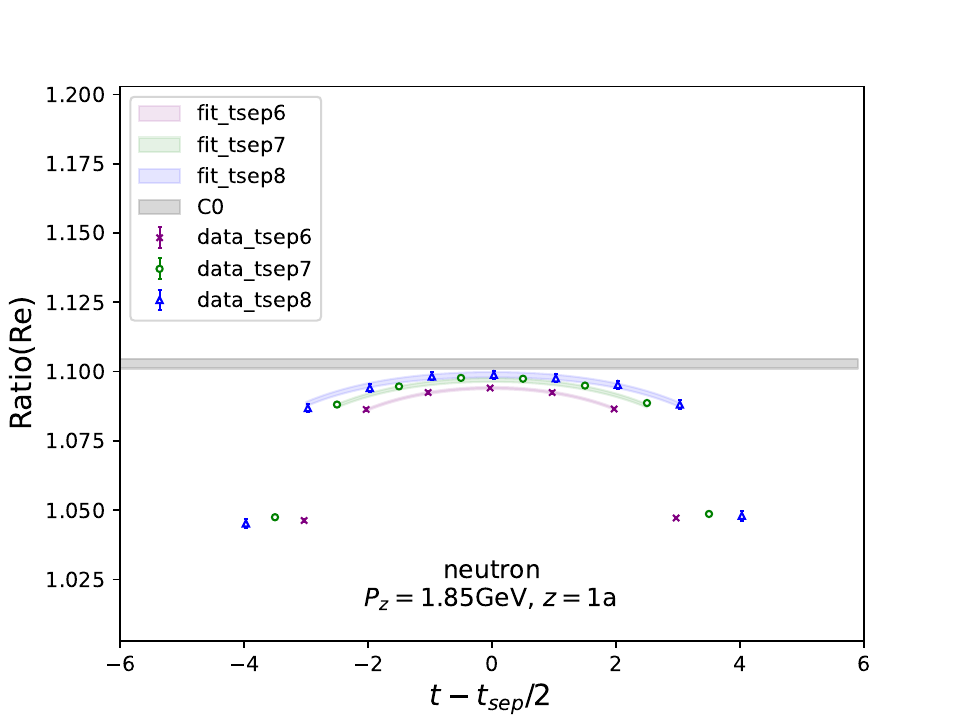}
\includegraphics[width=.3\textwidth]{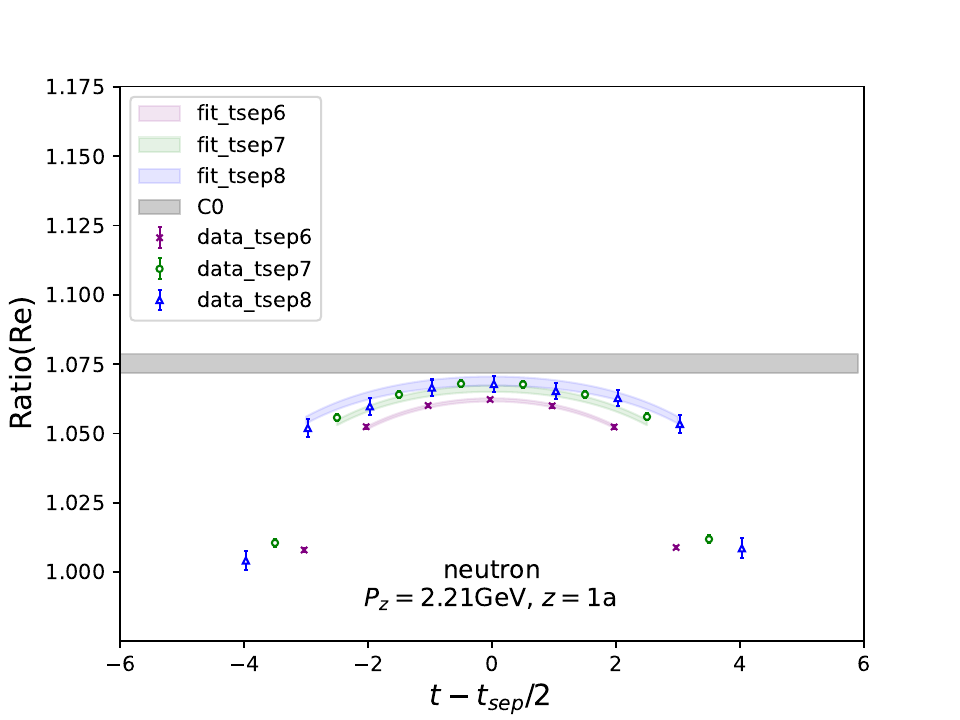}\\
\includegraphics[width=.3\textwidth]{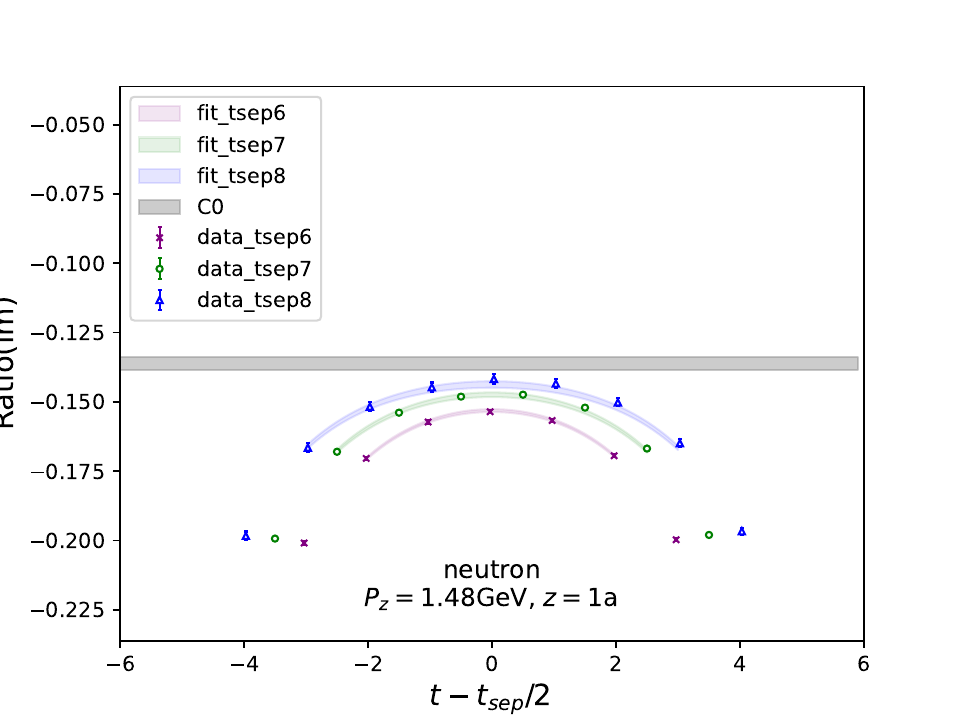}
\includegraphics[width=.3\textwidth]{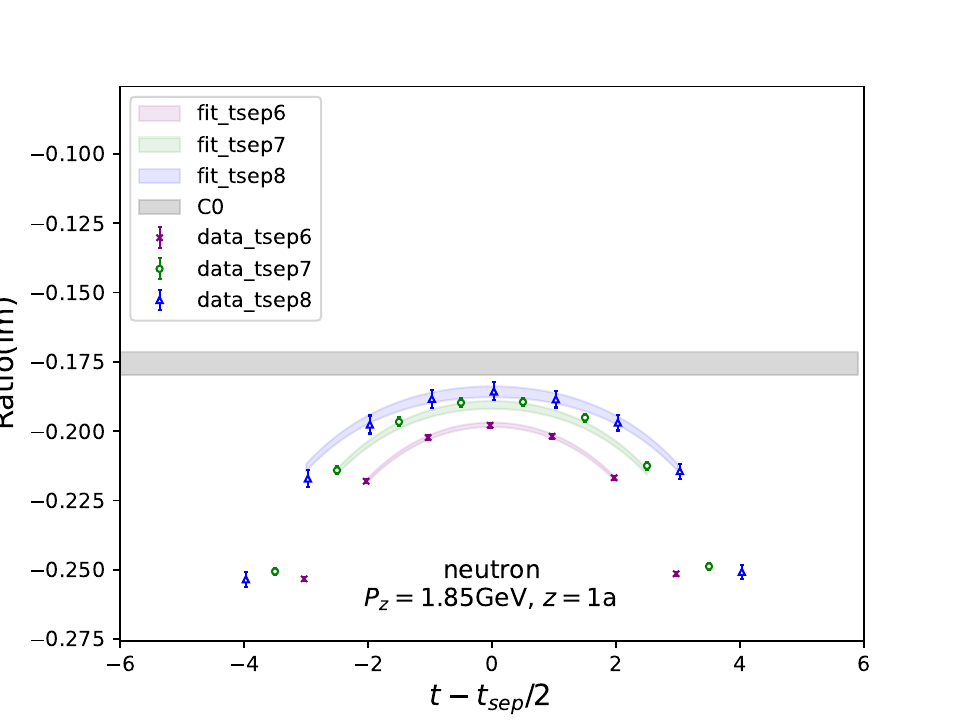}
\includegraphics[width=.3\textwidth]{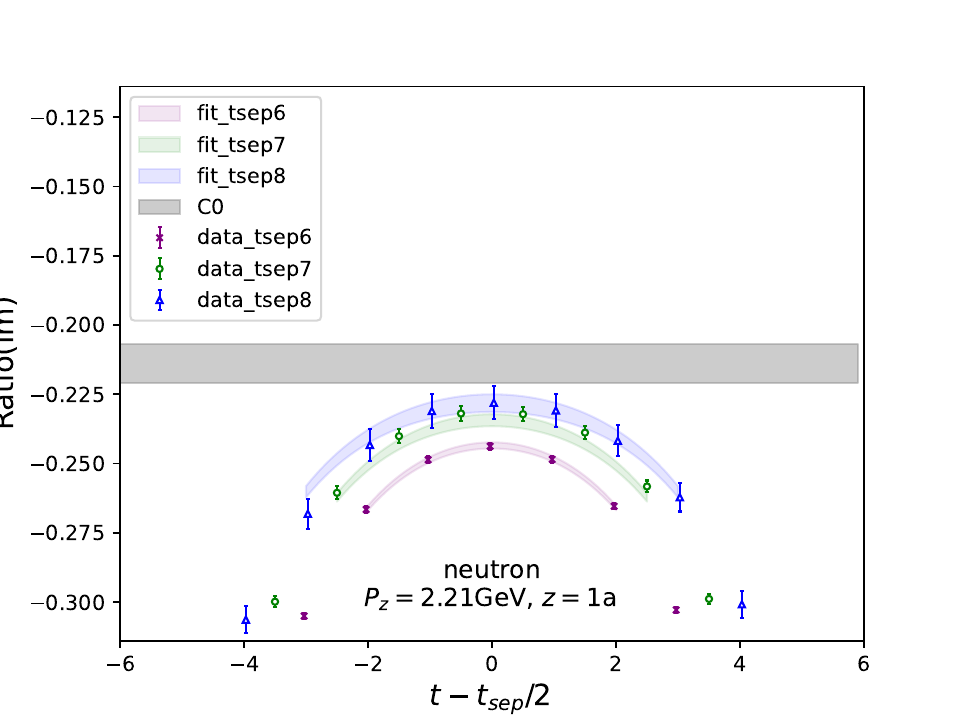}
\\
\caption{ Demonstration of    fitting the correlation function of ensemble C32P29. Here we represent the result of $z=1a$ of different momenta. We compare the lattice data of ratios (error bars)  and  results predicted by fitting (colored bands). The ground-state contribution $c_0$ obtained by the fitting is shown as the black band.}
\label{fig:fit_C32P29}
\end{figure*}

\begin{figure*}[htbp]
\includegraphics[width=.3\textwidth]{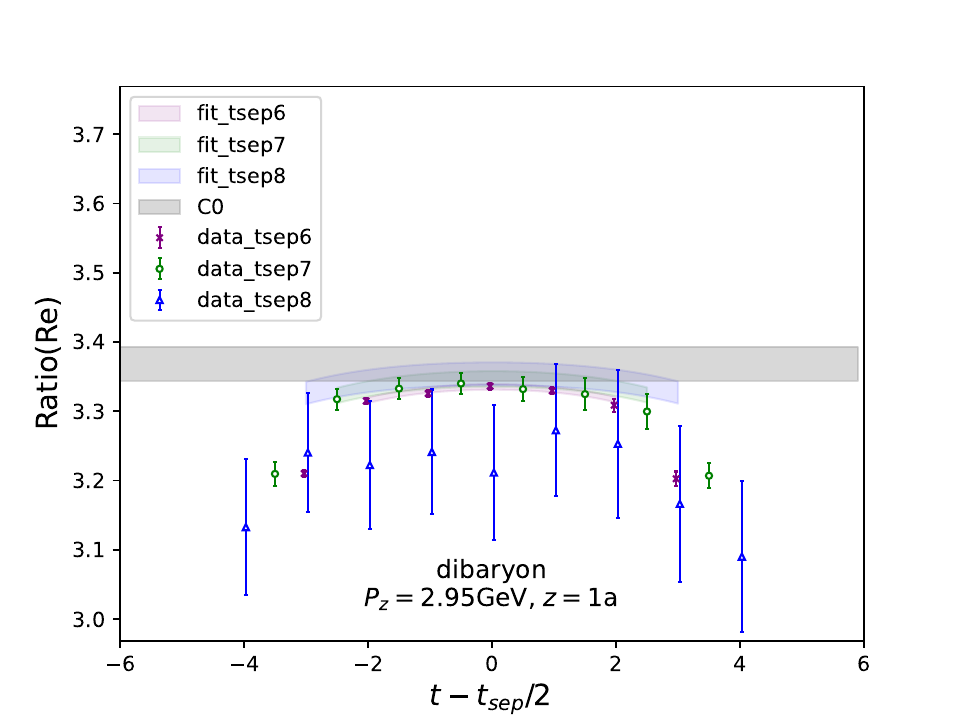}
\includegraphics[width=.3\textwidth]{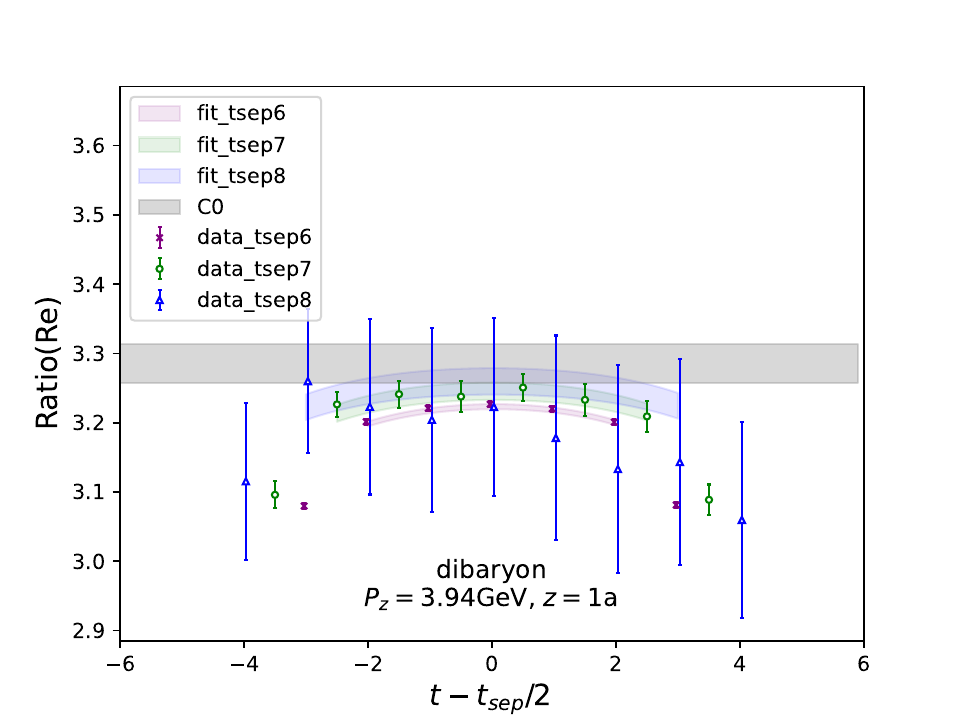}
\includegraphics[width=.3\textwidth]{figures/hybrid/P29S/fit_deu_mom5_z1_re.pdf}\\
  
\includegraphics[width=.3\textwidth]{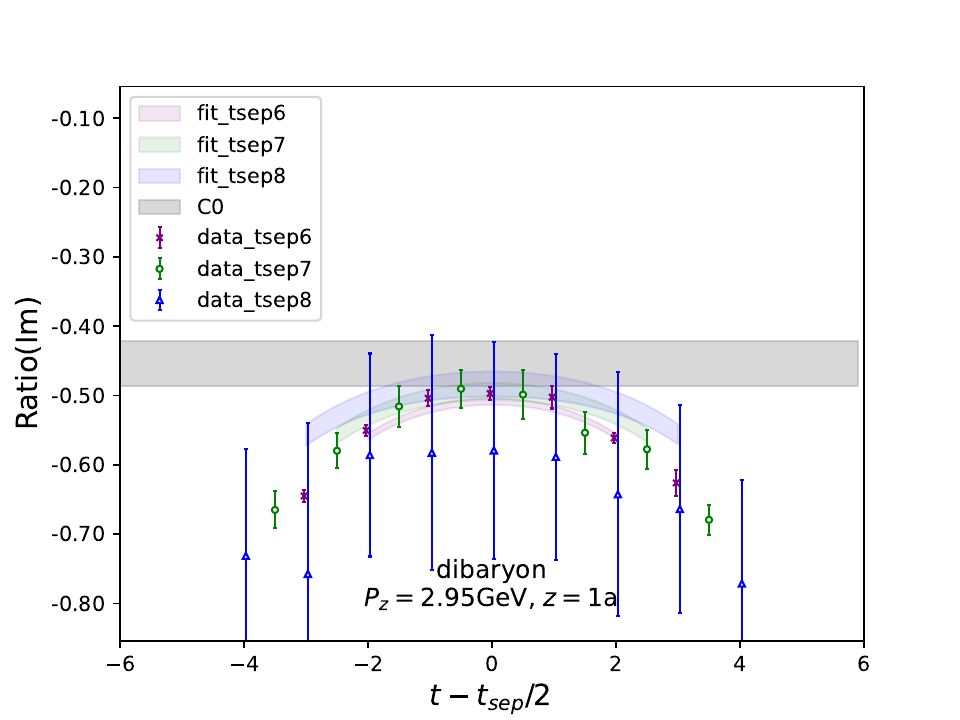}
\includegraphics[width=.3\textwidth]{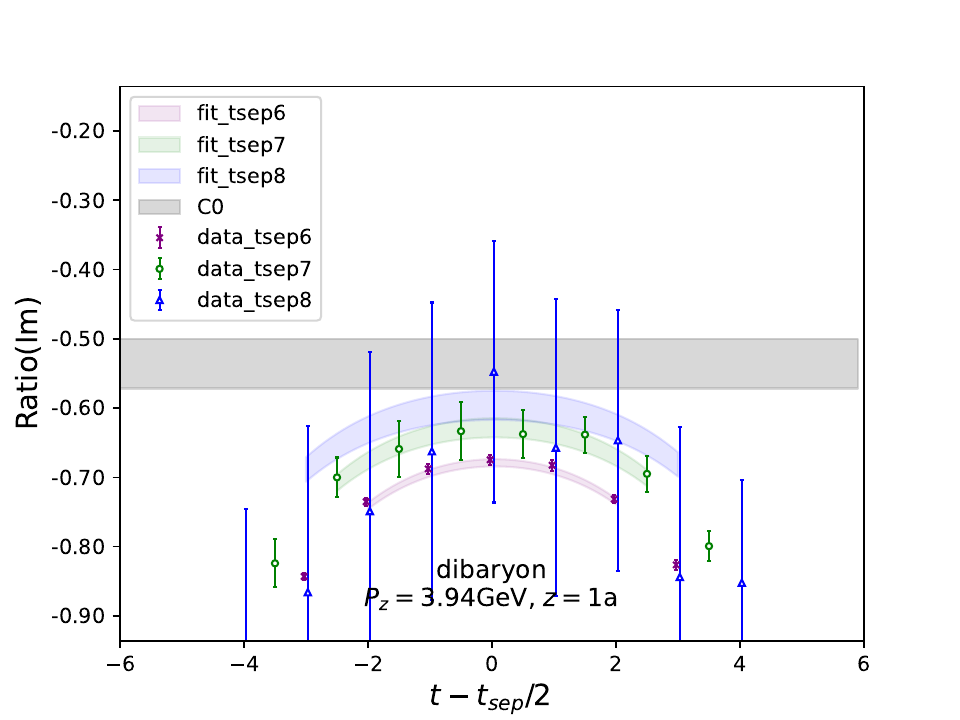}
\includegraphics[width=.3\textwidth]{figures/hybrid/P29S/fit_deu_mom5_z1_im.pdf}
\\
\includegraphics[width=.3\textwidth]{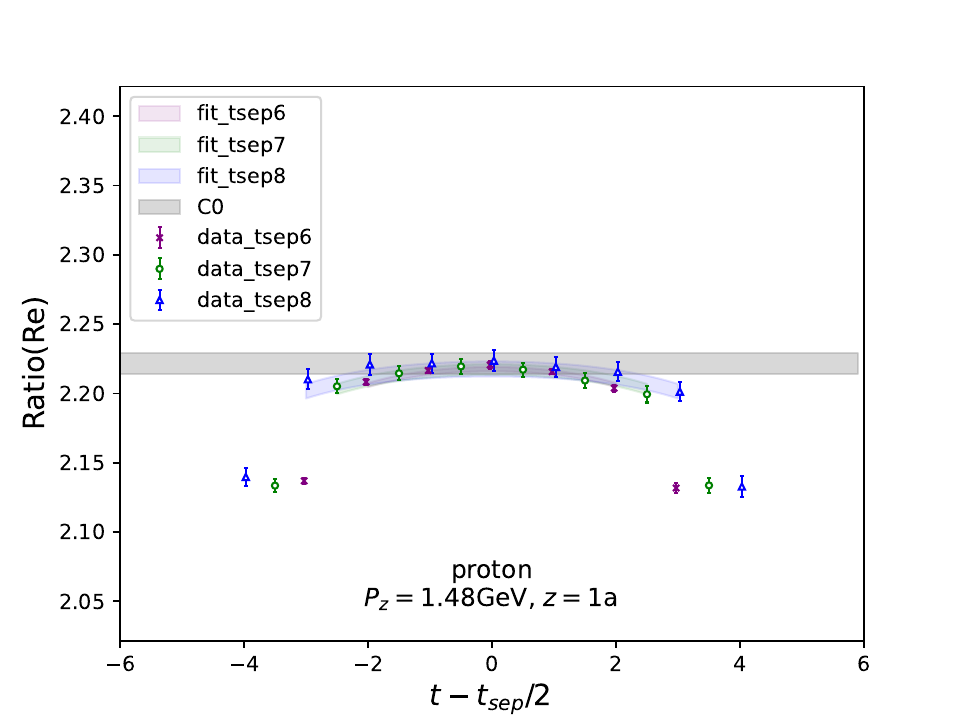}
\includegraphics[width=.3\textwidth]{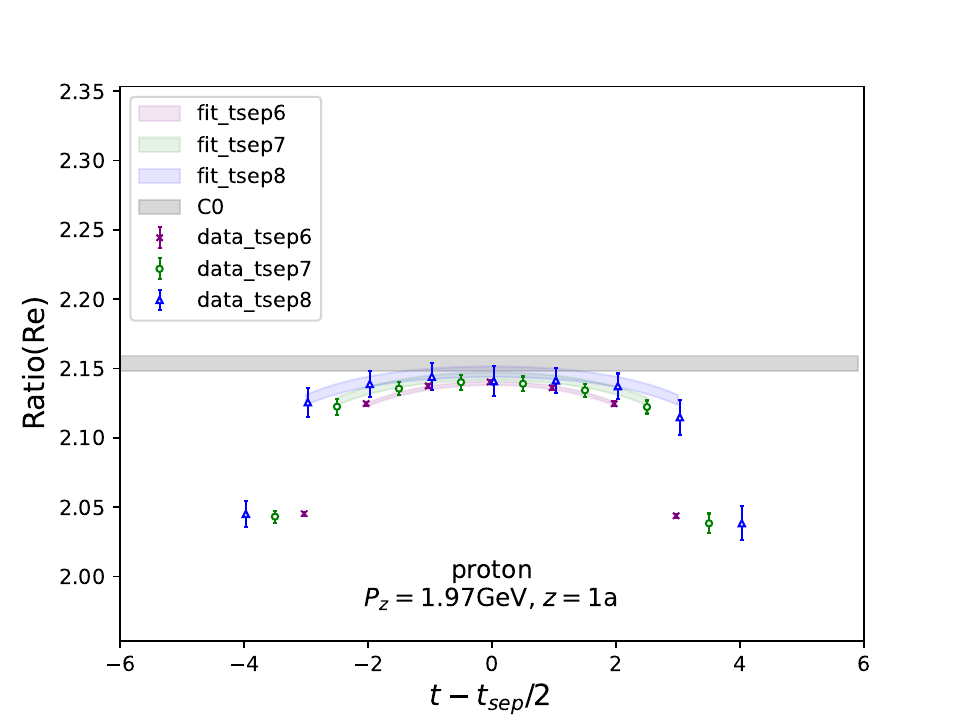}
\includegraphics[width=.3\textwidth]{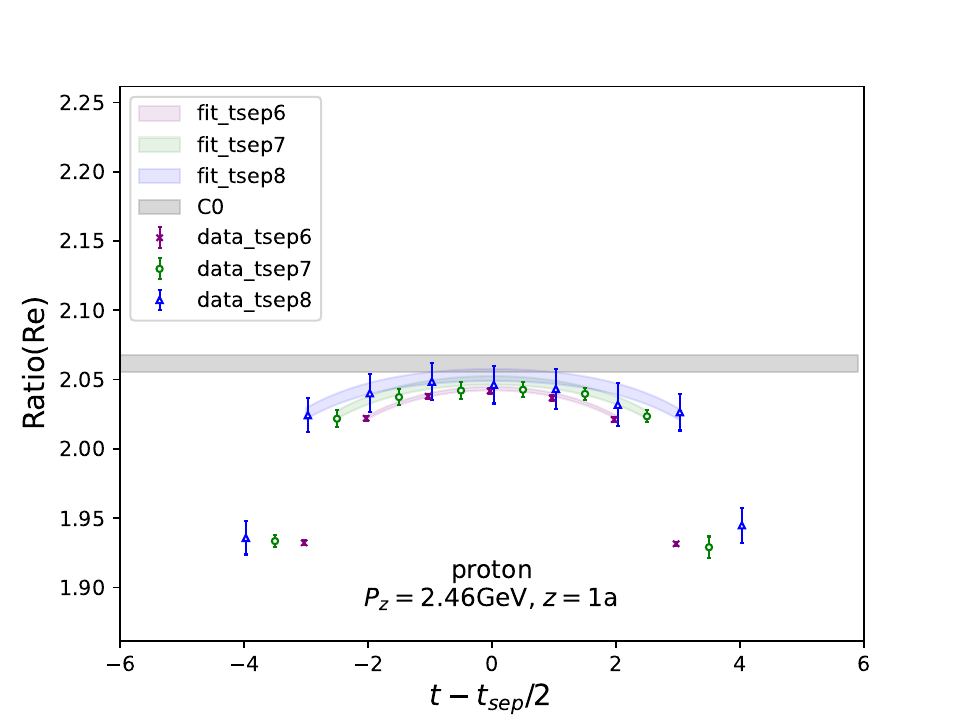}\\
\includegraphics[width=.3\textwidth]{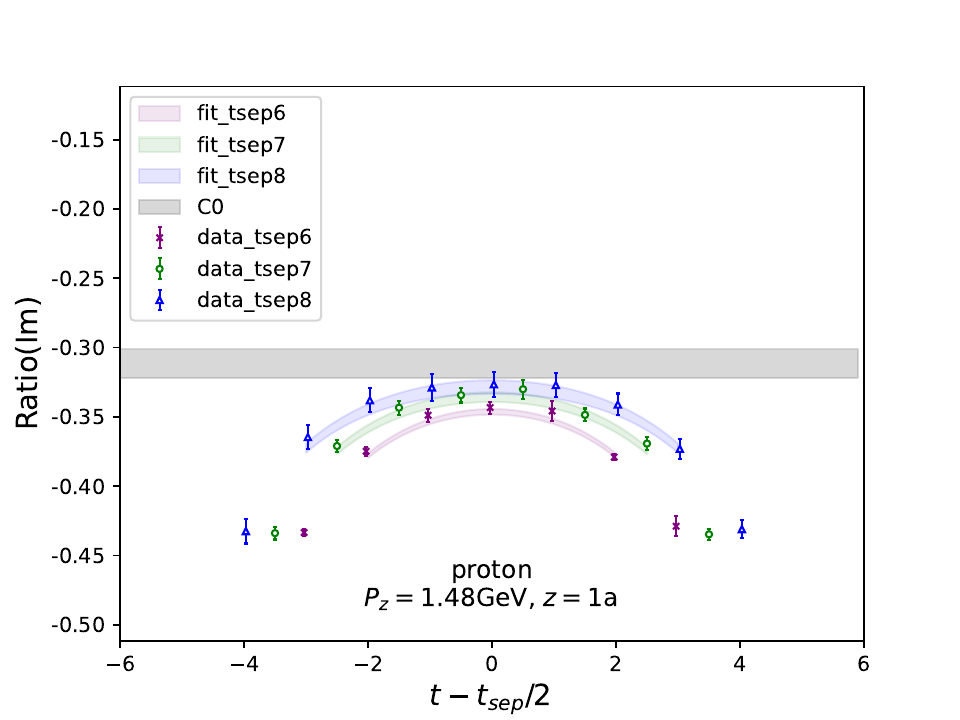}
\includegraphics[width=.3\textwidth]{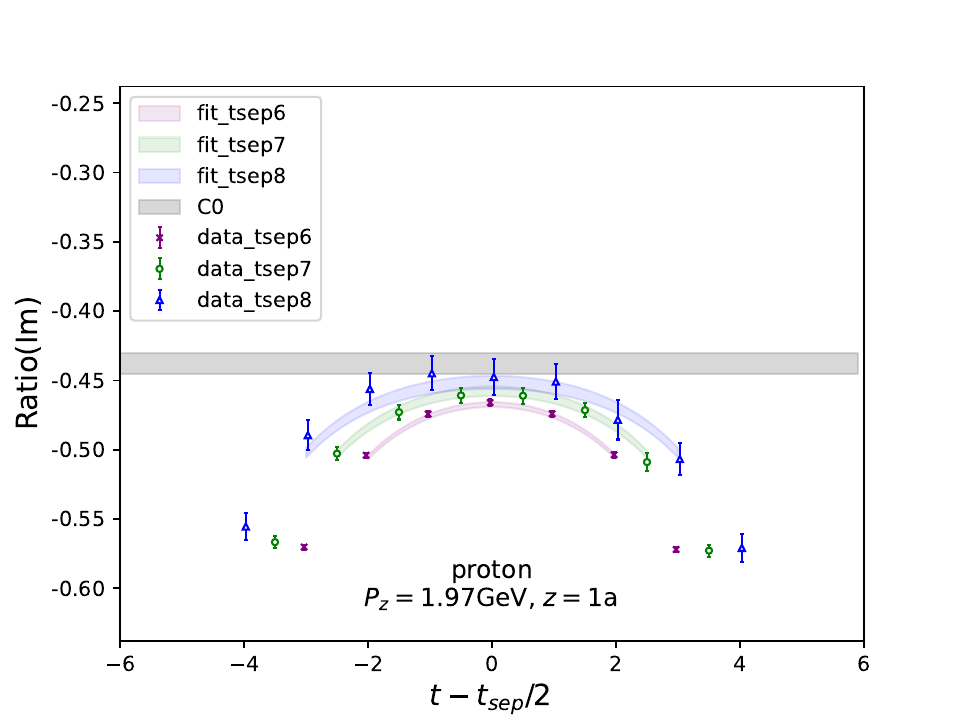}
\includegraphics[width=.3\textwidth]{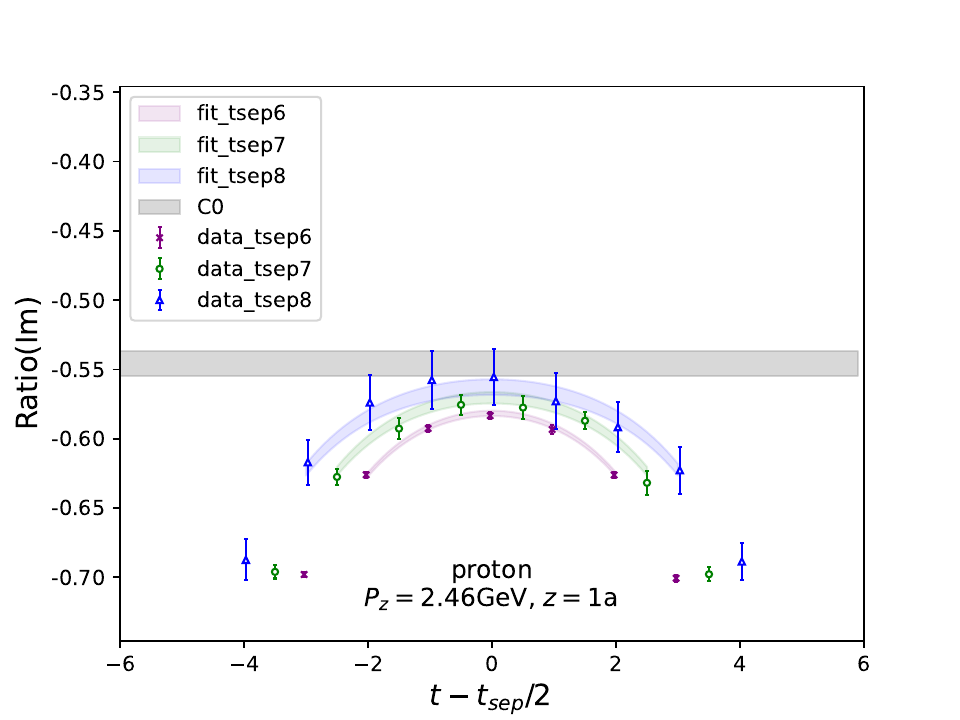}
\\
\includegraphics[width=.3\textwidth]{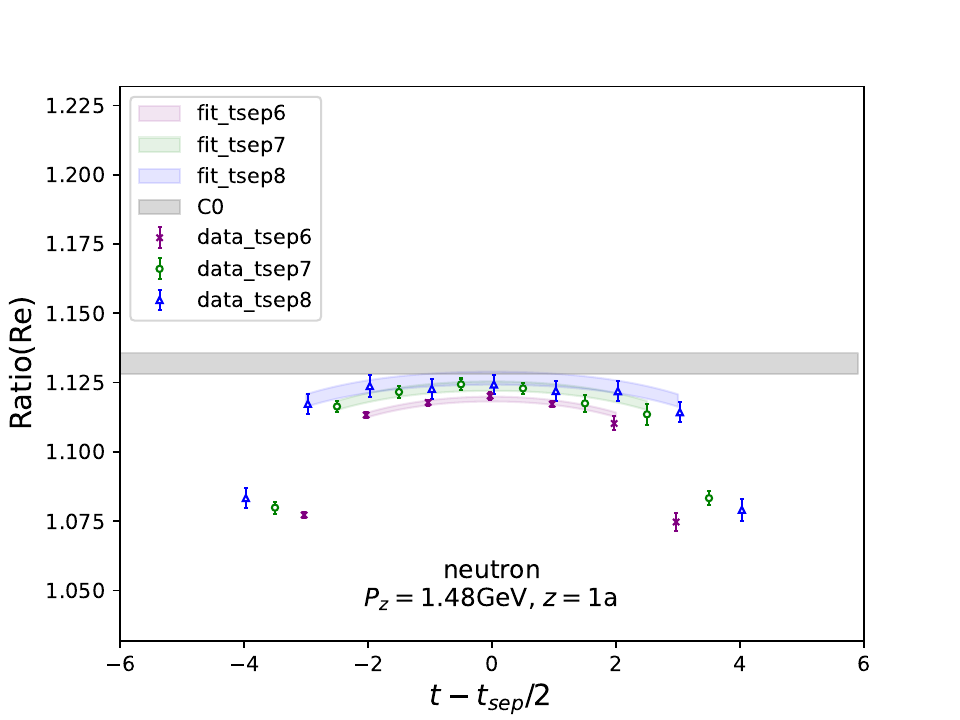}
\includegraphics[width=.3\textwidth]{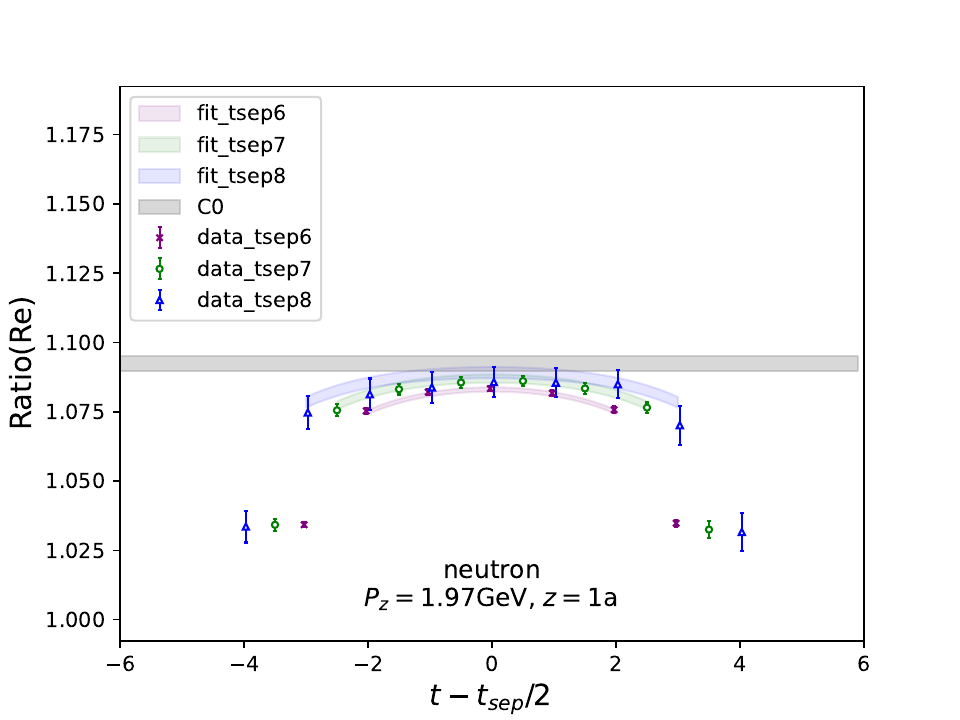}
\includegraphics[width=.3\textwidth]{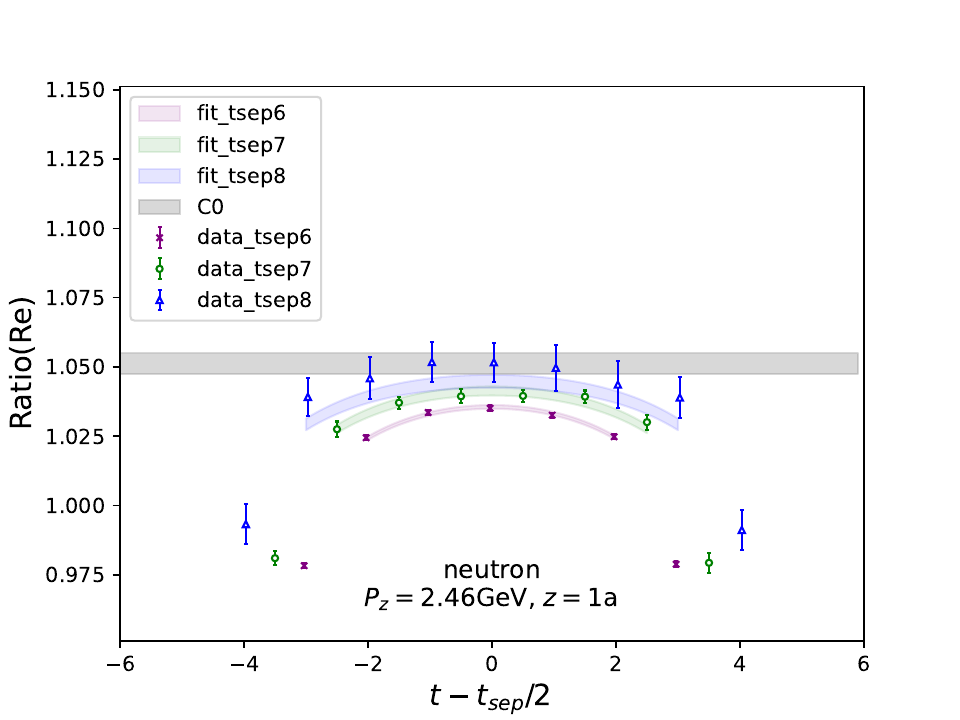}\\
\includegraphics[width=.3\textwidth]{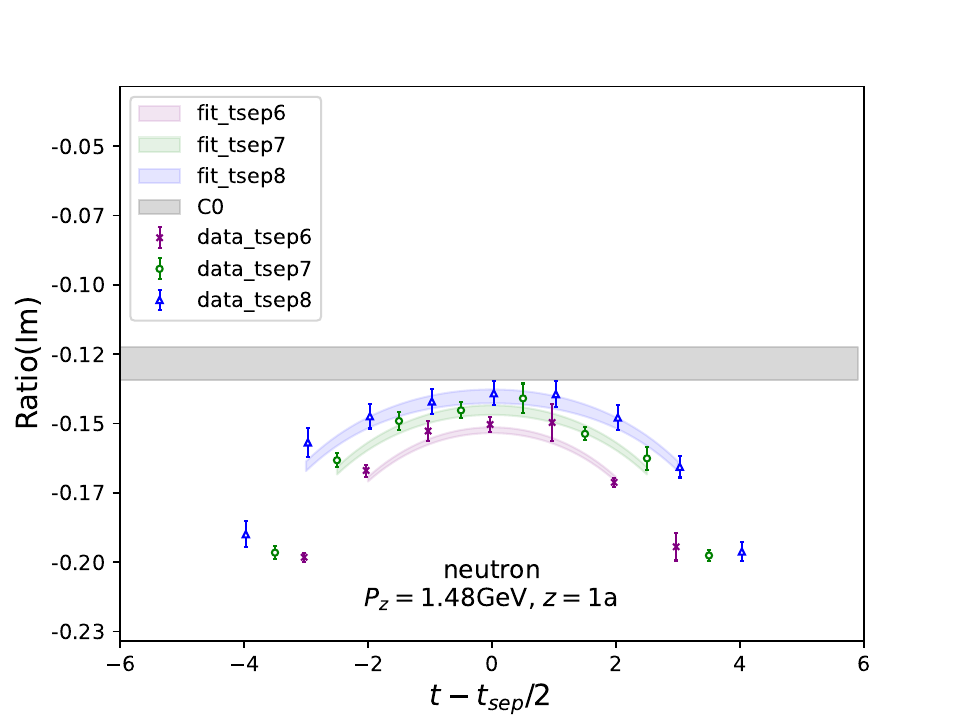}
\includegraphics[width=.3\textwidth]{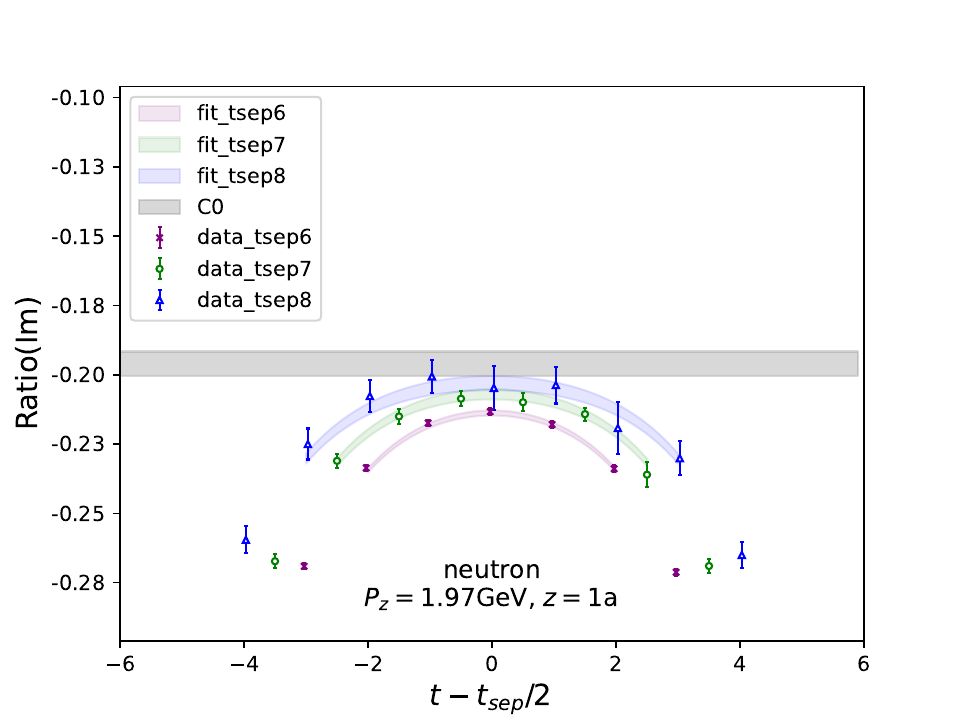}
\includegraphics[width=.3\textwidth]{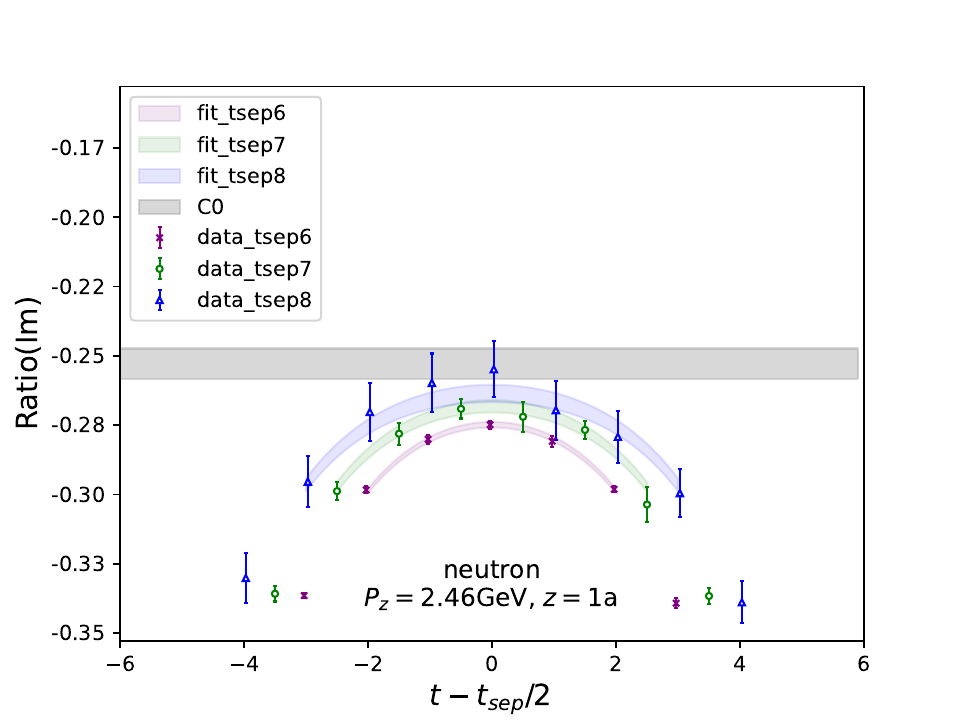}
\\
\caption{ Demonstration of    fitting the correlation function of ensemble C24P29. Here we represent the result of $z=1a$ of different momenta. }
\label{fig:fit_C24P29}
\end{figure*}

\begin{figure*}[htbp]
\includegraphics[width=.3\textwidth]{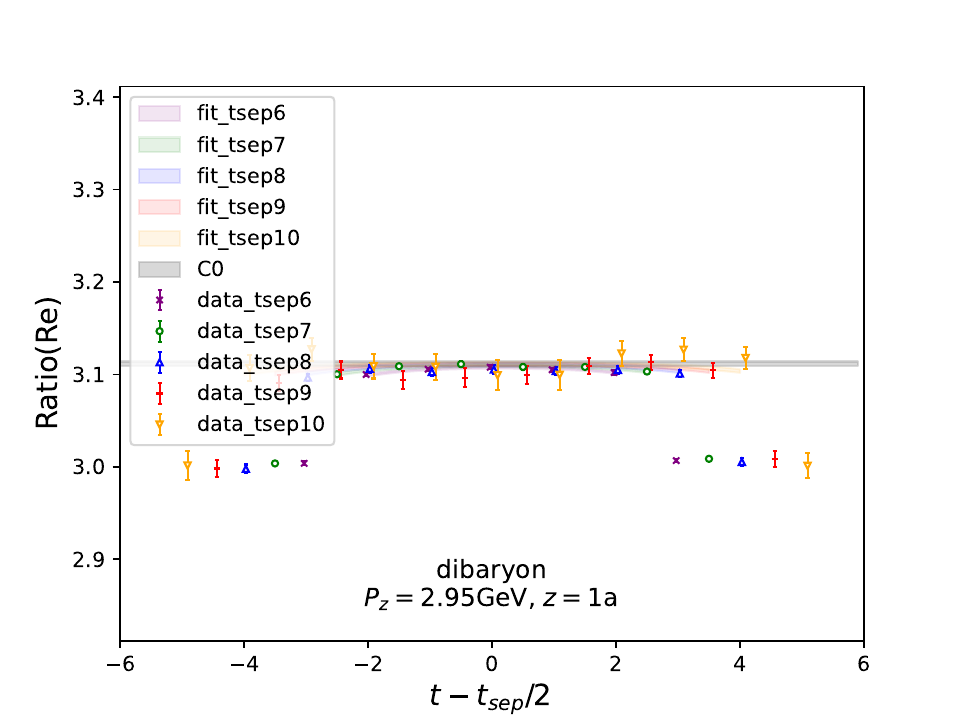}
\includegraphics[width=.3\textwidth]{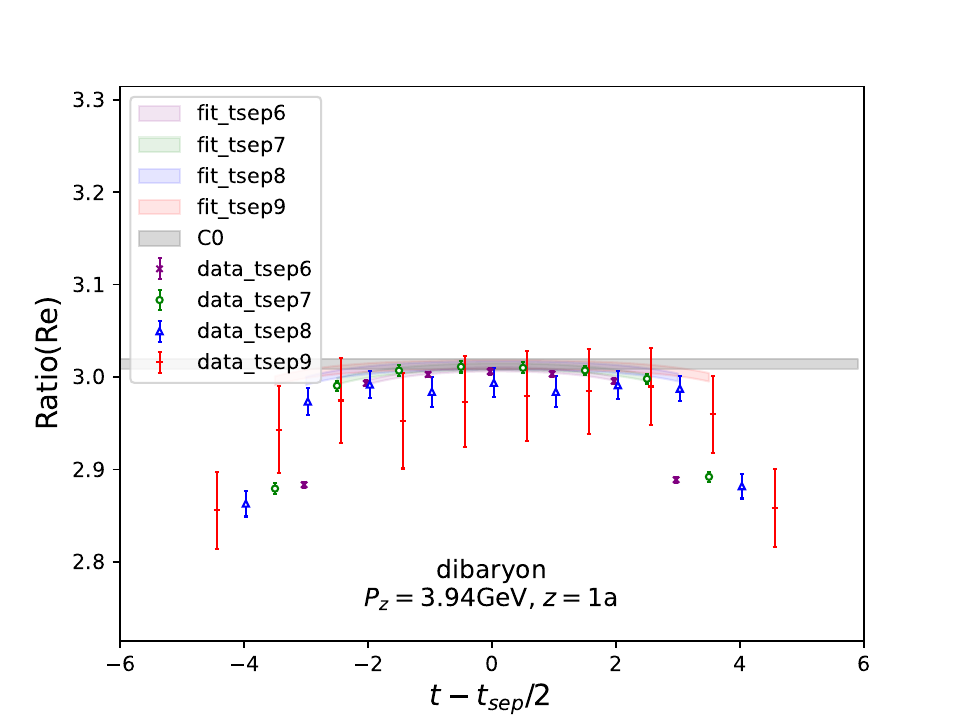}
\includegraphics[width=.3\textwidth]{figures/hybrid/P90S/fit_deu_mom5_z1_re.pdf}\\
  
\includegraphics[width=.3\textwidth]{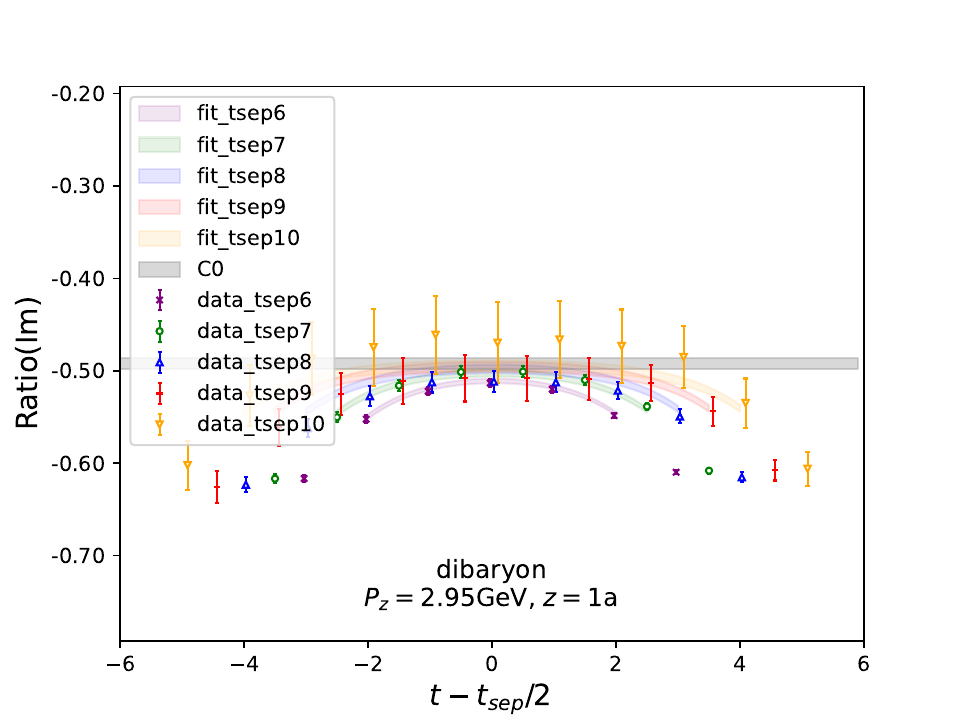}
\includegraphics[width=.3\textwidth]{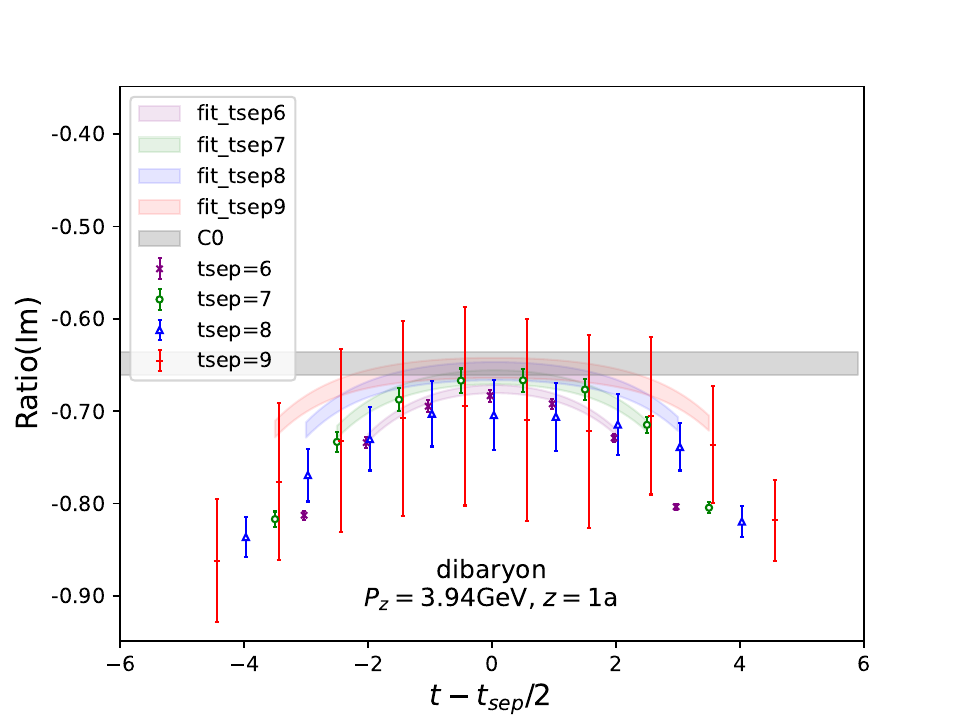}
\includegraphics[width=.3\textwidth]{figures/hybrid/P90S/fit_deu_mom5_z1_im.pdf}
\\
\includegraphics[width=.3\textwidth]{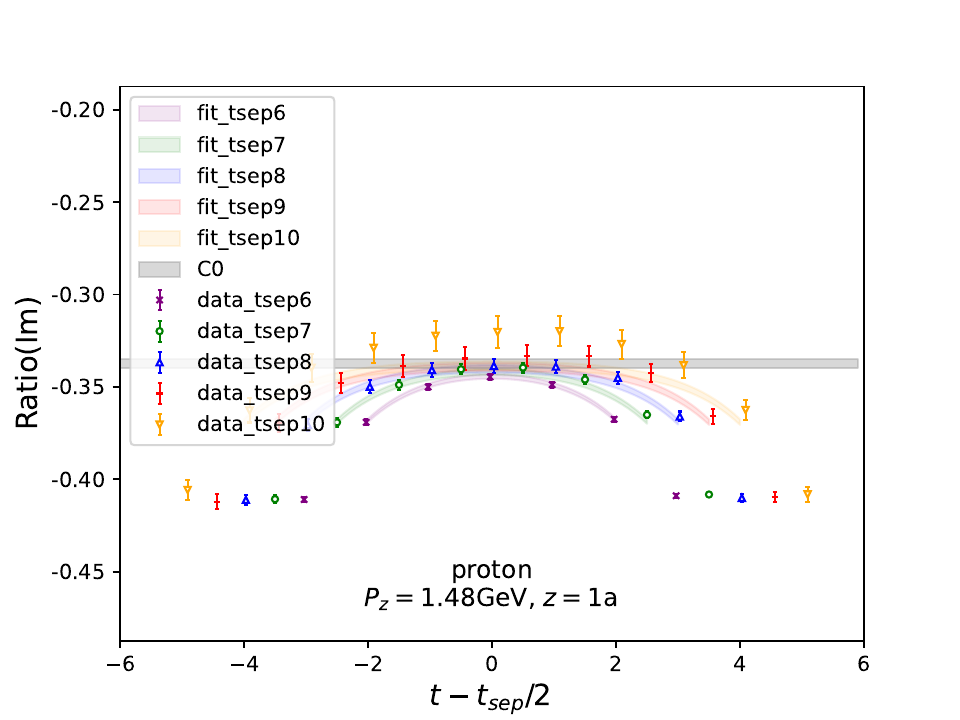}
\includegraphics[width=.3\textwidth]{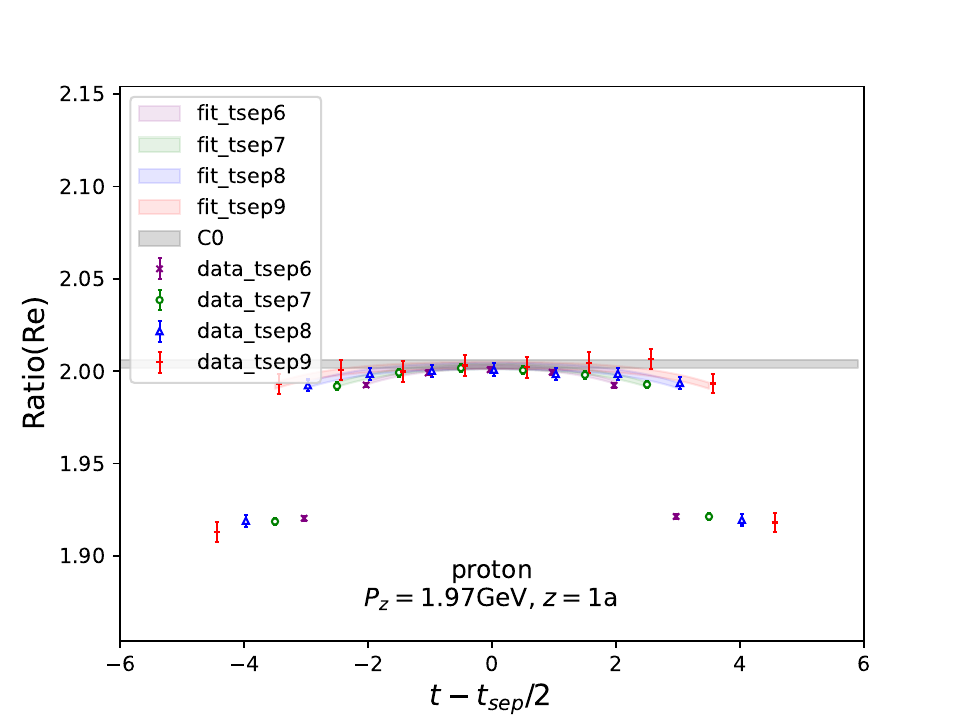}
\includegraphics[width=.3\textwidth]{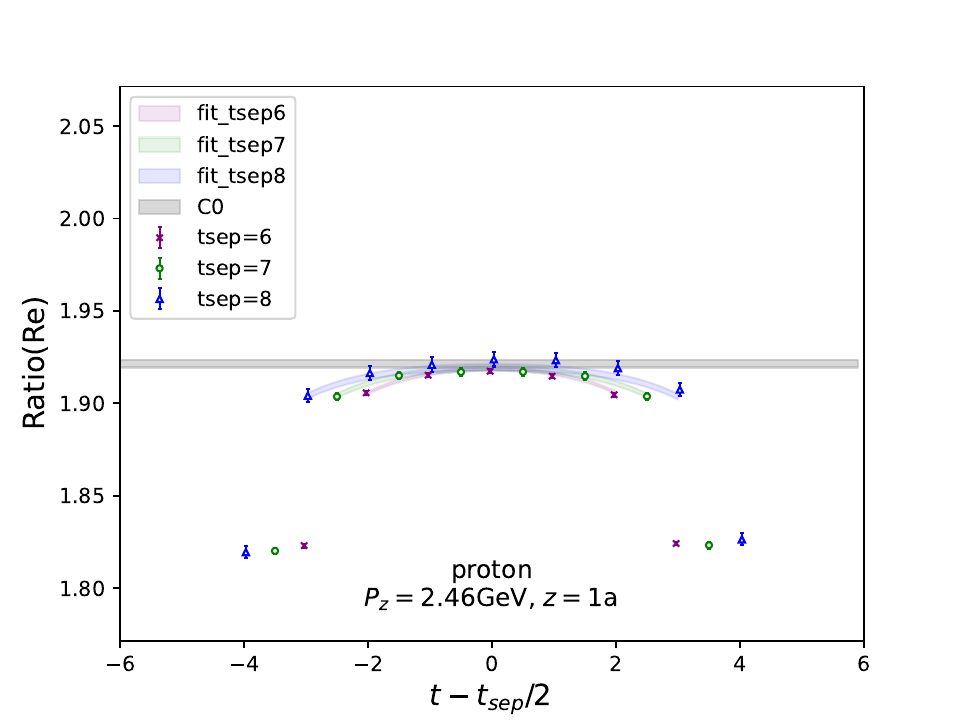}\\
\includegraphics[width=.3\textwidth]{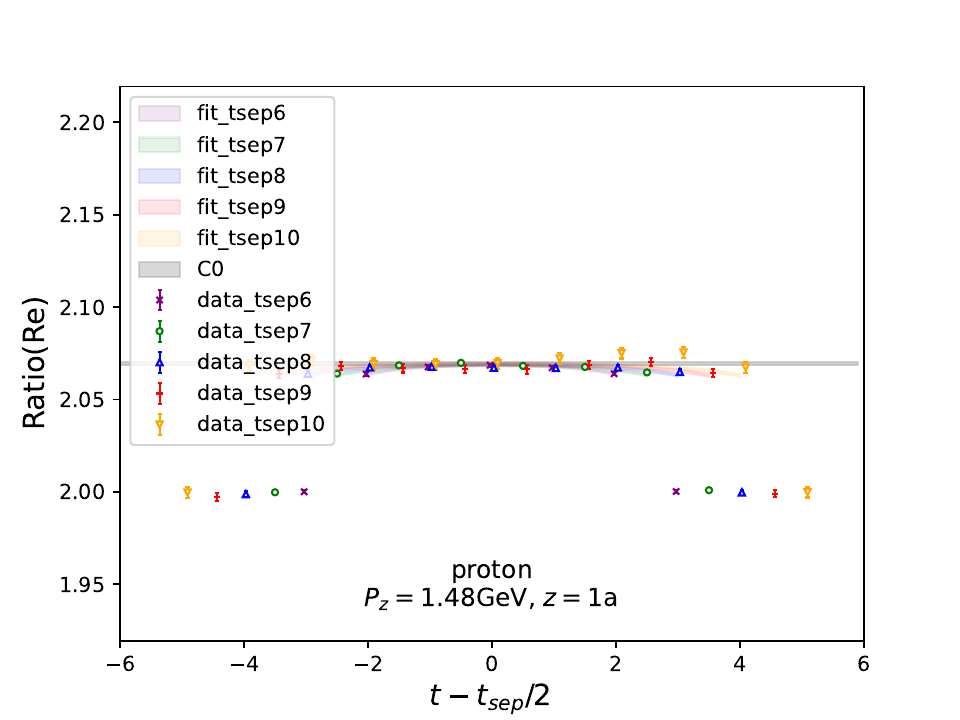}
\includegraphics[width=.3\textwidth]{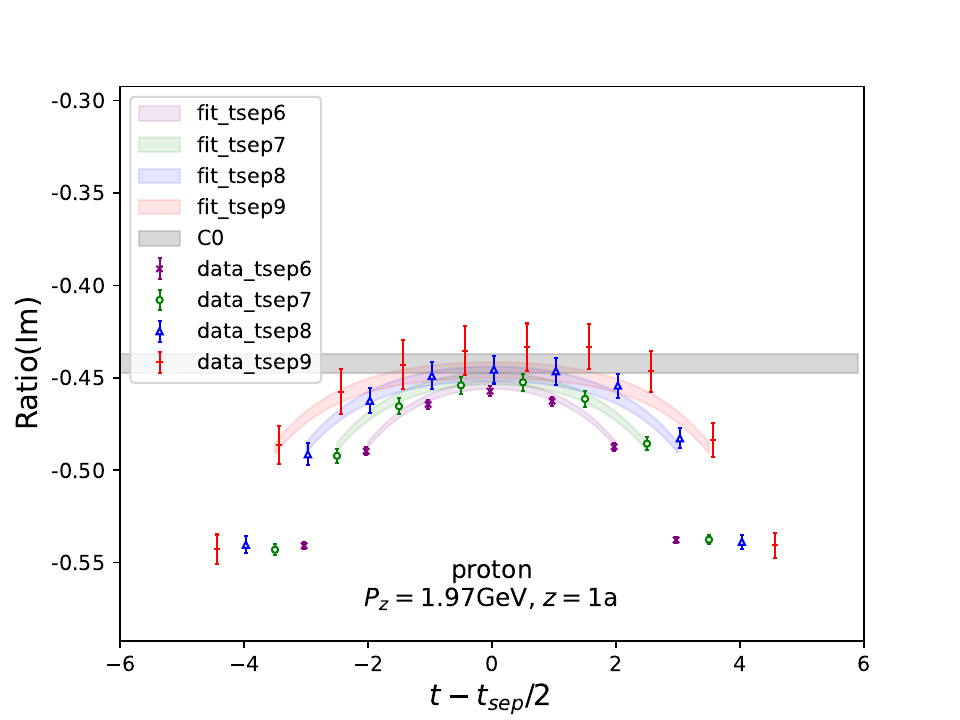}
\includegraphics[width=.3\textwidth]{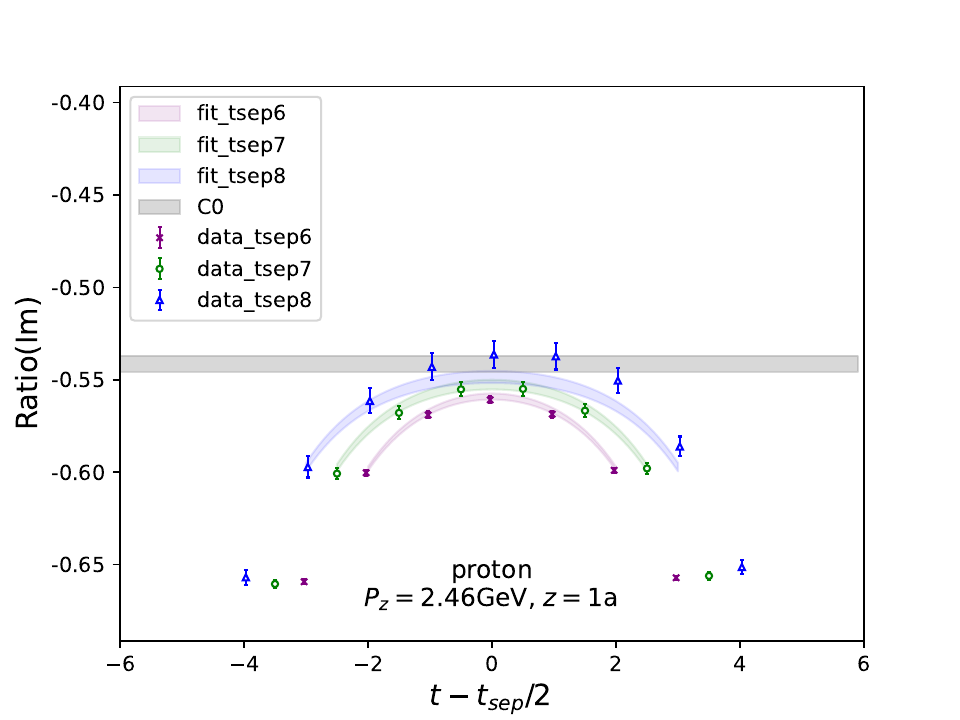}
\\
\includegraphics[width=.3\textwidth]{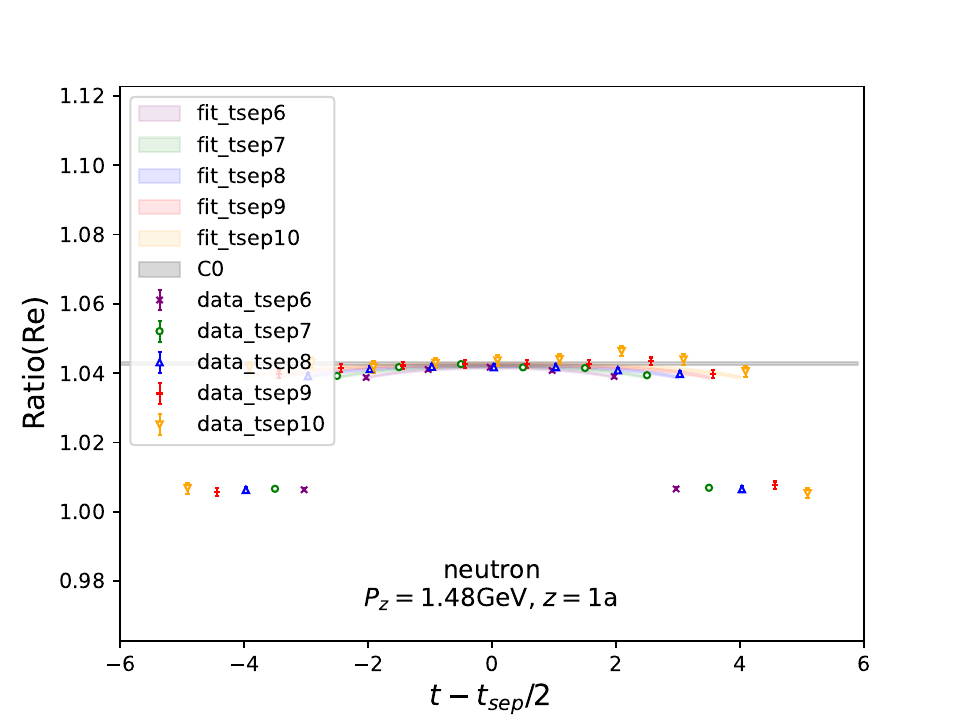}
\includegraphics[width=.3\textwidth]{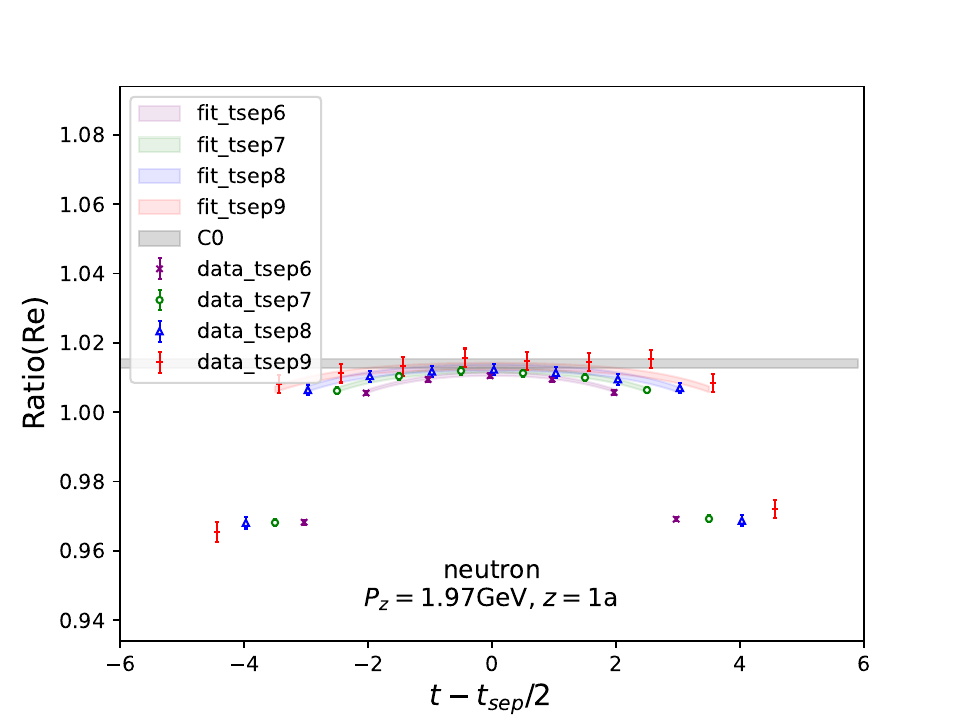}
\includegraphics[width=.3\textwidth]{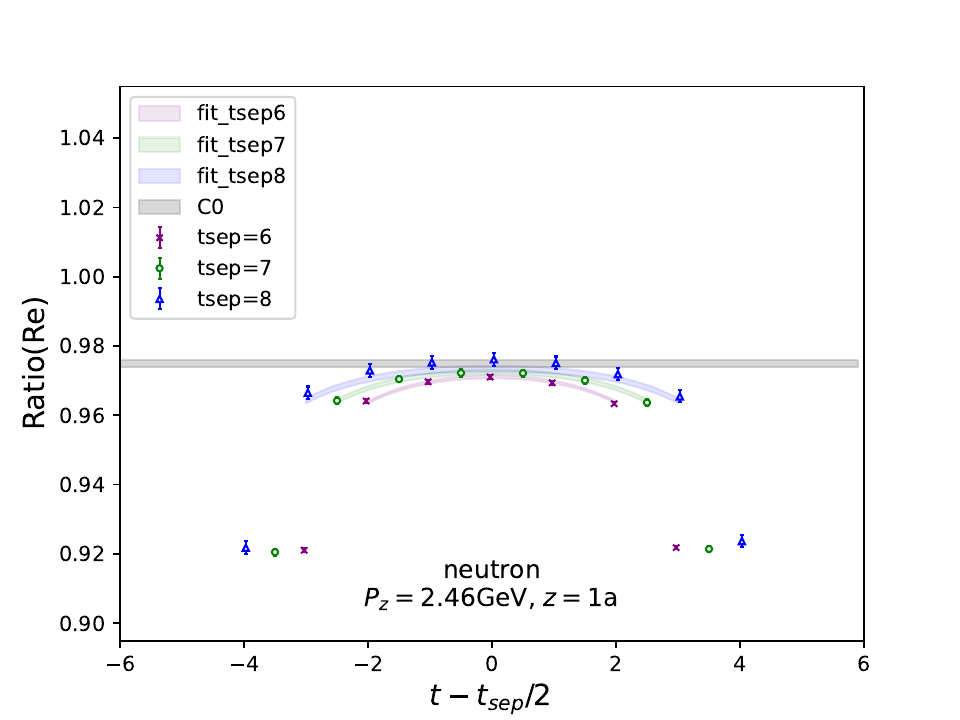}\\
\includegraphics[width=.3\textwidth]{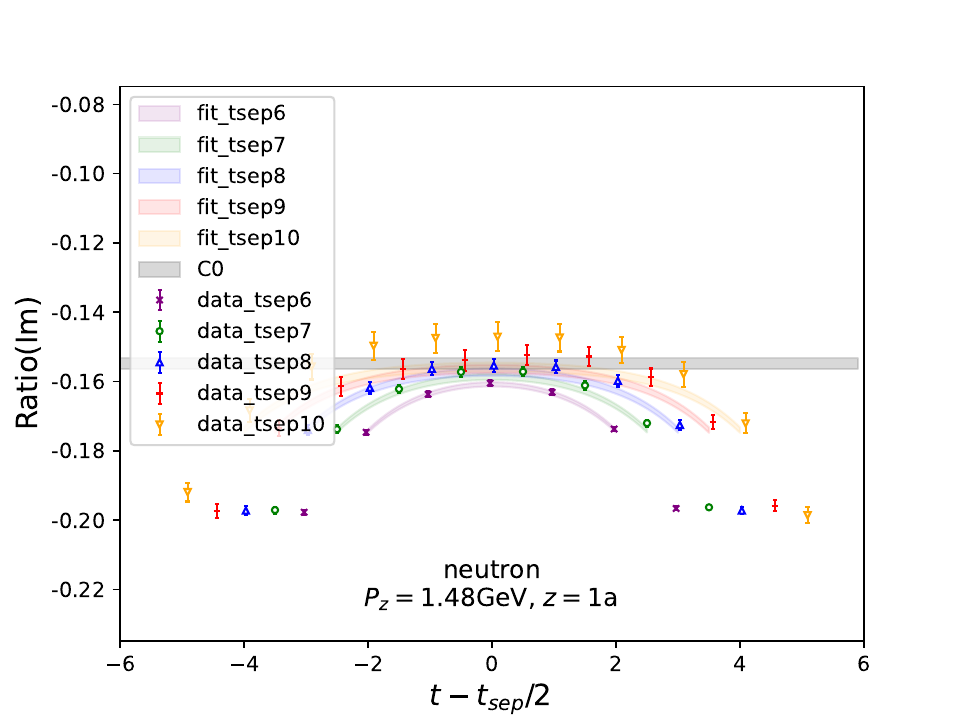}
\includegraphics[width=.3\textwidth]{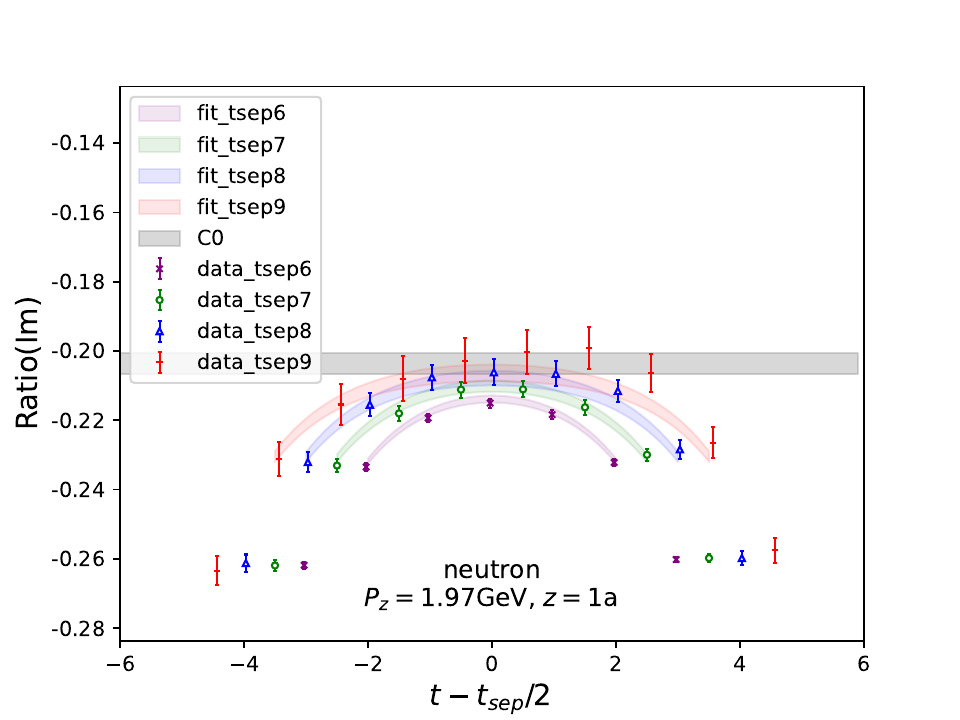}
\includegraphics[width=.3\textwidth]{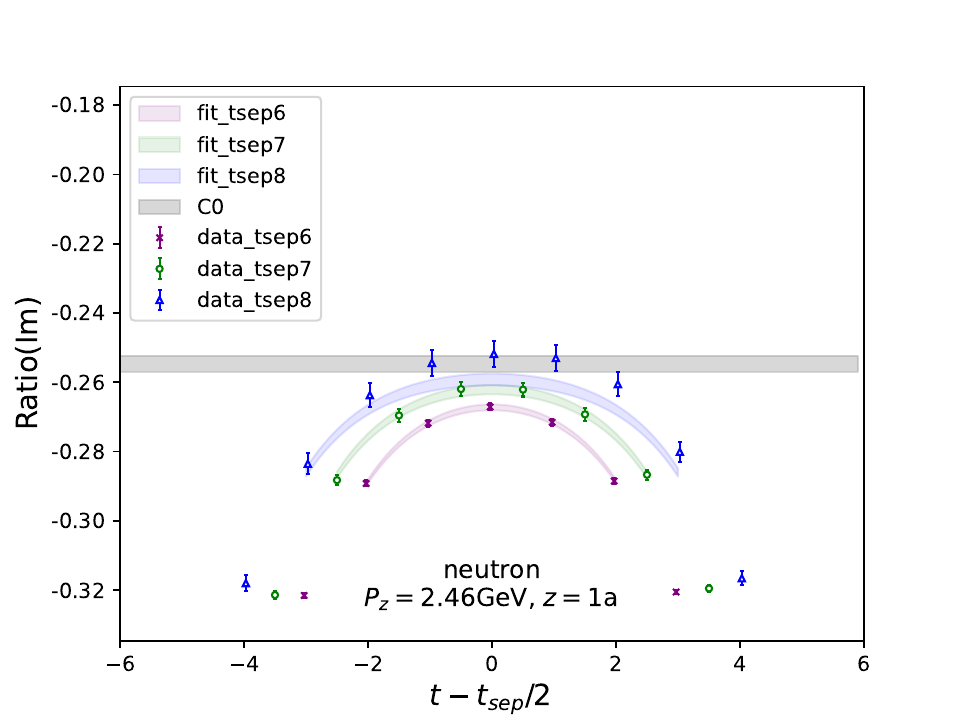}
\\
\caption{ Demonstration of    fitting the correlation function of ensemble C24P90. Here we represent the result of $z=1a$ of different momenta. }
\label{fig:fit_C24P90}
\end{figure*}

In Fig.~\ref{fig:fit_C32P29}, Fig.~\ref{fig:fit_C24P29} and 
 Fig.~\ref{fig:fit_C24P90} we plotted the ratio of 3pt to 2pt, comparing the fitting results with original data. $z=1a$ results of dibaryon, proton and neutron under different momenta are illustrated in the plots.

\begin{figure*}[htbp]
\includegraphics[width=.31\textwidth]{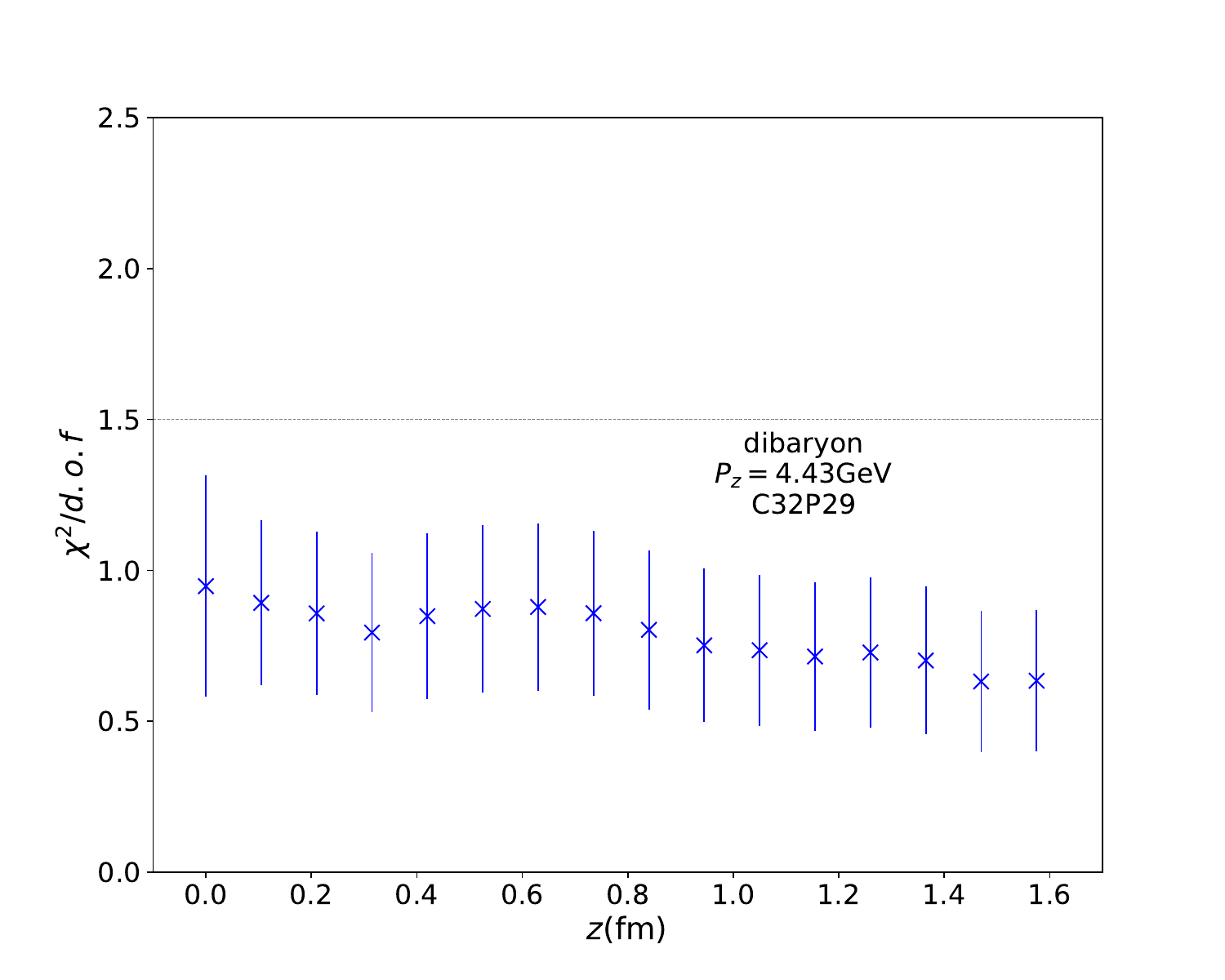}
\includegraphics[width=.31\textwidth]{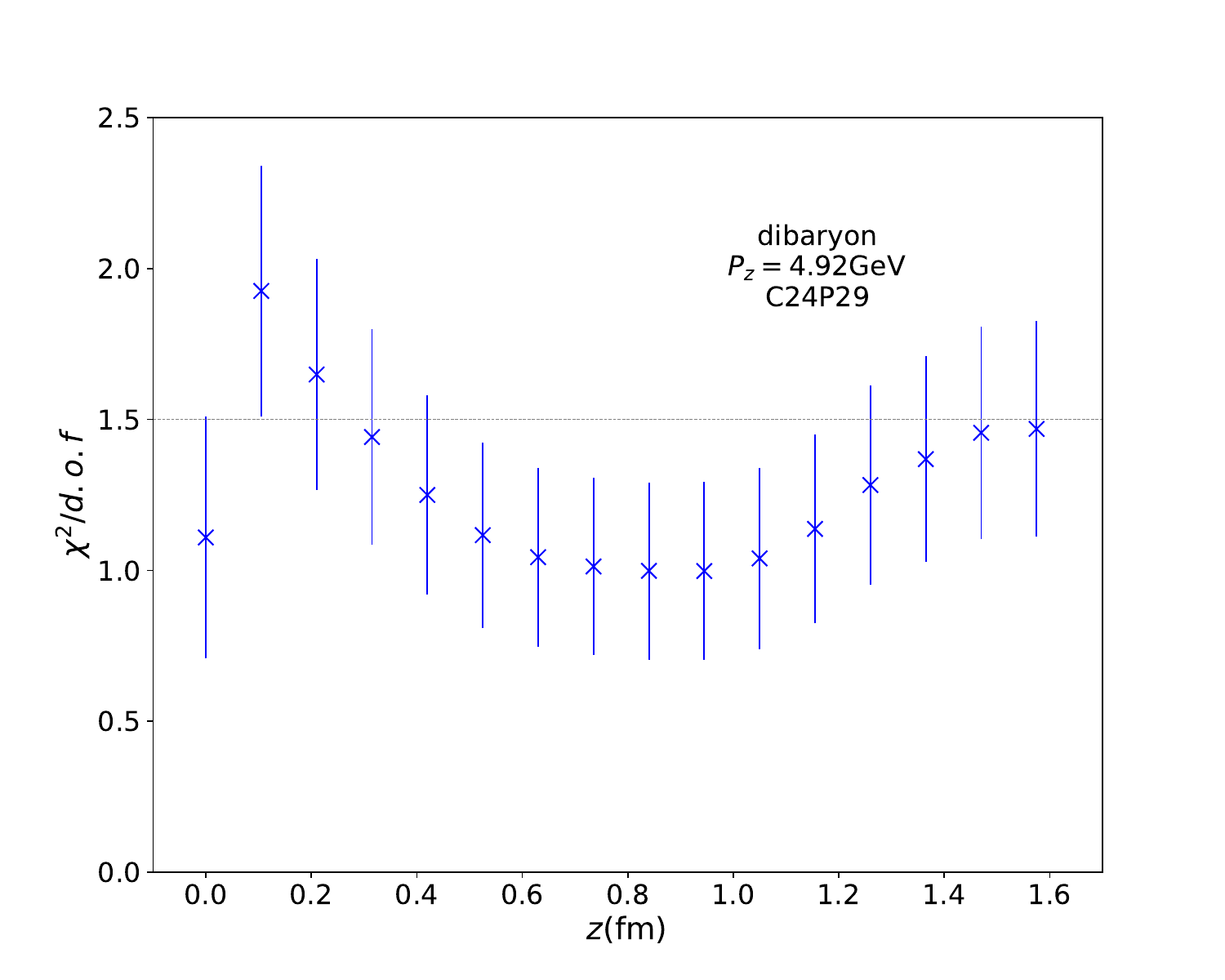}
\includegraphics[width=.31\textwidth]{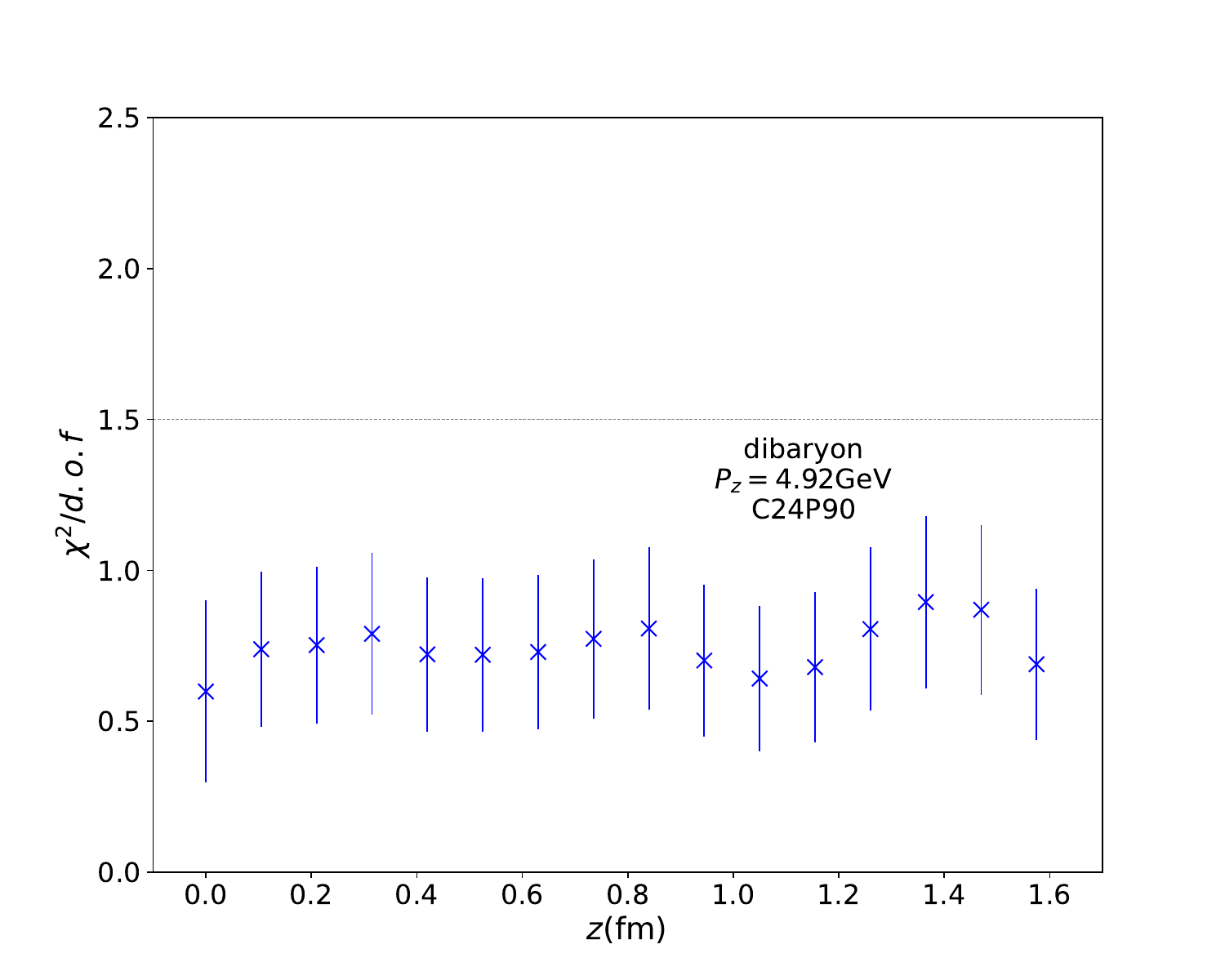}
\caption{ $\chi^2$/d.o.f. of correlated fits for  $c_0$, taking the largest momentum of the dibaryon  system for each ensemble as an example.}
\label{fig:chi2}
\end{figure*}

In Fig.~\ref{fig:chi2} we present the value of $\chi^2$/d.o.f. of correlated fits for $c_0$. We take the largest-momentum result  of the dibaryon system depending on different $z$ as an example.  Most of the $\chi^2$/d.o.f. are beneath 1.5, which illustrates the high quality of the fits.

\subsection{Renormalization group resummation and leading renormalon resummation}

\label{sec:LRR+RGR}
To determine the scheme dependence $m_0$, the method is the comparison between the renormalized matrix elements and the fixed-order Wilson coefficient shown in Eq.~(\ref{Zms}) at a short distance. In principle, we can resum the large logarithmic terms $\alpha_s^n(\mu)\text{ln}^n(z^2\mu^2)$ in $C_{0,\text{NLO}}$ to reduce the scale dependence.  Then the fitting parameters $m_0$ introduce small deviations from the fixed-order process, changing the $z$-dependence.
Here, the well-known leading infrared renormalon (IRR) arises from the long-distance contributions to the self-energy of the Wilson line, and the asymptotic series is $C(\alpha_s)=\sum_i p_{i}\alpha_s^i$. The divergence behavior for $c_{i}$ is the same as the quark ``pole" mass $m=\sum_n r_n \alpha_s^{n+1}$ that is well investigated~\cite{Bali:2013pla}.
Therefore, the strategy for IRR here is considering LRR proposed in Ref.~\cite{Zhang:2023bxs} by using state-of-the-art knowledge on the mass renormalon,
\begin{align}
r_n&\left(z,\mu\right)=N_m\,|z\mu|\left(\frac{\beta_0}{2\pi}\right)^n\frac{\Gamma(n+1+b)}{\Gamma(1+b)}\notag\\
&\times \left(1+\frac{b}{n+b}p_1+\frac{b(b-1)}{(n+b)(n+b-1)}p_2+...  \right)
\end{align}
where the parameters $b$, $p_1$, $p_2$ and $N_m$ are given in Ref.~\cite{Zhang:2023bxs}. One can obtain the leading renormalon contribution through a Borel transformation,
\begin{align}
\label{modifiedknl}
C_{0,\text{LRR}}^{\text{PV}}&\left(z,\mu\right)=N_m\,|z\mu|\frac{4\pi}{\beta_0}\int_{0,PV}^{\infty}du e^{-\frac{4\pi u}{\alpha_s(\mu) \beta_0}}\notag\\
&\times \frac{1}{(1-2u)^{1+b}}\left(1+p_1(1-2u)+p_2(1-2u)^2+... \right)
\end{align}
In practice, the modified Wilson coefficient in the continuum scheme is
\begin{align}\label{mdfdwlsncffcnt}
C_{0,\text{LRR}}&\left(z,\mu\right)=C_{0,\text{NLO}}\left( z,\mu\right)\notag\\
&+ \left[C_{0,\text{LRR}}^{\text{PV}}(z,\mu)-\alpha_s(\mu) r_0(z,\mu)\right].
\end{align}
where the last term $\alpha_s(\mu) r_0(z,\mu)$ is the NLO expansion of $C_{0,\text{LRR}}^{\text{PV}}(z,\mu)$. Now,  the result contains the fixed-order contribution and the higher-order leading renormalon contribution. In this work, we resum the leading divergent contributions to all orders at $\mu=z^{-1}$ performing RGR ~\cite{Su:2022fiu}.

\begin{figure}[htbp]
\includegraphics[width=.45\textwidth]{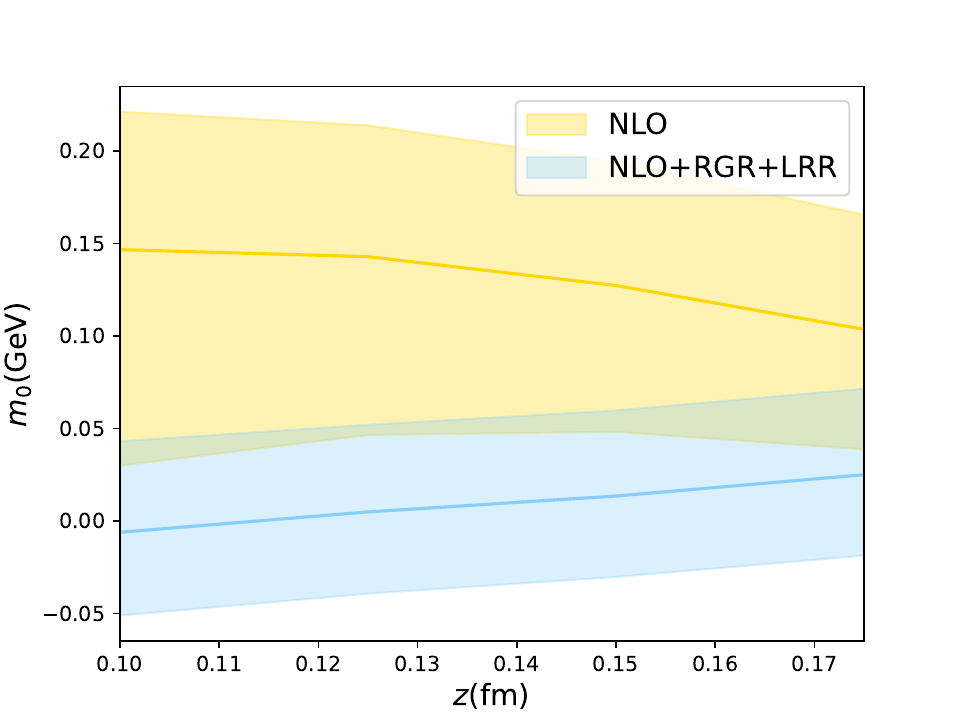}
\caption{$m_0$ determined by using fixed-order (NLO) and modified (NLO+LRR+RGR) Wilson coefficient as a function of $z$.  The main values are obtained at $\mu=\sqrt{10}$ GeV. The bands contain both statistical and systematic uncertainties with the latter determined by varying $\mu$ to 2 and 4 GeV.}
\label{fig:plt_m0}
\end{figure}

In this way, the $m_0$ determined using Eq.~(\ref{modifiedknl}) are shown in Fig.~\ref{fig:plt_m0} as a function of $z$. 
The main values are obtained at $ \mu=\sqrt{10}$ GeV by doing global fit.  Both statistical and systematic uncertainties are taken into account. The statistical uncertainties remain relatively consistent, regardless of whether LRR and RGR are included.   The systematic uncertainties are estimated by varying $\mu$ to $2$ and $4$ GeV.  As illustrated in the figure, including the leading renormalon reduces the scale dependence of $m_0$. 
By including LRR and RGR, we improve twist-three power accuracy and high-order correction for describing the matrix element in the rest frame, further enhancing the accuracy of the renormalization factor extraction.


In our process, after the renormalization and Fourier transformation, we can
extract the unpolarized PDF by applying
perturbative matching. It is worth noting that we should add the term $\Delta{C}_{LRR}$ provided in Ref.~\cite{Zhang:2023bxs} to the matching kernel $C_{\text{hybrid}}$ for LRR correction, and also perform the RGR to reduce the scale dependence as done in Ref.~\cite{Su:2022fiu}.

\begin{figure*}[htbp]
\includegraphics[width=.31\textwidth]{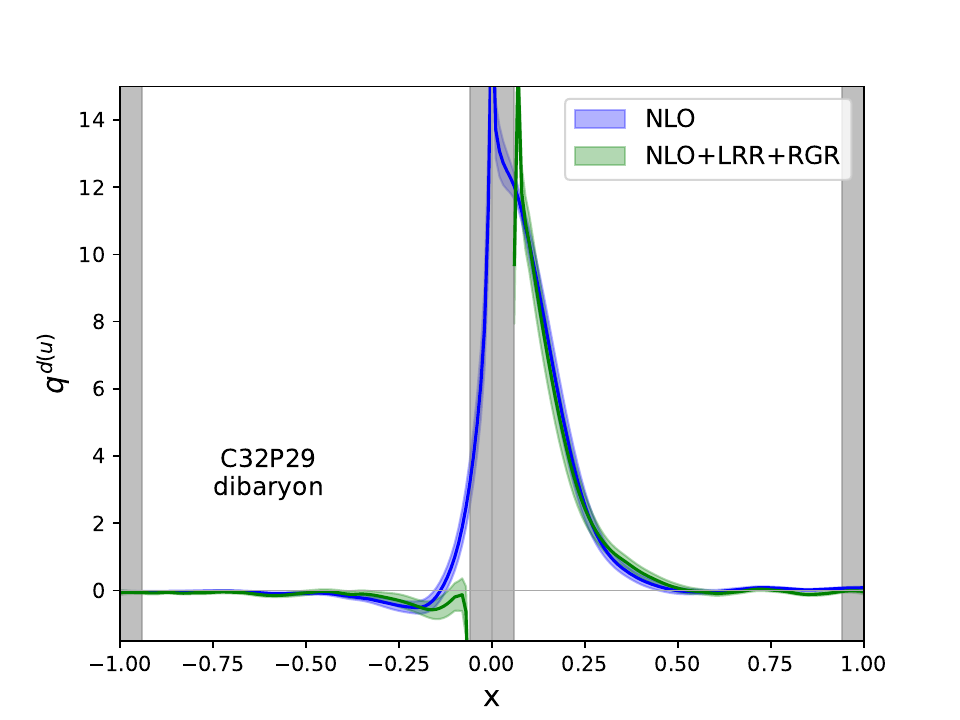}
\includegraphics[width=.31\textwidth]{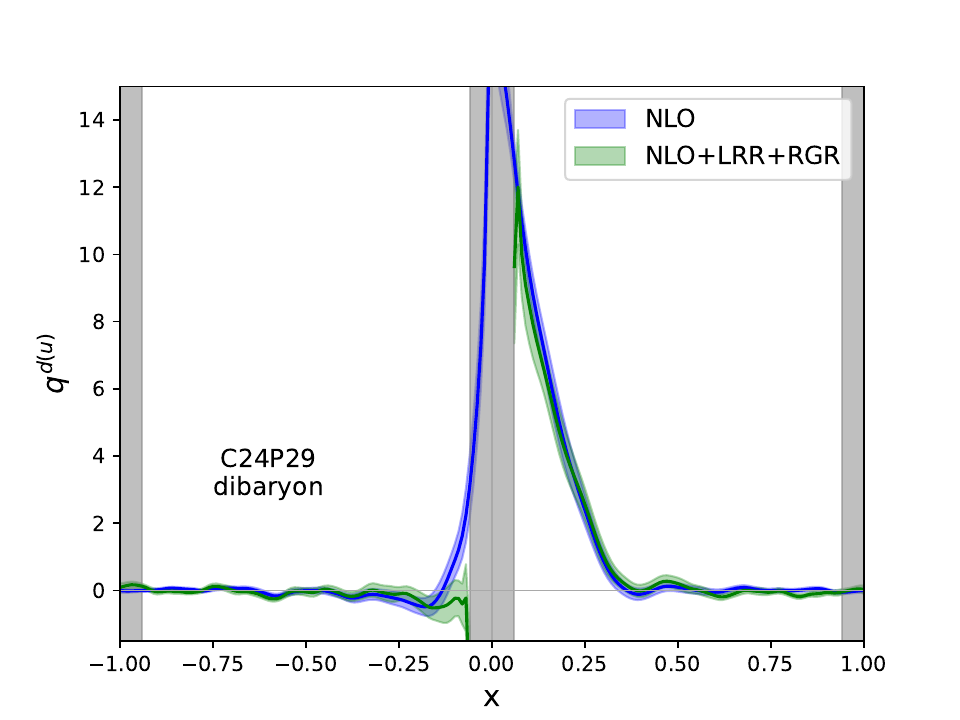}
\includegraphics[width=.31\textwidth]{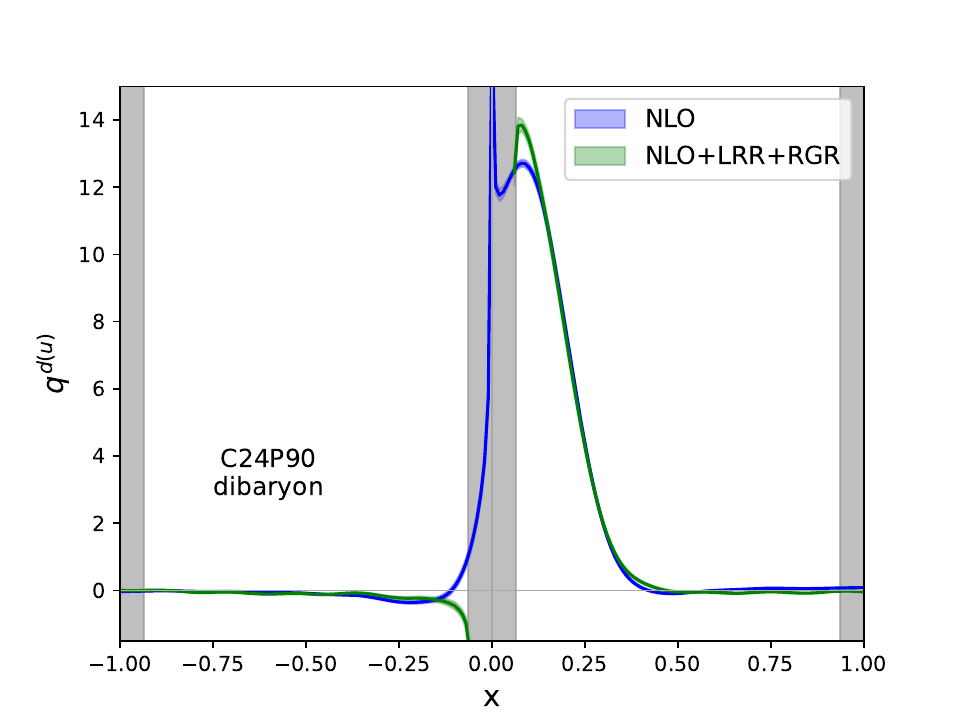}
\caption{ Comparison between fixed-order matching results of the dibaryon PDF  (blue bands) and that incorporating the LRR+RGR (green bands). The LRR+RGR results blow up at the small $x$ region ( $-0.065<x<0.065$), indicating that LaMET is unreliable at this region. Only statistical uncertainties are considered in the plots. }
\label{fig:extraZR_NLO+LRR+RGR}
\end{figure*}

In Fig.~\ref{fig:extraZR_NLO+LRR+RGR}, we present the light-cone PDF of the dibaryon system after applying  LRR and RGR, with only statistical
uncertainties taken into consideration. As
can be seen from the figure, the LRR+RGR-improved unpolarized
PDF blows up in the small x region ,
 indicating that LaMET is unreliable in this region. In the reliable region, the NLO+LGR+RGR result shows good agreement with fixed-order matching.

